\title{\bf Models of Neutrino Mass, Mixing and CP Violation\footnote{Invited Topical Review for Journal of Physics G: Nuclear and Particle Physics}}

\author{ 
  {Stephen F.~King\footnote{\tt king@soton.ac.uk}},
\\
  {\small \it Physics and Astronomy, University of Southampton, Southampton, SO17 1BJ, U.K.}\\
       }

\date{}
\documentclass[11pt]{article}

\usepackage[margin=2cm]{geometry}
\usepackage{color}
\usepackage{graphicx}
\usepackage{amssymb}
\usepackage{amsmath} 
\usepackage{pifont} 
\usepackage{hyperref}
\usepackage[noadjust]{cite}
\usepackage[center,footnotesize,hang]{subfigure}
\usepackage{multirow}

\usepackage{braket}

\usepackage{mathtools}
\usepackage{amssymb}
\usepackage{amsfonts}
\usepackage[footnotesize]{caption}
\usepackage{color}
\usepackage{braket}
\usepackage{cite}
\usepackage{hyperref}
\usepackage{url}
\usepackage{multirow}
\usepackage{relsize}
\usepackage{fullpage}
\usepackage{makecell}
\usepackage{fullpage}

\newcommand{\overbar}[1]{\mkern 1.5mu\overline{\mkern-1.5mu#1\mkern-1.5mu}\mkern 1.5mu}
\newcommand{\pmatr}[1]{\begin{pmatrix} #1 \end{pmatrix}}

\newcommand{\CP}{$\mathcal{CP}$\,}

\newcommand{\Tr}{\mathrm{Tr}\,}

\newcommand{\phiatm}{\phi_{\mathrm{atm}}}
\newcommand{\phisol}{\phi_{\mathrm{sol}}}
\newcommand{\phidec}{\phi_{\mathrm{dec}}}

\begin{document}

\maketitle

\begin{abstract} 
In this topical review we argue that 
neutrino mass and mixing data motivates extending the Standard Model to include a non-Abelian discrete flavour symmetry
in order to accurately predict the large leptonic mixing angles and \CP violation.
We begin with an overview of the Standard Model puzzles, followed by a description of some classic lepton mixing patterns. Lepton mixing may be regarded as a deviation from tri-bimaximal mixing,
with charged lepton corrections leading to solar mixing sum rules,
or tri-maximal lepton mixing leading to atmospheric mixing rules.
We survey neutrino mass models, using a roadmap based on 
the open questions in neutrino physics.
We then focus on the seesaw mechanism with right-handed neutrinos, 
where sequential dominance (SD) can account for large lepton mixing angles
and \CP violation, with precise predictions emerging from 
constrained SD (CSD).
We define the flavour problem and discuss progress towards a theory of favour
using GUTs and discrete family symmetry. 
We classify models
as direct, semidirect or indirect, according to the 
relation between the Klein symmetry of the mass matrices and the discrete family symmetry,
in all cases focussing on spontaneous \CP violation.    
Finally we give two examples of realistic and highly predictive indirect models with CSD,
namely an A to Z of flavour with Pati-Salam and a fairly complete 
$A_4\times SU(5)$ SUSY GUT of flavour, where both models 
have interesting implications for leptogenesis. 
 \end{abstract}

\begin{center}
{\it Dedicated to Guido Altarelli}
\end{center}

\tableofcontents

\section{Introduction}

The Nobel Prize in Physics for 2015 has just been awarded to 
Takaaki Kajita (Super-Kamiokande Collaboration, University of Tokyo, Japan) and to Arthur B. McDonald (Sudbury Neutrino Observatory Collaboration, SNO, Canada) ``for the discovery of neutrino oscillations, which shows that neutrinos have mass'' and ``for their key contributions to the experiments which demonstrated that neutrinos change identities. This metamorphosis requires that neutrinos have mass. The discovery has changed our understanding of the innermost workings of matter and can prove crucial to our view of the universe.'' In 1998 Takaaki Kajita presented to the world the discovery that neutrinos produced in the atmosphere switch between two identities on their way to Earth. Arthur McDonald subsequently led the Canadian collaboration which demonstrated that neutrinos from the Sun do not disappear on their way to Earth, but change identity by the time of arrival to the SNO detector. Since then there have been many developments in neutrino physics. This topical review will focus on the most recent developments since 2012.

The year 2012 was important in physics for two quite different reasons: the discovery of the Higgs boson 
and the measurement of the neutrino reactor angle.
While the Higgs boson discovery supports the electroweak symmetry breaking sector of the Standard Model (SM), the reactor angle marks the completion of the \CP conserving part of the leptonic mixing matrix.
While the Higgs discovery made a big splash in network TV headlines across the world, the reactor angle measurement only made a small plop in physics blogs and scientific journals, even though a year earlier neutrinos had been globally reported to {\it travel faster than light} ({\it sic}).
However truth is stranger than fiction, since although neutrinos are not superluminal,
their mass and mixing requires physics beyond the SM, making them
``ghostly beacons of new physics''~\cite{Hirsch}. By contrast, the Higgs discovery serves only to confirm
the SM, with its properties being exactly as predicted to increasing levels of accuracy.

While all physicists agree that the measured reactor angle opens up the prospect of measuring \CP violation in neutrino experiments in the forseeable future, the theoretical significance of the reactor angle discovery splits the community. Consider the early history of neutrino model building, from 1998 onwards,
summarised in the reviews
\cite{Altarelli:1999gu,King:2003jb,Mohapatra:2006gs,Strumia:2006db}. 
Following the large atmospheric and solar mixing discoveries in 2002, 
many of the models in \cite{King:2003jb} involved sequential dominance
(SD) \cite{King:1998jw,King:2002nf},
which predicts a normal neutrino mass hierarchy $m_1\ll m_2 \ll m_3$
and a large reactor angle 
$\theta_{13}\lesssim m_2/m_3$.
Models incorporating SD were constructed with $SU(3)$ family symmetry providing an explanation
of maximal atmospheric mixing via vacuum alignment \cite{King:2001uz}.
There then followed the age of tri-bimaximal mixing with an explosion of 
models involving a zero reactor angle, enforced by discrete family symmetry
\cite{Altarelli:2005yp}, as reviewed in \cite{Altarelli:2010gt,Ishimori:2010au}. 
A polar opposite approach called Anarchy was also put forward early on \cite{Hall:1999sn}, the idea being 
that lepton mixing is determined randomly as if God plays dice with the neutrino mass matrix.
According to Anarchy the reactor angle is on the same footing as the atmospheric and solar angles,
and hence was generally expected to be large.
Following the measurement of a large reactor angle in 2012,
many people have jumped to the conclusion that Anarchy is preferred to discrete family symmetry.
However, unlike that other dice-throwing theory (quantum mechanics),
Anarchy is intrinsically untestable. Instead one is led to 
rather sterile arguments about whether Anarchy is statistically better than models with
family symmetry \cite{Hirsch:2001he}. By contrast discrete family symmetry models 
are highly testable, indeed many of them were excluded by the measurement 
of the reactor angle. Moreover, the model builders have been hard at work, showing how
discrete family symmetry models could be modified to account for the observed reactor angle, as discussed in recent reviews \cite{King:2013eh,King:2014nza}. Before concluding that model builders are serial 
revisionists, it is worth remembering that SD
predicted a large reactor angle $\theta_{13}\lesssim m_2/m_3$, a decade before it was measured
\cite{King:2002nf},
although understanding why this 
bound is saturated requires further input. 
This topical review, then, summarises the
model building developments since 2012, 
including the latest progress in spontaneous \CP violation and versions of
CSD which explain why $\theta_{13}\sim m_2/m_3$,
as well as Grand Unified Theories (GUTs) with discrete flavour symmetry which incorporate these ideas.

We have mentioned that the discovery of 
neutrino mass and mixing in 1998, unlike the Higgs boson discovery, 
requires new physics beyond the SM. 
To understand why, it is enough to realise that the origin of neutrino mass remains unknown.
Although many types of new physics beyond the SM allow neutrino mass,
since no BSM physics has been found, many people favour adding just
right-handed (RH) neutrinos and nothing else. If we do this then immediately we face
the questions of how many RH neutrinos and what are their Majorana masses,
which may range from zero to the Planck scale? Alternatively, we may just add effective
operators with some cut-off mass scales (proposed by Weinberg) but if we do this we soon learn that 
some of these operators must be associated with a mass scale below the Planck scale,
and probably also below the scale of Grand Unification, so the SM must break down at some scale.
It is clear then that the origin of neutrino mass requires the first (and so far only) new physics beyond the SM
of Particle Physics.
What is the nature of the new physics?
Although the origin of tiny neutrino mass is unknown, it could imply
some sort of see-saw mechanism at a high mass scale, or maybe new particles associated with loop models of neutrino mass, or perhaps R-Parity violating supersymmetry (SUSY) - maybe even extra dimensions?
Any example of such new physics would have implications for the 
unification of matter, forces and flavour GUTs, and the extra 7 parameters (or 9 parameters if neutrinos
are Majorana) associated with neutrino mass and mixing 
makes the flavour problem of the SM less ignorable since any version of an extended SM must now have around 30 parameters in total, surely too many for a satisfactory theory of particle physics?
The presence of such a large number of parameters associated with flavour, as well as the phenomenon of large lepton mixing, very unlike that witnessed in the quark sector,
has stimulated attempts to address the flavour problem based on family symmetry.
Since atmospheric and solar mixing continues to display the tri-bimaximal form,
indicative of two and three-fold permutation symmetries, 
this continues to motivate the use of discrete non-Abelian family symmetry.
An important point to realise is that discrete non-Abelian family symmetry
does not imply zero reactor angle, as we discuss at length in this review.

In the realm of Cosmology, neutrinos could be responsible for our very existence, since leptogenesis is 
now the leading candidate for the origin of matter-antimatter asymmetry 
\cite{Fukugita:1986hr}
(the SM gives too small a value for
such asymmetry). Neutrino mass tends to wash out galaxy structures,
since eV neutrinos represent a hot dark matter component.
On the other hand warm dark matter, for example from keV sterile neutrinos, could be responsible for all the dark matter
in the universe~\cite{Merle:2013gea}.
More speculatively, neutrinos could play a role in inflating the universe from Planck scale size
to its present size via sneutrino inflation \cite{Murayama:1992ua}.
And it remains an intruiging possibility that neutrino mass
is somehow related to dark energy since the scales happen to be the same order of magnitude
(see e.g. \cite{Barbieri:2005gj} and references therein).

Turning to neutrino phenomenology, 
the observed pattern of neutrino masses and lepton mixing is quite remarkable.
Neutrinos have tiny masses (for all three neutrinos, much less than the electron mass) which are not very hierarchical.
Such masses break the separate lepton numbers $L_e , L_{\mu} , L_{\tau}$, but may or may not preserve the total 
lepton number $L=L_e + L_{\mu} + L_{\tau}$ (depending on whether neutrinos are Dirac or Majorana).
Neutrinos certainly mix a lot (unlike the quarks). 
As mentioned already, neutrino mass implies at least 7 new parameters as compared to the
minimal SM:
3 neutrino masses, 3 lepton mixing angles, 1 \CP violating phase.
If the 3 neutrino masses are Majorana in nature, there are 2 further \CP violating phases.
As discussed above, the origin of neutrino mass is unknown.
For example, heavy right-handed neutrino exchange could be responsible for the Weinberg operators.
Such heavy right-handed Majorana masses play a crucial role in generating the matter-antimatter asymmetry via leptogenesis \cite{Fukugita:1986hr}, mentioned above.
Alternatively, the origin of neutrino mass 
(and matter-antimatter asymmetry) could be something completely different.

\begin{figure}[t]
\centering
\includegraphics[width=0.80\textwidth]{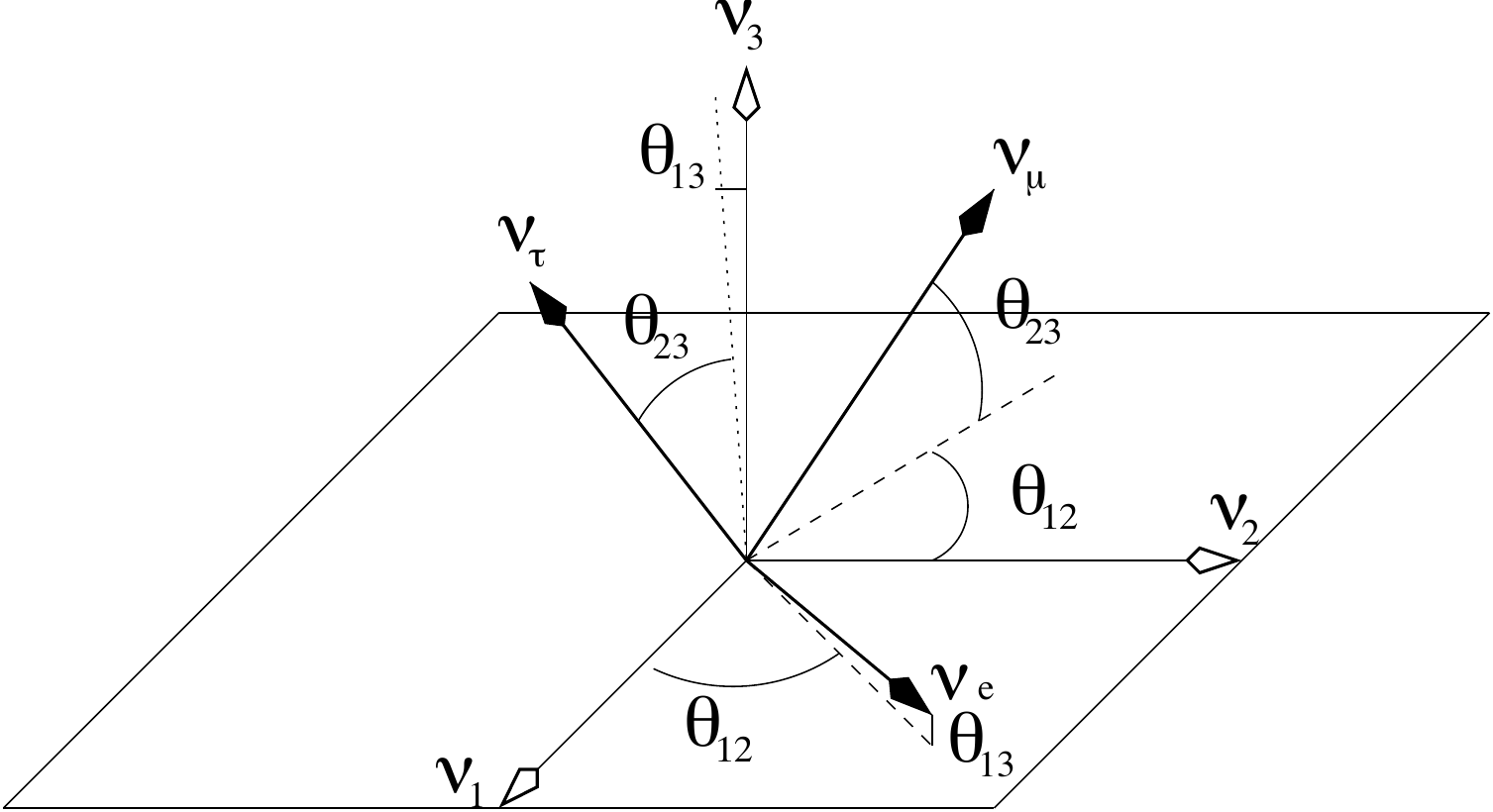}
    \caption{Neutrino mixing angles (assuming zero \CP violation) may be represented as Euler angles relating the charged lepton mass basis states $(\nu_e, \nu_{\mu}, \nu_{\tau})$ to the mass eigenstate basis states
    $(\nu_1, \nu_2, \nu_3)$.} \label{angles}
\end{figure}

\begin{figure}[htb]
\centering
\includegraphics[width=0.5\textwidth]{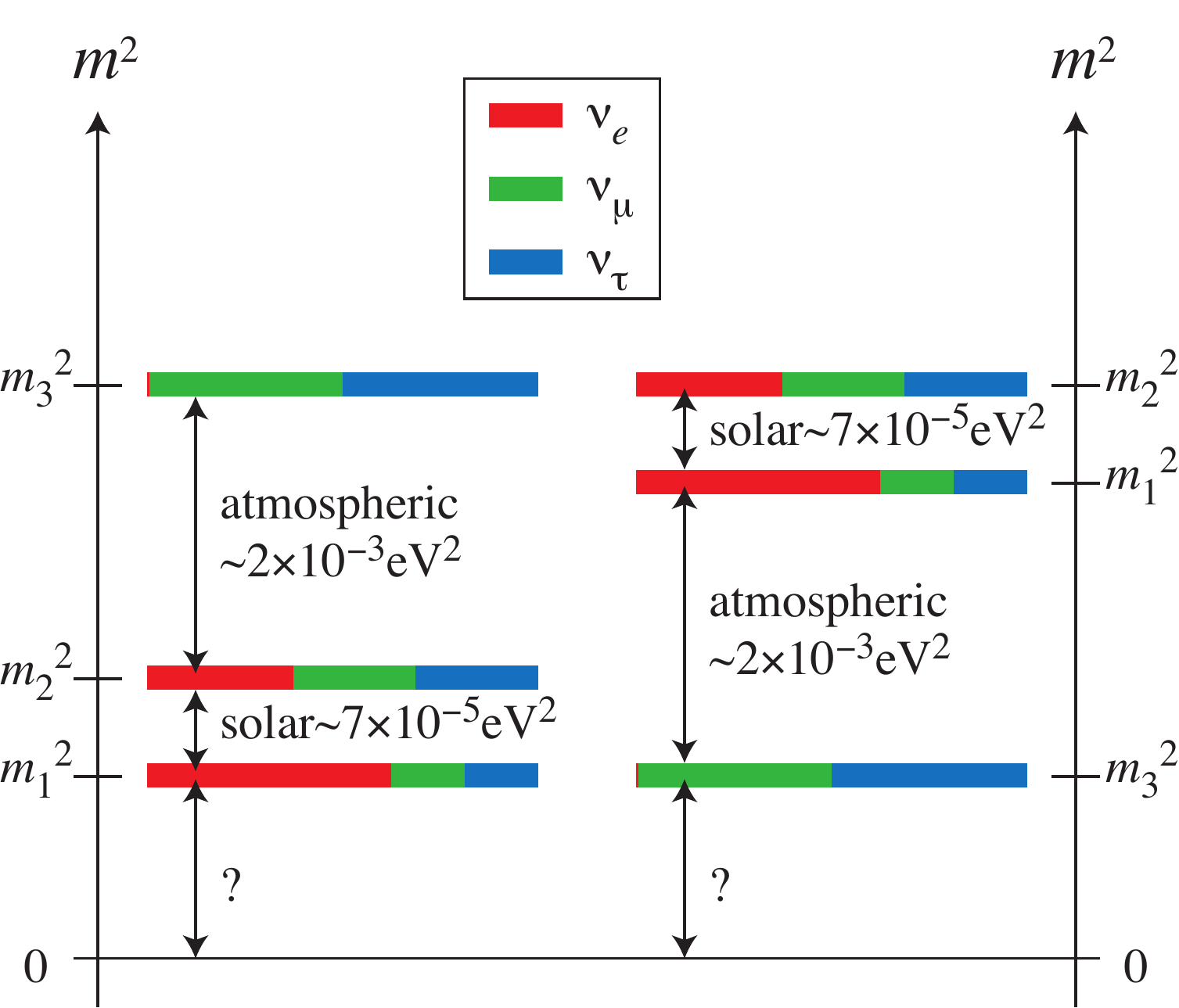}
\caption{\label{mass}\small{The probability that a particular neutrino
mass state $\nu_i$ with mass $m_i$ contains a particular charged lepton mass basis state 
$(\nu_e, \nu_{\mu}, \nu_{\tau})$ is represented by colours.
The left and right panels of the figure are 
referred to as normal or inverted mass squared ordering, respectively,
referred to as NO or IO.
The value of the lightest neutrino mass is presently unknown.}}
\end{figure}

\begin{figure}[t]
\centering
\includegraphics[width=0.30\textwidth]{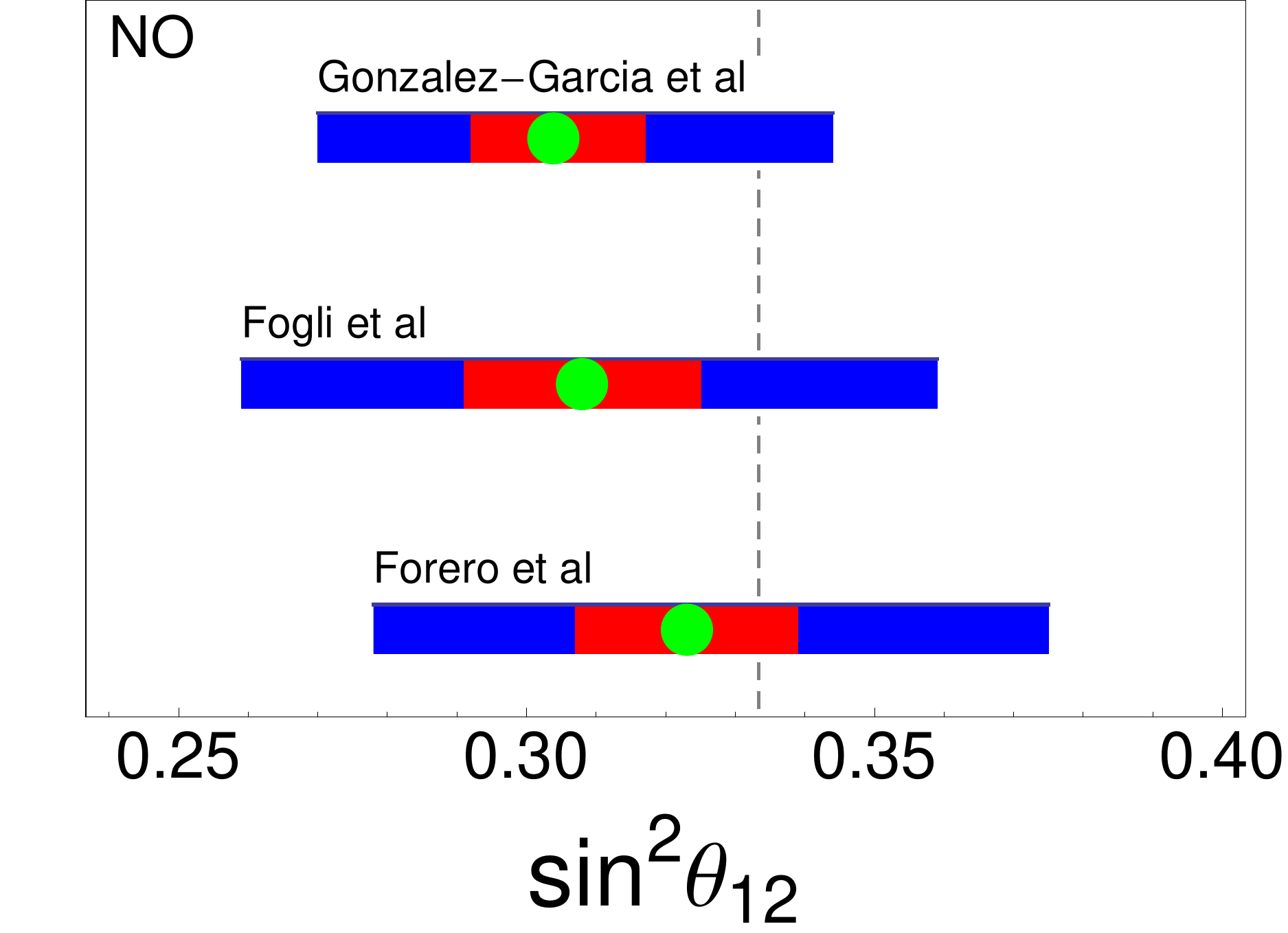}
\includegraphics[width=0.30\textwidth]{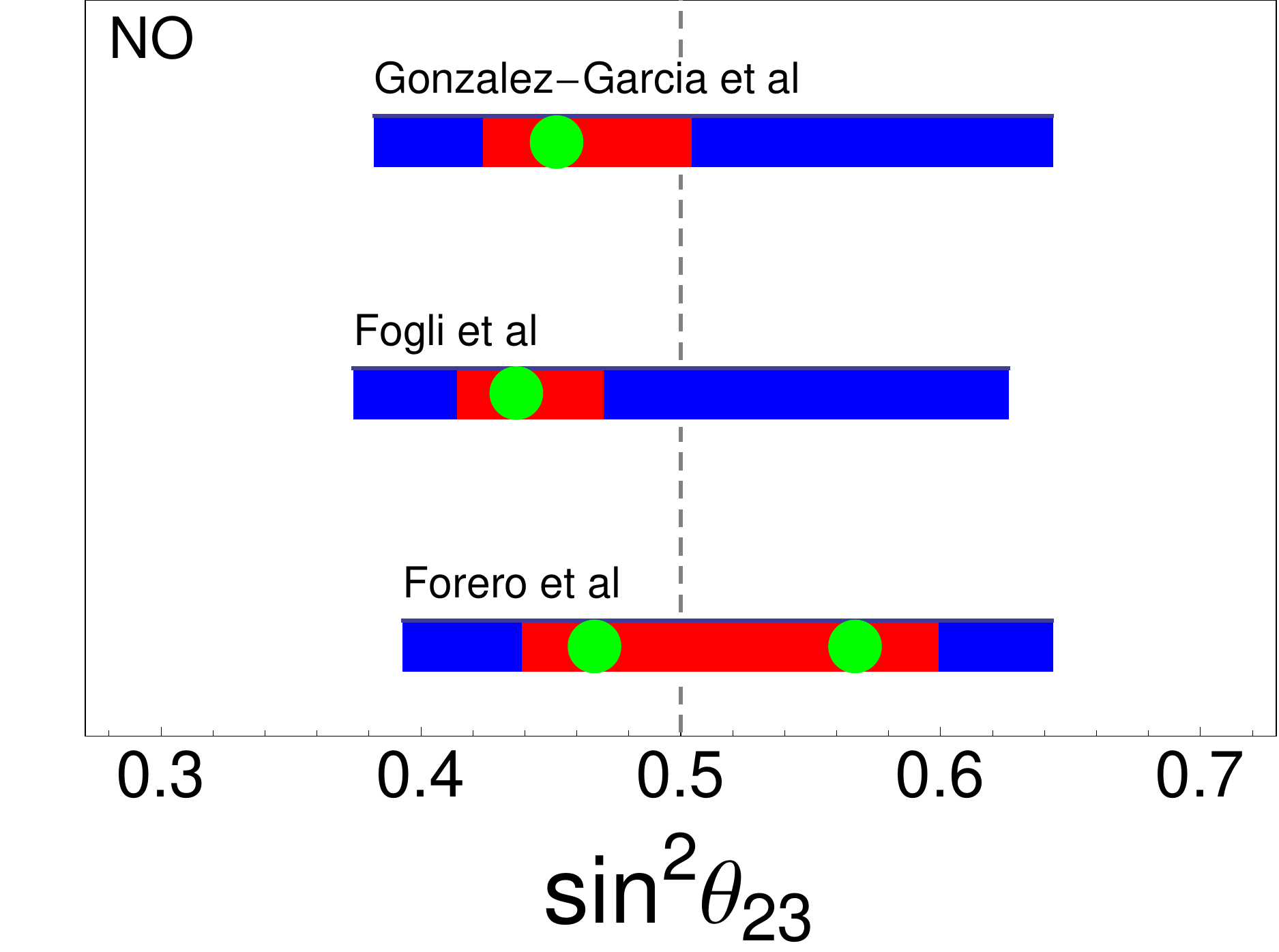}
\includegraphics[width=0.30\textwidth]{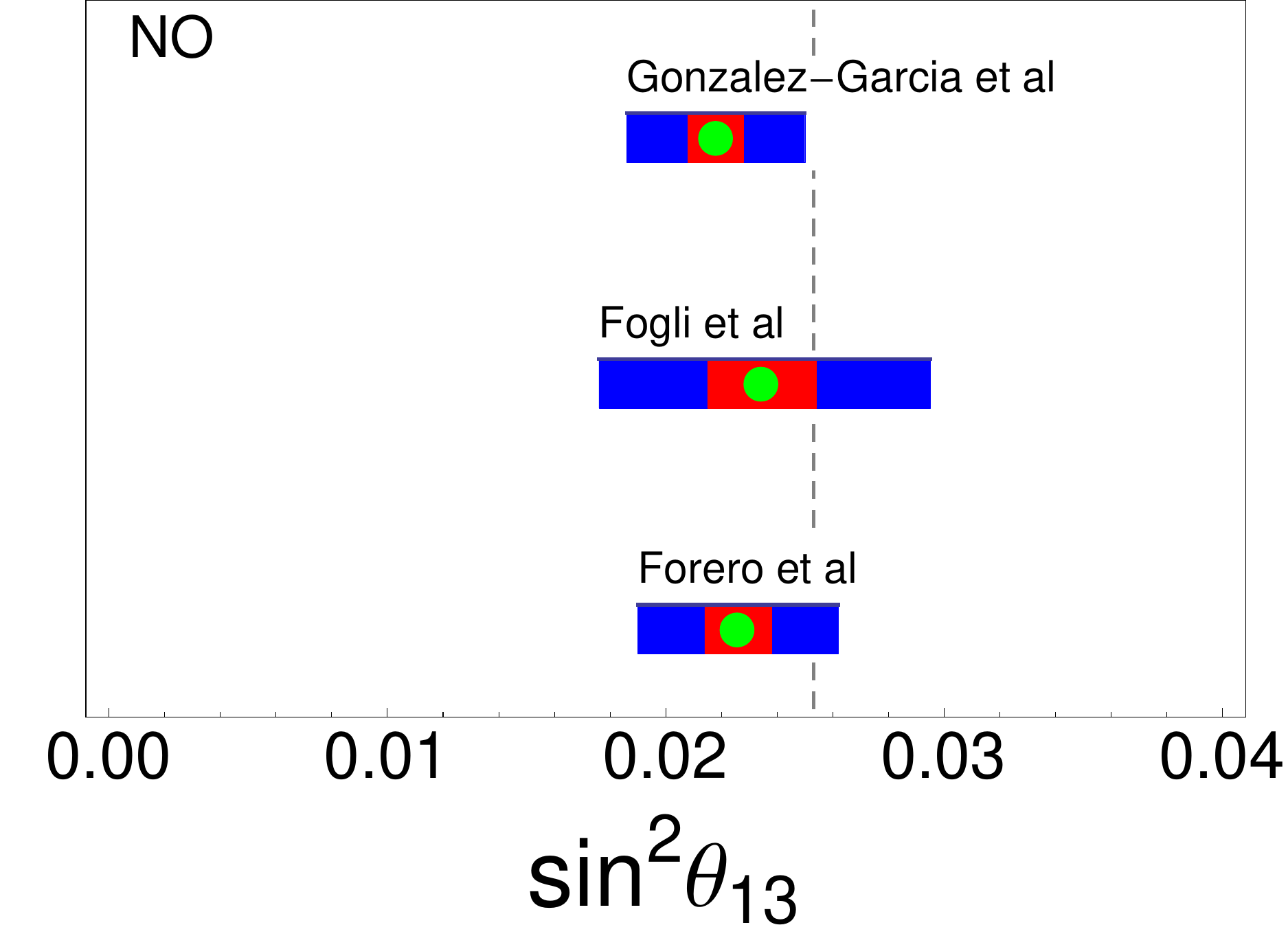}
    \caption{Global fits \cite{Gonzalez-Garcia:2014bfa,Capozzi:2013csa,Forero:2014bxa}
    to the lepton mixing angles for the case of normal neutrino mass squared ordering.
    The green dots are the best fit points, the red (blue) areas indicate the one (three) sigma ranges.
    The dashed lines indicate tri-bimaximal-Cabibbo (TBC) mixing \cite{King:2012vj}, namely the values: 
    $s^2_{12}=1/3$, $s^2_{23}=1/2$
    $s^2_{13}=\theta^2_C/2$. 
    The Fogli et al fits do not include the latest Daya Bay results, whereas the other two do. 
    This is a modified version of a figure provided privately by Stefano Morisi.
     } \label{global}
\vspace*{-2mm}
\end{figure}

The ground breaking neutrino oscillation milestones may be summarised as
(for original experimental references see e.g.\cite{Verma:2015tpa}):
\begin{itemize}
\item 1998 Atmospheric $\nu_{\mu}$ disappear, implying large $\theta_{23}$ (SuperKamiokande) 
\item  2002 Solar $\nu_{e}$ disappear, implying large $\theta_{12}$ (SuperKamiokande, following the classic Homestake and Gallium experiments.) 
\item 2002 Solar $\nu_{e}$ converted to $\nu_{\mu}$ and $\nu_{\tau}$ (Sudbury Neutrino Observatory).
\item 2004 Reactor $\overline{\nu_{e}}$ seen to disappear and reappear (KamLAND)
\item 2004 Accelerator $\nu_{\mu}$ first seen to disappear (K2K)
\item 2006 Accelerator $\nu_{\mu}$ disappearance studied in detail (MINOS)
\item 2010 Accelerator $\nu_{\mu}$ converted to an observed $\nu_{\tau}$ (OPERA) 
\item 2011 Accelerator $\nu_{\mu}$ converted to $\nu_{e}$ giving a hint for $\theta_{13}$ (T2K, MINOS)
\item 2012 Reactor $\overline{\nu_{e}}$ disappear, $\theta_{13}$ accurately measured (Daya Bay, RENO)
\end{itemize}
The fast pace of neutrino physics is well illustrated by the reactor angle which was unmeasured before
2012 but is now measured to incredible accuracy:
$\theta_{13}\approx 8.4^{\circ}\pm 0.2^{\circ}$ (see \cite{Hu:2015gva} and references therein). 
The other angles are determined from global fits
\cite{Gonzalez-Garcia:2014bfa,Capozzi:2013csa,Forero:2014bxa}
 to be: $\theta_{12}\approx 34^{\circ}\pm 1^{\circ}$ and $\theta_{23}\approx 45^{\circ}\pm 3^{\circ}$,
and first hints of the \CP-violating (CPV) phase $\delta \sim -\pi /2$ have been reported,
however with a large error $\pm \pi/3$. The meaning of the angles is given in Fig.\ref{angles}.
Two possible mass squared orderings are possible as explained in Fig.\ref{mass}.
The above quoted angles are extracted from the global fits which are displayed in Fig.\ref{global}
for the normal ordering case. 


Despite the great pace of progress in neutrino physics, there are still several unanswered experimental questions, as follows:
\begin{itemize}
\item Is the atmospheric neutrino angle $\theta_{23}$ in the first or second octant?
\item Do neutrino mass squared eigenvalues have a normal ordering (NO) or inverted ordering (IO)?
\item What is the value of the lightest neutrino mass?
\item Are neutrinos Dirac or Majorana?
\item Is \CP violated in the leptonic sector and if so by how much?
\end{itemize}
What is the \CP violating phase $\delta$? Is the current hint $\delta \sim -\pi/2$
going to hold up? It is common but incorrect 
to refer to the mass squared ordering question as the
``neutrino mass hierarchy''. However the ``ordering'' question
is separate from whether neutrinos are
hierarchical in nature or approximately degenerate,
which is to do with the lightest neutrino mass.
Many neutrino experiments are underway or in the planning stages to address these questions such as T2K, NO${\nu}$A, Daya Bay, JUNO, RENO, KATRIN, LBNE/DUNE and many neutrinoless double beta decay experiments running and planned \cite{expts}.

The layout of the remainder of this topical review is as follows.
In section \ref{puzzles} we give an overview of the Standard Model puzzles, followed 
in section \ref{patterns} 
by a description of some classic lepton mixing patterns. Lepton mixing may be regarded as a deviation of tri-bimaximal mixing, as discussed in section \ref{deviations},
with charged lepton corrections leading to solar mixing sum rules,
while tri-maximal lepton mixing leads to atmospheric mixing rules.
Motivated by the open questions in neutrino physics, in section \ref{origin}
we discuss a roadmap of the origin of neutrino mass, 
focussing on the seesaw mechanism with right-handed neutrinos, leading to sequential dominance, whose constrained form can lead to predictions for mixing angles.
In section \ref{ToF}
we survey the flavour problem and its possible resolution based on symmetry,
using both GUTs and family symmetry, then classify flavour models
as direct, semidirect or indirect, according to the 
relation between the Klein symmetry of the mass matrices and the underlying discrete family symmetry,
focussing on the possibility of spontaneous \CP violation.    
In section \ref{realistic}
we give two examples of realistic and highly predictive indirect models, 
namely an A to Z of flavour with Pati-Salam and a fairly complete 
$A_4\times SU(5)$ Supersymmetric (SUSY) Grand Unified Theory (GUT) of flavour, where both models 
incorporate constrained sequential dominance and 
have interesting implications for leptogenesis.
In section~\ref{F} we speculate about the possible F-theory origin 
of SUSY GUTs with discrete family symmetry.
Section \ref{conclusion} concludes this topical review.

\section{The Standard Model Puzzles}
\label{puzzles}

Even though the Standard Model (SM) is essentially complete, following the Higgs boson discovery,
we are far from satisfied since it offers no solutions to the cosmological
puzzles of matter-antimatter asymetry, dark matter and dark energy.
It therefore cannot be the final answer. In addition it leaves three unresolved puzzles in its wake:

{\bf The origin of mass}  - the origin of the Higgs vacuum expectation value, its stability under radiative corrections, and the solution to the hierarchy problem (most urgent problem of LHC).

{\bf The quest for unification} - the question of whether the three known forces of the standard model may be related into a grand unified theory, and whether such a theory could also include a unification with gravity.

{\bf The problem of flavour} - the problem of the undetermined fermion masses and mixing angles (including neutrino masses and lepton mixing angles) together with the \CP violating phases, in conjunction with the observed smallness of flavour changing neutral currents and very small strong \CP violation.
In particular the unknown origin of the extremely small neutrino masses {\it for all three families} 
may offer a clue as to what lies beyond the SM.

The differences between quark and lepton mixing may also offer clues
concerning the flavour problem. Certainly the flavour problem has now become much richer, following the discovery of neutrino mass and mixing, so we shall discuss more about this in section~\ref{ToF}.
We now digress slightly to discuss an alternative point of view that frequently is voiced. 
Namely, all that is required for neutrino masses is to 
add two or three right-handed neutrinos with zero Majorana mass due to a conserved $B-L$,
and Yukawa couplings for all
neutrino families of about $10^{-11}$ and that no new physics beyond this is required.
However this conservative point of view involves a new mystery, namely why the third family Yukawa couplings are of order unity for the top quark, and not very small for the 
bottom quark and $\tau$ lepton, 
but are of order $10^{-11}$ for the third family of neutrinos.
The see-saw mechanism \cite{seesaw}, i.e. large third family neutrino Yukawa couplings, with 
physical neutrino masses suppressed by 
heavy right-handed Majorana masses with $B-L$
broken at a high scale, provides an elegant solution to this mystery, and opens the door to leptogenesis.
However, the see-saw mechanism by itself does not account for the observed large lepton mixing,
so we need to go further.

It has been one of the long standing goals of theories of particle
physics beyond the Standard Model to predict quark and lepton
masses and mixings. With the discovery of neutrino mass and
mixing, this quest has received a massive impetus. Indeed, perhaps
the greatest advance in particle physics over the past decade has
been the discovery of neutrino mass and mixing involving large
mixing. The largeness of the lepton mixing angles contrasts with the smallness of the
quark mixing angles, and this observation, together with the
smallness of neutrino masses, provides new and tantalising clues
in the search for the origin of quark and lepton flavour.
For example, it is intruiging that the smallest lepton mixing may be related to the largest quark mixing,
$|U_{e3}|\approx \theta_C/\sqrt{2}$ where $\theta_C$ is the Cabibbo angle,
although this relation is in tension with the latest Daya Bay results.
The quest to understand the origin of the three families of quarks and leptons
and their pattern of masses and mixing parameters is called the flavour puzzle,
and motivates the introduction of family symmetry. In particular, as we shall see,
lepton mixing provides a motivation for discrete family symmetry, which will
form the central part of this review. As we shall also see, such theories demand
a high precision knowledge of the lepton mixing angles, beyond that currently achieved.

The PDG \cite{pdg} advocates CKM and the PMNS mixing matrices being parameterised by:
\begin{eqnarray}
 \label{eq:matrix}
\!\!\!\!\!\!\!\!\!\!\!\!\!\!\!\!\!
\left(\begin{array}{ccc}
    c_{12} c_{13}
    & s_{12} c_{13}
    & s_{13} e^{-i\delta}
    \\
    - s_{12} c_{23} - c_{12} s_{13} s_{23} e^{i\delta}
    & \hphantom{+} c_{12} c_{23} - s_{12} s_{13} s_{23}
    e^{i\delta}
    & c_{13} s_{23} \hspace*{5.5mm}
    \\
    \hphantom{+} s_{12} s_{23} - c_{12} s_{13} c_{23} e^{i\delta}
    & - c_{12} s_{23} - s_{12} s_{13} c_{23} e^{i\delta}
    & c_{13} c_{23} \hspace*{5.5mm}
    \end{array}\right)
   \end{eqnarray}
where $\delta \equiv \delta_{CP}$ is the \CP violating phase in each sector (quark and lepton) 
and $s_{13}=\sin \theta_{13}$, etc. with (very) different angles for quarks and leptons.
In the quark sector the mixing angles are all small, with 
\begin{equation}
s_{12}= \lambda , \ \  s_{23}\sim \lambda^2, \ \  s_{12}\sim \lambda^3
\end{equation}
where $\lambda = 0.226\pm 0.001$ is the Wolfenstein parameter \cite{pdg}.
The \CP violating phase in the quark sector is roughly $\delta \sim (\pi/2)/ \sqrt{2}$.
\footnote{Interestingly, in the original KM parametrisation, the \CP violating phase is 
roughly maximal $\delta \sim \pi/2$, as is the angle $\alpha \sim \pi/2$ in the standard unitarity triangle
representing the orthogonality of the first and third columns of the CKM matrix  \cite{pdg}.}
The large lepton mixing, discussed in further in the next section, must arise in conjunction with the mechanism responsible
for the smallness of neutrino mass, which however is unknown.
In the case of Majorana neutrinos, the PMNS matrix also involves the phase matrix
 \cite{pdg}:
$\textrm{diag}(1,e^{i\frac{\alpha_{21}}{2}},e^{i\frac{\alpha_{31}}{2}})$ which post-multiplies the above matrix. 
It is a puzzle why the quark mixing angles are so small while the lepton mixing angles are so large.

\section{Patterns of Lepton Mixing}
\label{patterns}
The origin of neutrino mass is unknown, as discussed above, meaning that there is not a unique electroweak description,
as for the quarks. There are basically two possibilities, either neutrinos are Dirac or Majorana.
Here we shall exclusively focus on the Majorana case. Majorana neutrino masses must be generated in such a way that, 
below the electroweak breaking scale,
we should obtain the leptonic Lagrangian,
\begin{equation}
	{\cal L}^{\rm lepton} = 
	-v_dY^e_{ij}\overline e^i_{\mathrm{L}} e^j_{\mathrm{R}}  
	-\frac{1}{2} m^{\nu_e}_{ij} \overline{{\nu}_e^i}_{L} {\nu}_{eL}^{cj} 
	+ \mathrm{H.c.}
\label{lepton}
\end{equation}
The resulting matrices are diagonalised by unitary transformations,
\begin{eqnarray}
U_{e_L}Y^eU_{e_R}^{\dagger}=
\left(\begin{array}{ccc}
y_e&0&0\\
0&y_{\mu}&0\\
0&0&y_{\tau}
\end{array}\right), \ \ \ \ 
U_{{\nu_e}_L}m^{\nu_e}U_{{\nu_e}_L}^{T}=
\left(\begin{array}{ccc}
m_1&0&0\\
0&m_2&0\\
0&0&m_3
\end{array}\right).
\end{eqnarray}
The couplings to $W^-$ are given by 
$-\frac{g}{\sqrt{2}}\overline{e}^i_L\gamma^{\mu}W_{\mu}^-{\nu}^i_{eL}$, hence
the charged currents in terms of mass states are, 
\begin{eqnarray}
{\cal L}^{CC}_{\rm lepton}= -\frac{g}{\sqrt{2}}
\left(\begin{array}{ccc}
\overline{e}_L  & \overline{\mu}_L &  \overline{\tau}_L
\end{array}\right)
U_{\rm PMNS}
\gamma^{\mu}W_{\mu}^-
\left(\begin{array}{c}
{\nu}_{1L} \\ 
{\nu}_{2L} \\ 
{\nu}_{3L}
\end{array}\right)+H.c.
\end{eqnarray}
where we identify the PMNS matrix as,
\begin{equation}
\label{enu}
U_{\rm PMNS}=U_{e_L}U_{{\nu}_{eL}}^{\dagger}.
\end{equation}
Now only three of the six phases can be removed since each of the three charged lepton mass terms such as 
$m_e\overline e_{\mathrm{L}}e_{\mathrm{R}}$,
etc., is left unchanged by a phase rotation $e_{\mathrm{L}}\rightarrow e^{i\phi_e}e_{\mathrm{L}}$
and $e_{\mathrm{R}}\rightarrow e^{i\phi_e}e_{\mathrm{R}}$, etc., where the three phases $\phi_e$, etc., are chosen to
leave three physical (irremovable) phases in $U_{\rm PMNS}$.
There is no such phase freedom in the Majorana mass terms 
$-\frac{1}{2} {m_i} \overline{{\nu}_{iL}} {\nu}_{iL}^{c}$ where $m_i$ are real and positive.

We already discussed the PDG parametrisation of the PMNS matrix $U_{\rm PMNS}$ in Eq.\ref{eq:matrix}.
We now discuss three simple ansatze for $U_{\rm PMNS}$ which have been proposed.
Although each of them involves a zero reactor angle and is hence excluded,
they will serve to motivate approaches which involve a non-zero reactor angle.

\subsection{Bimaximal Mixing}

An early suggested pattern of lepton mixing is known as bimaximal (BM) mixing
with $s^2_{13}=0$ and $s^2_{12}=s^2_{23}=1/2$
which could originate from the discrete group $S_4$ (see later). It has a
maximal solar mixing angle \cite{Barger:1998ta}, and is given by a matrix of
the form 
\begin{equation}\label{BM}
U_{\rm BM} =
\left(
\begin{array}{ccc}
\frac{1}{\sqrt{2}} & \frac{1}{\sqrt{2}} & 0\\
-\frac{1}{2} & \frac{1}{2} & \frac{1}{\sqrt{2}}\\
\frac{1}{2} & -\frac{1}{2} & \frac{1}{\sqrt{2}}
\end{array}
\right).
\end{equation}%

\subsection{Tri-bimaximal Mixing}

A second pattern of lepton mixing which came to dominate the model building community until the measurement of the reactor angle is the tribimaximal (TB) mixing matrix \cite{Harrison:2002er}. This has been
associated with models based on the flavour symmetries A$_4$ and S$_4$ (see later). 
Like BM mixing it predicts $s^2_{13}=0$ and $s^2_{23}=1/2$ but differs in that it
predicts a solar mixing angle given by $s_{12}=1/\sqrt{3}$,
i.e. $\theta_{12}\approx 35.3^\circ$. The mixing matrix is given explicitly by
\begin{equation}\label{TB}
U_{\rm TB} =
\left(
\begin{array}{ccc}
\sqrt{\frac{2}{3}} &  \frac{1}{\sqrt{3}}
&  0 \\ - \frac{1}{\sqrt{6}}  & \frac{1}{\sqrt{3}} &  \frac{1}{\sqrt{2}} \\
\frac{1}{\sqrt{6}} & -\frac{1}{\sqrt{3}} &  \frac{1}{\sqrt{2}}    
\end{array}
\right).
\end{equation}%

\subsection{Golden Ratio Mixing}

Another pattern of lepton mixing which was viable until the reactor angle measurement associates the golden ratio
$\varphi=\frac{1+\sqrt{5}}{2}$ with the solar mixing angle.
 The original golden ratio (GR) mixing pattern
is related to the flavour symmetry $A_5$~\cite{Datta:2003qg}. As above, it predicts
$s^2_{13}=0$ and $s^2_{23}=1/2$ but differs by having a solar mixing angle given by
$t_{12}^\nu=1/\varphi$, i.e. $\theta^\nu_{12}\approx 31.7^\circ$, resulting in
the mixing matrix
\begin{equation}\label{GR}
U_{\rm GR} =
\left(
\begin{array}{ccc}
\frac{\varphi}{\sqrt{2+\varphi}} &
\frac{1}{\sqrt{2+\varphi}} &  0 \\ -\frac{1}{\sqrt{4+2\varphi}}&
\frac{\varphi}{\sqrt{4+2\varphi}} & \frac{1}{\sqrt{2}} \\
\frac{1}{\sqrt{4+2\varphi}} & -\frac{\varphi}{\sqrt{4+2\varphi}} &
\frac{1}{\sqrt{2}}
\end{array}
\right).
\end{equation}%

\section{Deviations from TB mixing}
\label{deviations}
\subsection{Deviation parameters}
After the measurement of the reactor angle, TB mixing is excluded. However, TB mixing 
still remains a reasonable approximation to lepton mixing for the solar and atmospheric angles.
It therefore makes sense to expand the angles about their TB values \cite{King:2007pr,Pakvasa:2008zz}:
\begin{eqnarray}
\sin\theta_{12} &=& \frac{1}{\sqrt{3}} (1+s),\\
\sin\theta_{23} &=& \frac{1}{\sqrt{2}} (1+a),\\
\sin\theta_{13} &=& \frac{r}{\sqrt{2}},
\end{eqnarray}
where $s,\,a$, and $r$ are the ($s$)olar, ($a$)tmospheric and ($r$)eactor deviation parameters
such that TB mixing  \cite{Harrison:2002er} is recovered for $s=a=r=0$. For example, TBC mixing 
\cite{King:2012vj} corresponds to $s=a=0$
and $r=\theta_C$, where $\theta_C$ is the Cabibbo angle, which is consistent with data at three sigma as shown
in Fig.\ref{global}.

\subsection{Tri-maximal mixing and atmospheric sum rules}
\label{atmospheric}
\begin{figure}[t]
\centering
\includegraphics[width=0.60\textwidth]{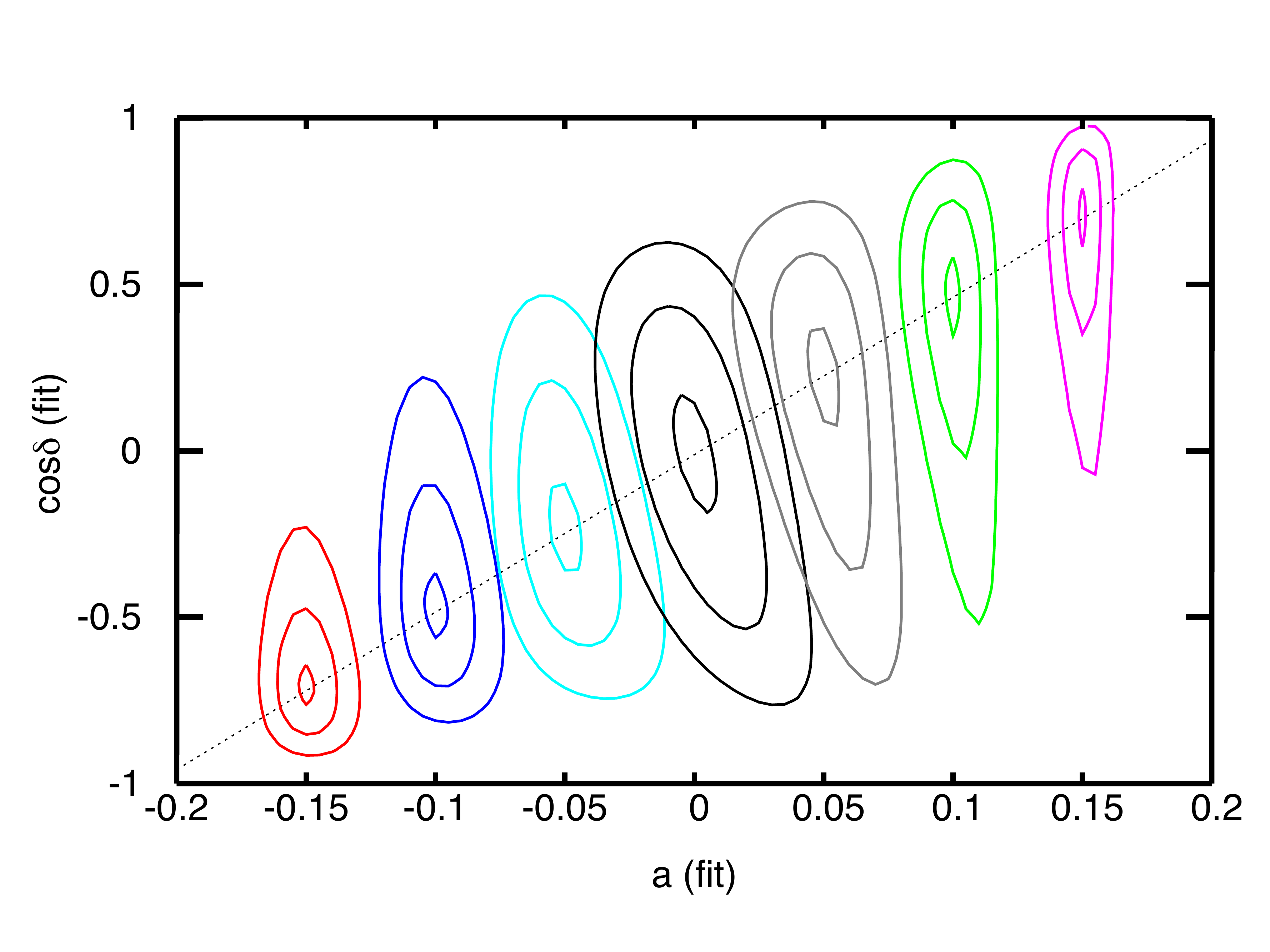}
\vspace*{-4mm}
    \caption{Expectation for the determination of the TM1 atmospheric mixing sum rule 
     $a\approx  r\cos\delta$ at one, three and five sigma for a low energy neutrino factory with a
    magnetised iron detector (for more details see \cite{Ballett:2013wya}).  } \label{potatoes}
\vspace*{-2mm}
\end{figure}

Tri-maximal mixing is a variation which preserves either the first or the second column of the TB mixing 
mixing matrix in Eq.\ref{TB}, leading to two versions called TM1 or TM2,
{\small
\begin{equation}\label{TMM}
\!\!\!\!\!\!\!\!
U_{\rm TM1} =
\left(
\begin{array}{ccc}
\frac{2}{\sqrt{3}} &  - &  - \\ 
-\frac{1}{\sqrt{6}} &  - &  - \\
\frac{1}{\sqrt{6}} &  - &  -  
\end{array}
\right),\ \ \ \ 
U_{\rm TM2} =
\left(
\begin{array}{ccc}
- &  \frac{1}{\sqrt{3}} &  - \\ 
- & \frac{1}{\sqrt{3}} &  - \\
- & -\frac{1}{\sqrt{3}} &  -  
\end{array}
\right).
\end{equation}%
}
The dashes indicate that the other elements are undetermined. However these are fixed once the reactor angle is specified (it is a free parameter here). These imply the relations
\begin{eqnarray} 
\!\!\!\!\!\!\!\!{\rm TM1}:\ \ \ \ \left | U_{e1} \right | = 
\sqrt{\frac{2}{3}} \ \  {\rm and}\ \   \left|U_{\mu1}\right|
=\left|U_{\tau1}\right| =\frac{1}{\sqrt{6}}\ ;  \label{TM1}\\
\!\!\!\!\!\!\!\!\!\!\!\!\!\!\!\!{\rm TM2}:\ \ \ \ \left|U_{e2}\right| =  \left|U_{\mu2}\right| = \left|U_{\tau2}\right| =
\frac{1}{\sqrt{3}}\ .  \label{TM2} 
\end{eqnarray}
The {\it atmospheric mixing sum rule}
\begin{equation} 
a= \lambda r\cos\delta + \mathcal{O}(a^2,r^2), \ \  {\rm with }\ \ s= \mathcal{O}(a^2,r^2),
\end{equation} 
was first derived in~\cite{King:2007pr} by expanding the PMNS matrix to first order in $r,s,a$. 
It also follows from a first order expansion of Eqs.\ref{TM1},\ref{TM2}, where
$\lambda=1$ for TM1 and $\lambda=-1/2$ for TM2. The study of correlations of
this type, and their application to the discrimination between underlying
models, has been shown to be a realistic aim for a next-generation superbeam
experiment \cite{Ballett:2013wya}, see for example Fig.\ref{potatoes}.

\subsection{Charged lepton corrections and solar sum rules}
\label{solar}

\begin{figure}[t]
\centering
\includegraphics[width=0.60\textwidth]{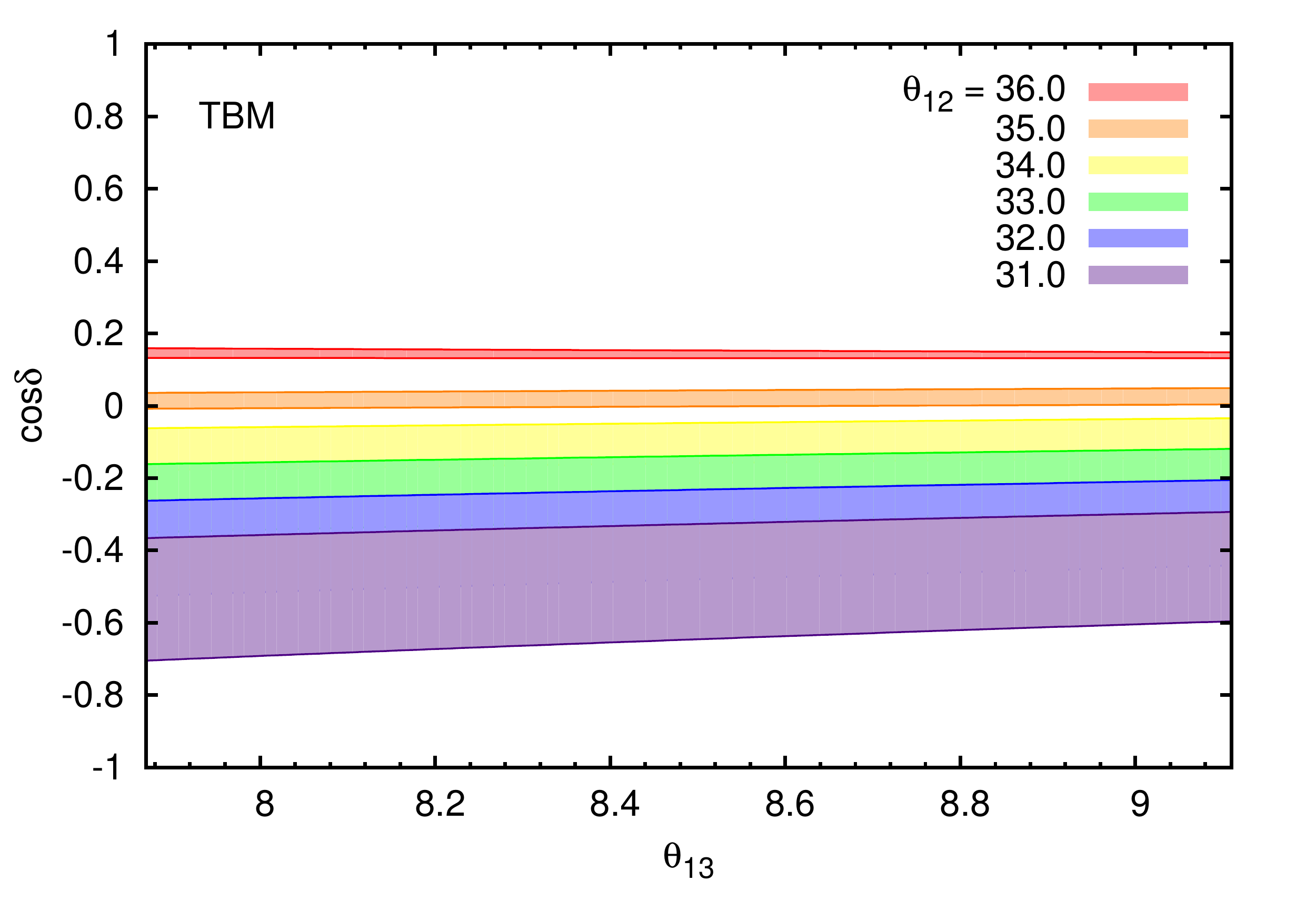}
\vspace*{-4mm}
    \caption{Solar sum rules prediction for 
    $\cos \delta$ using the exact result in 
    Eq.\ref{sol4} for TB neutrino mixing (for more details see \cite{Ballett:2014dua}).  } \label{sum}
\vspace*{-2mm}
\end{figure}

Now suppose that neutrino mixing $U^{\nu}_{\rm TB}$ obeys TB exactly
so that the PMNS matrix according to Eq.\ref{enu} is given
by $U_{\rm PMNS}= U^e U^{\nu}_{\rm TB}$ where 
$U^{\nu}_{\rm TB}$ is equated to Eq.\ref{TB} while
$U^e$ encodes some unknown charged lepton corrections
which must be small since $U_{\rm PMNS}$ is not far from TB mixing.
The {\it solar mixing sum rule}
\cite{King:2005bj,Masina:2005hf,Antusch:2005kw}
then follows from the assumption that $\theta^e_{23}=\theta^e_{13}=0$.
If the charged lepton mixing matrix involves a Cabibbo-like mixing,
then the PMNS matrix is given by,
\begin{equation}
U_{\mathrm{PMNS}} = \left(\begin{array}{ccc}
\!c^e_{12}& s^e_{12}e^{-i\delta^e_{12}}&0\!\\
\!-s^e_{12}e^{i\delta^e_{12}}&c^e_{12} &0\!\\
\!0&0&1\!
\end{array}
\right)
\left( \begin{array}{ccc}
\sqrt{\frac{2}{3}} & \frac{1}{\sqrt{3}} & 0 \\
-\frac{1}{\sqrt{6}}  & \frac{1}{\sqrt{3}} & \frac{1}{\sqrt{2}} \\
\frac{1}{\sqrt{6}}  & -\frac{1}{\sqrt{3}} & \frac{1}{\sqrt{2}}
\end{array}
\right)
= \left(\begin{array}{ccc}
\! \cdots & \ \ 
\! \cdots&
\! \frac{s^e_{12}}{\sqrt{2}}e^{-i\delta^e_{12}} \\
\! \cdots
& \ \
\! \cdots
&
\! \frac{c^e_{12}}{\sqrt{2}}
\!\\
\frac{1}{\sqrt{6}}  & -\frac{1}{\sqrt{3}} & \frac{1}{\sqrt{2}}
\end{array}
\right)
\label{Ucorr}
\end{equation}
Comparing to the PMNS parametrisation in Eq.\ref{eq:matrix} we identify,
\begin{eqnarray}
\label{Eq:Sumrule4}  |U_{e3}|= s_{13} &=&  \frac{s^e_{12}}{\sqrt{2}}  \; , \\
\label{Eq:Sumrule1} |U_{\tau 1}|= |s_{23}s_{12}-s_{13}c_{23}c_{12}e^{i\delta} |    &=&     \frac{1}{\sqrt{6}}\; , \\
\label{Eq:Sumrule2} |U_{\tau 2}|=  |- c_{12} s_{23} - s_{12} s_{13} c_{23} e^{i\delta}| &=&     \frac{1}{\sqrt{3}}\; , \\
|U_{\tau 3}|=c_{13}c_{23} &=& \frac{1}{\sqrt{2}}.   
\end{eqnarray}
The first equation predicts a reactor angle $\theta_{13}\approx 9.2^{\circ}$
if $\theta_e\approx \theta_C \approx 13^\circ$. 
The second and fourth equations allow to eliminate $\theta_{23}$ to
give a new relation between the PMNS parameters, $\theta_{12}$, $\theta_{13}$ and $\delta$
called a {\em solar sum rule},
which may be expanded to first order to give the approximate relations,
\begin{equation} 
\!\!\!\!\!\!\!\!\!\!
\theta_{12}\approx  35.26^o + \theta_{13}\cos\delta 
\ \ \ \  {\rm or} \ \ \ \  \cos\delta \approx \frac{\theta_{12}-35.26^o}{\theta_{13}}
\label{eq:linearSSR} 
\end{equation}
where $35.26^o=\sin ^{-1}\frac{1}{\sqrt{3}}$,
which can be recast as \cite{King:2007pr},
\begin{equation} 
s = r \cos \delta + \mathcal{O}(a^2,r^2).
\label{eq:linearSSR2} 
\end{equation}
Recently it has been realised that, keeping $\theta^e_{13}=0$,
but allowing $\theta^e_{23}\neq 0$,
the following exact result can be obtained by generalising 
Eq.\ref{Ucorr} to allow both $\theta^e_{12}, \theta^e_{23}\neq 0$ \cite{Ballett:2014dua}:
\begin{equation}
\frac{\left | U_{\tau 1} \right |}{\left | U_{\tau 2} \right |
}=\frac{|s_{12} s_{23} - c_{12} s_{13} c_{23} e^{i\delta}|}
    {|- c_{12} s_{23} - s_{12} s_{13} c_{23} e^{i\delta}|}
= \frac{1}{\sqrt{2}}\;, \label{sol3}
\end{equation}
For the previous case $\theta^e_{23}=\theta^e_{13}=0$ this result follows trivially by taking the ratio
of the two equations \ref{Eq:Sumrule1} and \ref{Eq:Sumrule2}. Therefore this result applies to that case also.
However it turns out that $\theta^e_{23}$ cancels in this ratio which is therefore a more general 
sum rule (though not completely general since it still assumes $\theta^e_{13}=0$).
After some algebra, Eq.\ref{sol3} leads to an exact prediction for $\cos \delta$ in terms of the other 
physical lepton angles,
\begin{equation}
\cos \delta =\frac
{t_{23}s^2_{12}+s^2_{13}c^2_{12}/t_{23}-\frac{1}{3}(t_{23}+s^2_{13}/t_{23})}
{\sin 2\theta_{12}s_{13}}, \label{sol4}
\end{equation}
as displayed in Fig.\ref{sum}.

When expanded to first order, Eq.\ref{sol4} 
reduces to the leading order sum rule in Eq.\ref{eq:linearSSR}.
This is not too surprising since the previous sum rule case also satisfies Eq.\ref{sol3}.
The leading order sum rule in Eq.\ref{eq:linearSSR} offers a simple way to understand the results in 
Fig.\ref{sum}. For example from Fig.\ref{sum} it seems that 
TB neutrino mixing predicts 
$\cos \delta \approx 0$ if $\theta_{12}\approx 35.26^o$,
which is obvious from Eq.\ref{eq:linearSSR}
\footnote{Note that $\cos \delta \sim 0$ is consistent with 
the current hint $\delta \sim -\pi /2$.}. 
This can also be understood from 
Eq.\ref{sol4} where we see that for $s_{12}^2=1/3$
the leading terms $t_{23}s^2_{12}$ and $\frac{1}{3}t_{23}$ in the numerator cancel, leaving $\cos \delta = s_{13}/(2\sqrt{2}t_{23})\approx 0.05$
which is consistent with the numerical estimates of the error given above
for a range of $\theta_{12}$.

Solar sum rules can also be obtained for 
different types of neutrino mixing such as
$U^{\nu}_{\rm BM}$ (which is almost excluded) and $U^{\nu}_{\rm GR}$ (which gives similar results
to the case $U^{\nu}_{\rm TB}$ considered here). The general formula given in \cite{Ballett:2014dua} is,
\begin{equation}
\cos \delta =\frac
{t_{23}s^2_{12}+s^2_{13}c^2_{12}/t_{23}-s^{\nu 2}_{12}(t_{23}+s^2_{13}/t_{23})}
{\sin 2\theta_{12}s_{13}}, \label{sol5}
\end{equation}
where $s^{\nu 2}_{12}=\frac{1}{3},\frac{1}{2}$ for $U^{\nu}_{\rm TB,BM}$ and so on.
The prospects for 
studying {\it solar sum rules} at JUNO and LBNF is discussed in \cite{Ballett:2014dua}.
A slightly more lengthy but equivalent 
formula to Eq.\ref{sol4} had been previously derived \cite{Marzocca:2013cr} by an alternative method 
involving an auxiliary phase $\phi$ without using the elegant result Eq.\ref{sol3},
\begin{equation}
\!\!\!\!\!\!\!\!\!\!\!\!\cos \delta =\frac{t_{23}}{\sin 2\theta_{12}s_{13}}
\left[\cos 2\theta^{\nu}_{12}+
(s^2_{12}-c^{\nu 2}_{12})(1-\cot^2\theta_{23}s^2_{13})
\right] . \label{sol6}
\end{equation}
We prefer the simpler form in Eq.\ref{sol5} which involves $\theta^{\nu}_{12}$
in only one place since it exhibits the cancellation between the terms
$t_{23}s^2_{12}$ and $s^{\nu 2}_{12}t_{23}$ when $s^2_{12}=s^{\nu 2}_{12}$
responsible for the prediction $\cos \delta \approx 0$ in this case
We also advocate the much simpler derivation of Eq.\ref{sol5} given in \cite{Ballett:2014dua}.

Finally we give a word of caution that when
comparing leading order sum rules to the exact results
the ratio $(\cos \delta)_{\rm exact}/(\cos \delta)_{\rm LO}$
used in  \cite{Girardi:2014faa} will lead to
misleading results when $(\cos \delta)_{\rm LO} \approx 0$. 
In general it is safer to compare them using the
experimentally relevant quantity 
$\Delta(\cos\delta)=(\cos \delta)_{\rm exact}-(\cos \delta)_{\rm LO}$
defined in \cite{Ballett:2014dua}. For example we find 
$\Delta(\cos\delta) \lesssim 0.1$ for TB neutrino mixing corrected by charged lepton mixing.
It will take experiment a long time to reach this level of precision, so for present purposes
the linear approximation in Eq.\ref{eq:linearSSR} is adequate for TB neutrino mixing.

\section{Neutrino Mass Models}
\label{origin}
\subsection{The open questions from neutrino physics}
Despite the great progress coming from neutrino oscillation experiments 
there are still some outstanding questions.
Are the lepton mixing angles consistent with TBC mixing?
If not then is the atmospheric angle in the first or second octant?
What is the leptonic \CP violating phase $\delta$? Is the current hint $\delta \sim -\pi/2$
going to hold up? Maybe there is no \CP violation in the lepton sector?
Are neutrino mass squared ordered 
normally or inverted? \footnote{It is common but incorrect 
to refer to this question as the
``neutrino mass hierarchy'' since the ``ordering'' question
is separate from whether neutrinos are
hierarchical in nature or approximately degenerate,
which is to do with the lightest neutrino mass.}
What is the lightest neutrino mass?
Are neutrino masses Majorana or Dirac in nature?
Many neutrino experiments are underway or in the planning stages to address these questions such as T2K, NO${\nu}$A, Daya Bay, JUNO, RENO, KATRIN, DUNE and many neutrinoless double beta decay experiments running and planned \cite{expts}.

\subsection{Road Map of Neutrino Mass Models}
\begin{figure}[t]
\centering
\includegraphics[width=0.80\textwidth]{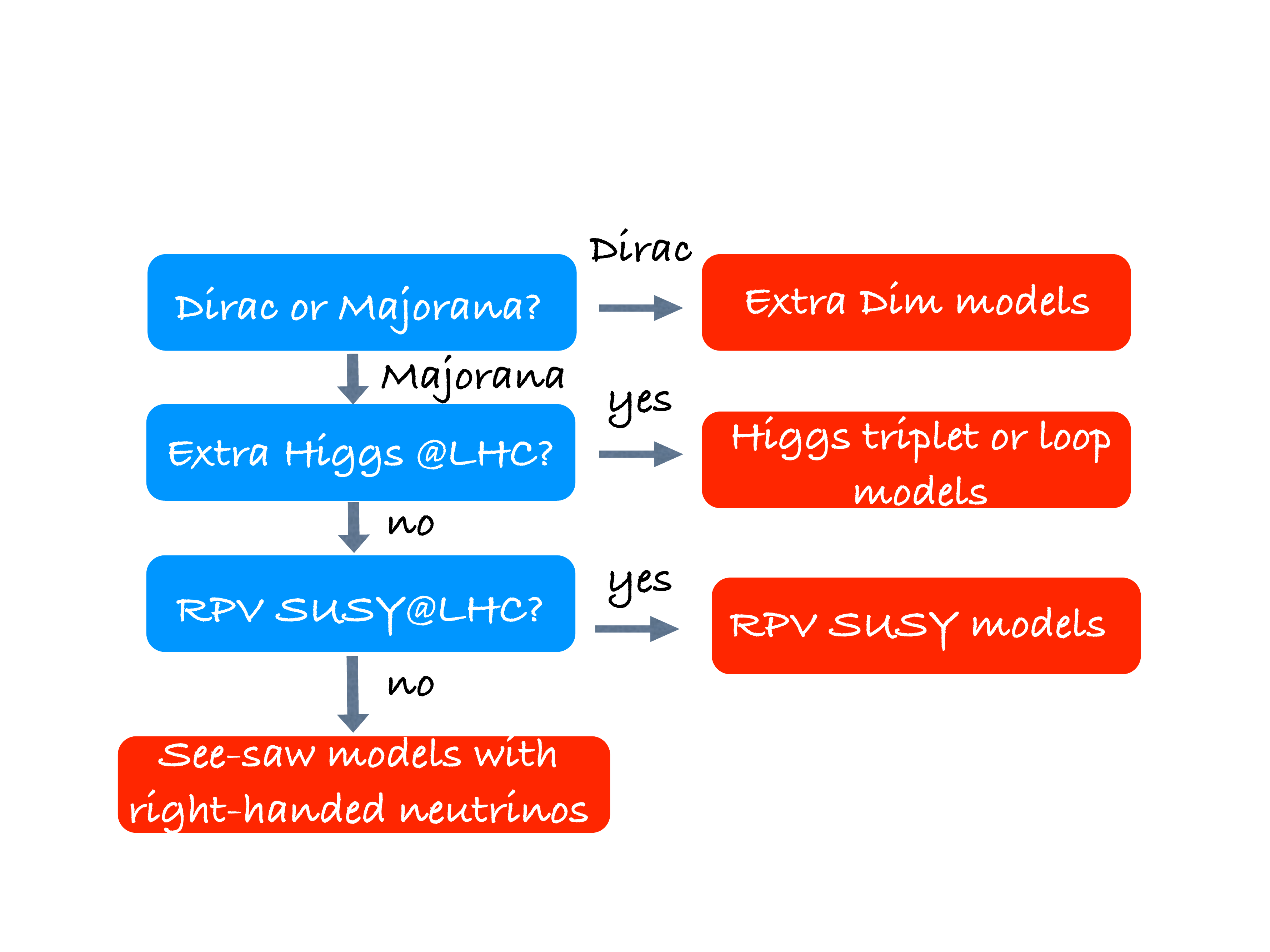}
\vspace*{-4mm}
    \caption{Roadmap of neutrino mass models.    
    } \label{roadmap}
\vspace*{-2mm}
\end{figure}

Everyone can invent her or his personal roadmap of neutrino mass
models, one example being that shown in Fig.\ref{roadmap}.
The blue boxes contain experimental questions and the red boxes
possible theoretical consequences. In this subsection we shall briefly describe the possible
theoretical options for producing neutrino mass.

It is worthwhile to first recall why the observation of non-zero neutrino mass and mixing
is evidence for new physics beyond the SM. The most intuitive way to
understand why neutrino mass is forbidden in the Standard Model, is to
understand that the Standard Model predicts that neutrinos
always have a ``left-handed'' spin - rather like rifle bullets which
spin counter clockwise to the direction of travel.
In fact this property was first experimentally measured in 1958,
two years after the neutrino was discovered, by
Maurice Goldhaber, Lee Grodzins and Andrew Sunyar.  More accurately,
the ``handedness'' of a particle describes the direction of its spin vector
along the direction of motion, and the neutrino being ``left-handed''
means that its spin vector always points in the opposite direction to its
momentum vector.  The fact that the neutrino is left-handed, written as 
$\nu_L$, implies that it
must be massless. If the neutrino has mass then, according to special
relativity, it can never travel at the speed of light. In principle,
a fast moving observer could therefore overtake the
spinning massive neutrino and would see it moving in the opposite
direction. To the observer, the massive neutrino would therefore
appear right-handed. Since the Standard Model predicts that neutrinos must
be strictly left-handed, it follows that neutrinos are massless
in the Standard Model. It also follows that the discovery of neutrino
mass implies new physics beyond the SM, with
profound implications for particle physics and cosmology.

Neutrinos are massless in the Standard Model for three independent reasons:
\begin{itemize}
\item There are no right-handed neutrinos $\nu_R$
\item There are only Higgs doublets (and no Higgs triplets) of $SU(2)_L$ 
\item There are only renormalisable terms
\end{itemize}
In the SM, the three massless neutrinos $\nu_e$, $\nu_{\mu}$, $\nu_{\tau}$ are
distinguished by separate lepton numbers $L_e$, $L_{\mu}$,
$L_{\tau}$. Neutrinos and antineutrinos are distinguished by total conserved
lepton number 
$L=L_e+L_{\mu}+L_{\tau}$. To generate neutrino mass we must relax
one or more of the above three conditions. For example, by adding
right-handed neutrinos the Higgs mechanism of the Standard Model
can give neutrinos the same type of mass as the Dirac electron mass or
other charged lepton and quark masses, which would generally break the separate
lepton numbers $L_e$, $L_{\mu}$, $L_{\tau}$, but preserve the total lepton
number $L$. However it is also possible for neutrinos to have a new type of mass
of a type first proposed by Majorana, which would also break $L$. 
There exists a special case where total lepton number $L$ is broken, but the
combination $L_e-L_{\mu}-L_{\tau}$ is conserved; such a symmetry would give
rise to a neutrino mass matrix with an inverted mass spectrum.

From the theoretical perspective, the main unanswered question is the origin 
of neutrino mass, and in particular the smallness of neutrino mass. 
The simplest possibility is that neutrinos have Dirac mass
just like the electron mass in the SM, namely due to a term like
$y_D\overline{L}H\nu_R$, where $L$ is a lepton doublet containing $\nu_L$, $H$
is a Higgs doublet and $\nu_R$ is a right-handed neutrino. 
The observed smallness of neutrino masses implies that the Dirac Yukawa coupling $y_D$ must be 
of order $10^{-12}$ to achieve a Dirac neutrino mass of about 0.1~eV. 
Advocates of Dirac masses point out that the electron mass already requires a
Yukawa coupling $y_e$ of about $10^{-6}$, so we are used to such small Yukawa
couplings. In this case, 
all that is required is to add right-handed neutrinos $\nu_R$ to the SM and we are done.
Well, almost. It still needs to be explained why the $\nu_R$ have zero Majorana mass,
after all they are gauge singlets and so nothing prevents them acquiring
(large) Majorana mass terms $M_{RR} \nu_R \nu_R$ where $M_{RR}$ could be as
large as the Planck scale. 
Moreover, Majorana masses offer a unique (and testable) way to generate
neutrino masses (since neutrinos do not carry electric charge) even without
right-handed neutrinos. 
The simplest way to generate Majorana mass is via $y_M \Delta L L $
where $\Delta$ is a Higgs triplet and $y_M$ is a Yukawa coupling associated with Majorana mass. 
Alternatively,
at the effective level, Majorana neutrino mass can result from some additional
dimension 5 operators which couple two lepton doublets $L$ to two Higgs
doublets $H$  first proposed by Weinberg~\cite{Weinberg:1980bf},
\begin{equation}
-\frac{1}{2}HL^T\kappa HL ,
\label{dim5}
\end{equation}
where $\kappa$ has dimension $[\mathrm{mass}]^{-1}$. 
This is a non-renormalisable operator, so it violates one of the tenets of the SM. 
In order to account for a neutrino mass of order 0.1 eV requires $\kappa \sim 10^{-14}$ GeV$^{-1}$.
This suggests a new high energy mass scale $M$ in physics, a small dimensionless coupling associated
with $\kappa$, or both. 

There are basically five different proposals for the origin of neutrino mass:
\begin{itemize}
\item The seesaw mechanisms~\cite{seesaw,type2,Foot:1989type3}, 
including low scale seesaw mechanisms \cite{Morisi:2012fg}
(Weinberg operator typically from large Majorana mass $M=M_{R}$ for right-handed neutrinos $\nu_R$) 
\item $R$-parity violating supersymmetry~\cite{Drees:1997id}
(Weinberg operator from TeV scale Majorana mass for neutralinos $\chi$)  
\item TeV scale loop mechanisms~\cite{Zee:1980ai,Ma:2012ez,King:2014uha}
(Majorana mass from extra Higgs doublets and singlets at the TeV scale)
\item Extra dimensions~\cite{Arkani-Hamed:1998vp} 
(Dirac mass with small $y_D$ due to right-handed neutrinos $\nu_R$ in the bulk)
\item String theory~\cite{Mohapatra:bd,Mohapatra:1986bd} (new mechanisms for generating large Majorana mass for right-handed neutrinos $\nu_R$
from Planck or string scale physics) 
\end{itemize}

These different mechanisms are reviewed in~\cite{Bandyopadhyay:2007kx}.
Returning to the roadmap in Fig.\ref{roadmap}, we see these mechanisms for neutrino mass
represented by the red boxes, being related to the experimental question in the blue boxes.

\begin{figure}[t]
\centering
\includegraphics[width=0.8\textwidth]{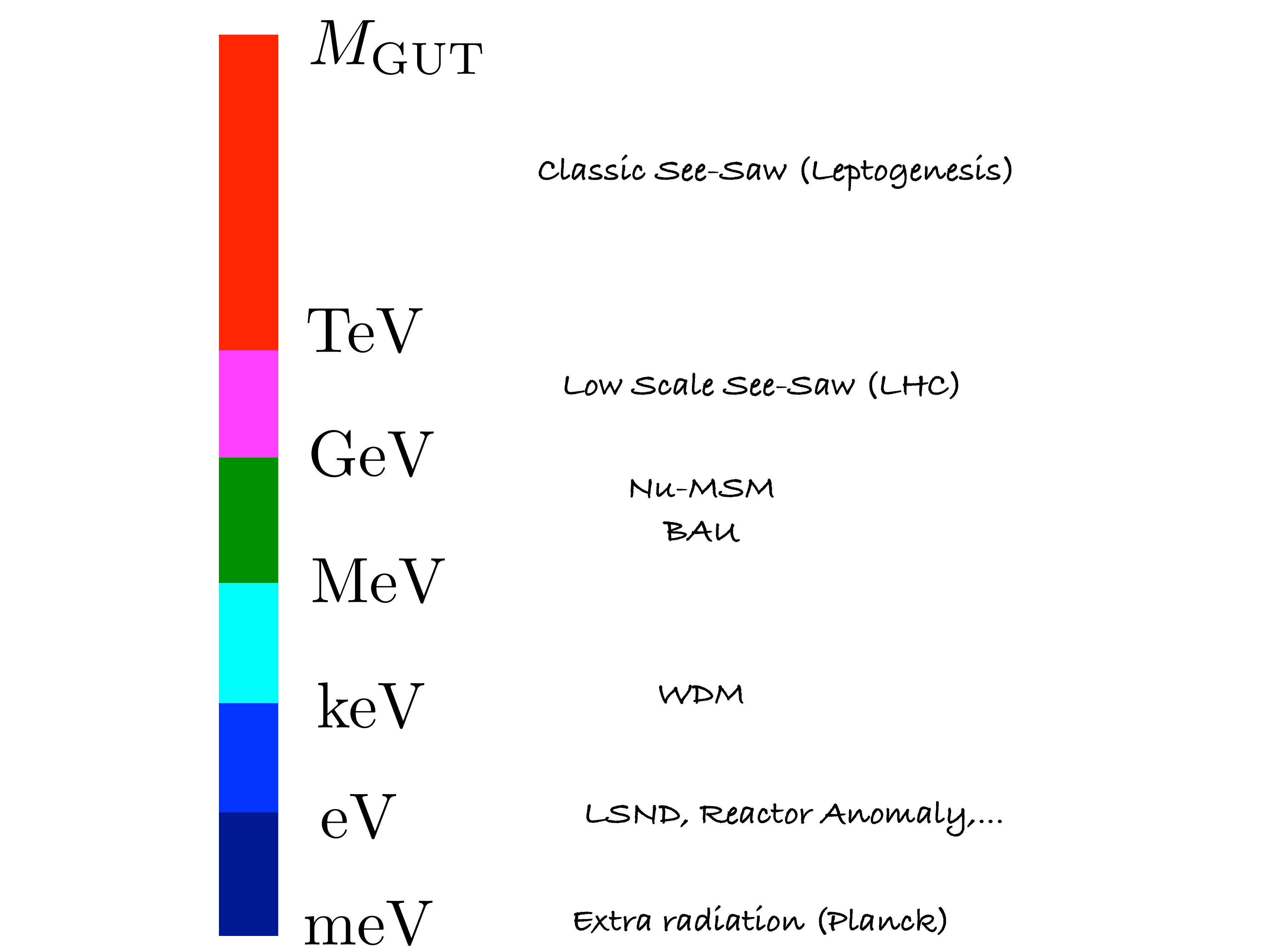}
\vspace*{-4mm}
    \caption{Possible mass spectrum of right-handed (or sterile) neutrinos
    corresponding to the physical implications as shown.} \label{steriles}
\vspace*{-2mm}
\end{figure}

Since no new physics has yet emerged from the LHC, we are led to consider the seesaw mechanism
with right-handed neutrinos. However even in this case
both the number of species and the mass spectrum of right-handed (or sterile) neutrinos is completely
unknown \cite{Drewes:2015jna}. As shown in Fig.\ref{steriles} the mass spectrum can 
cover the whole range with different physical consequences as indicated.
It is one of the goals of neutrino physics to determine this spectrum.
In this topical review we shall focus on 
the classic see-saw mechanism \cite{seesaw} with very heavy right-handed neutrinos,
with masses above the TeV scale,
which may be
incorporated into a theory of flavour, possibily related to string theory.

\subsection{See-saw mechanism with two right-handed neutrinos and sequential dominance}

In this subsection we consider the high scale (classic) {\em see-saw} neutrino model involving just 
{\em two} right-handed neutrinos
$\nu^{\rm sol}_R$ and $\nu^{\rm atm}_R$
with Yukawa couplings \cite{King:1998jw},
\footnote{We follow the notation of the third paper in \cite{King:1998jw},
which was the first paper to discuss a phenomenologically viable model with
two right-handed neutrinos (see also \cite{King:2002nf}).
Subsequently
two right-handed neutrino models with two texture zeros were proposed
in \cite{Frampton:2002qc}, however such two texture zero 
models are now phenomenologically excluded \cite{Harigaya:2012bw} for the case of a normal neutrino mass hierarchy considered here, while the one texture zero case $d=0$
(see the second paper in \cite{King:2002nf}) remains viable.}
\begin{equation}
(H_u/v_u)(a\overline L_e+ b\overline L_{\mu} +c\overline L_{\tau}){\nu^{\rm sol}_R}
+ (H_u/v_u)(d\overline L_e+ e\overline L_{\mu} +f\overline L_{\tau}){\nu^{\rm atm}_R}+H.c.,
\end{equation}
where $H_u$ is a Higgs doublet
and $v_u$ its vacuum expectation value (VEV).
The heavy right-handed Majorana masses are,
\begin{equation}
M_{\rm sol}\overline{\nu^{\rm sol }_R}({\nu^{\rm sol}_R})^c
+M_{\rm atm}\overline{\nu^{\rm atm }_R}({\nu^{\rm atm}_R})^c +H.c..
\end{equation}
In the basis, with rows
$(\overline \nu_{eL}, \overline \nu_{\mu L}, \overline \nu_{\tau L})$ and columns ${\nu^{\rm atm}_R}, {\nu^{\rm sol}_R}$, the resulting Dirac mass matrix is,
\begin{equation}
m^D=
\left( \begin{array}{cc}
d & a \\
e & b \\
f & c
\end{array}
\right)\equiv
(m^D_{\rm atm} \ \ m^D_{\rm sol}), \ \ \ \ 
m^D_{\rm atm}\equiv (d \ e \ f)^T, \ \ m^D_{\rm sol}\equiv (a\ b \ c)^T.
\label{mD}
\end{equation}

The (diagonal) 
right-handed neutrino heavy Majorana mass matrix $M_{R}$
with rows $(\overline{\nu^{\rm atm}_R}, \overline{\nu^{\rm sol}_R})^T$ and columns $(\nu^{\rm atm}_R, \nu^{\rm sol}_R)$
is,
\begin{equation}
M_{R}=
\left( \begin{array}{cc}
M_{\rm atm} & 0 \\
0 & M_{\rm sol}
\end{array}
\right)
\end{equation}

The see-saw formula is \cite{seesaw},
\begin{equation}
m^{\nu}=-m^DM^{-1}_{R}(m^D)^T,
\label{seesaw}
\end{equation}
where $m^{\nu}$ is the 
the light effective left-handed Majorana neutrino mass matrix
(i.e. the physical neutrino mass matrix),
$m^D$ is the Dirac mass matrix in LR convention and $M_R$ is the (heavy) Majorana
mass matrix. 
Using the see-saw formula 
dropping the overall minus sign which is physically irrelevant,
the light effective left-handed Majorana neutrino mass matrix
$m^{\nu}$
(i.e. the physical neutrino mass matrix) is, by multiplying the matrices,
\begin{equation}
m^{\nu}=m^DM^{-1}_{R}(m^D)^T=
\left( \begin{array}{ccc}
\frac{a^2}{M_{\rm sol}}+ \frac{d^2}{M_{\rm atm}}& \frac{ab}{M_{\rm sol}}+ \frac{de}{M_{\rm atm}} 
& \frac{ac}{M_{\rm sol}}+ \frac{df}{M_{\rm atm}}  \\
\frac{ab}{M_{\rm sol}}+ \frac{de}{M_{\rm atm}} & \frac{b^2}{M_{\rm sol}}+ \frac{e^2}{M_{\rm atm}}  
& \frac{bc}{M_{\rm sol}}+ \frac{ef}{M_{\rm atm}}  \\
\frac{ac}{M_{\rm sol}}+ \frac{df}{M_{\rm atm}} 
& \frac{bc}{M_{\rm sol}}+ \frac{ef}{M_{\rm atm}}
& \frac{c^2}{M_{\rm sol}}+ \frac{f^2}{M_{\rm atm}}  
\end{array}
\right)
\label{2rhn}
\end{equation}
The sequential dominance (SD) \cite{King:1998jw} assumptions are that $d\ll e,f$ and 
\begin{equation}
 \frac{(e,f)^2}{M_{\rm atm}} \gg \frac{(a,b,c)^2}{M_{\rm sol}}.
\label{SD0}
\end{equation}
By explicit calculation, one can check that $\det m^{\nu} = 0$.
Since the determinant of a Hermitian matrix is the product of mass eigenvalues 
$$
\det (m^{\nu}m^{{\nu}\dagger}) = m_1^2m_2^2m_3^2,
$$
one may deduce that one of the mass eigenvalues of the complex symmetric matrix above
is zero, which under the SD assumption is the lightest one $m_1=0$
with $m_3\gg m_2$ since the model approximates to a single right-handed neutrino model 
\cite{King:1998jw}.
Hence we see that {\it SD implies a normal neutrino mass hierarchy.}
Including the solar right-handed neutrino as a perturbation, it can be shown that,
for $d=0$, together with the assumption of a dominant atmospheric right-handed neutrino 
in Eq.\ref{SD0}, leads to the approximate results for the solar and atmospheric angles 
\cite{King:1998jw},
\begin{equation}
\tan \theta_{23}\sim \frac{e}{f}, \ \ \ \ \tan \theta_{12} \sim \frac{\sqrt{2}a}{b-c}.
\label{t12}
\end{equation}
Under the above SD assumption, 
each of the right-handed neutrinos contributes uniquely to a particular physical neutrino mass.
The SD framework above with $d=0$
leads to the relations in Eq.\ref{t12} together with the reactor angle bound \cite{King:2002nf},
\begin{equation}
\theta_{13} \lesssim m_2/m_3
\label{13}
\end{equation}
{\it This result shows that SD allows for large values of the reactor angle, consistent with the 
measured value.} Indeed the measured reactor angle, observed a decade after this 
theoretical bound was derived, approximately saturates the upper limit.
In order to understand why this is so, we must go beyond the SD assumptions
stated so far, and enter the realms of constrained sequential dominance (CSD).

\subsection{Constrained Sequential Dominance: the minimal predictive seesaw model}
\label{CSDn}
Let us 
return to Eq.\ref{mD} and 
set $d=0$ and $e=f$, with $b=a$ and $c=-a$ \cite{King:2005bj}.
The motivation is that from Eq.\ref{t12} one then approximately expects 
the good phenomenological relations $t_{23}\sim 1$ and $t_{12}\sim 1/\sqrt{2}$, although the value of the reactor angle bounded by Eq.\ref{13} remains to be seen.
With the above assumption, Eq.\ref{2rhn} becomes
\begin{equation}
m^{\nu}=
\left( \begin{array}{ccc}
\frac{a^2}{M_{\rm sol}} & \frac{a^2}{M_{\rm sol}}
& \frac{-a^2}{M_{\rm sol}}  \\
\frac{a^2}{M_{\rm sol}} & \frac{a^2}{M_{\rm sol}}+ \frac{e^2}{M_{\rm atm}}  
& \frac{-a^2}{M_{\rm sol}}+ \frac{e^2}{M_{\rm atm}}  \\
\frac{-a^2}{M_{\rm sol}} 
& \frac{-a^2}{M_{\rm sol}}+ \frac{e^2}{M_{\rm atm}}
& \frac{a^2}{M_{\rm sol}}+ \frac{e^2}{M_{\rm atm}}  
\end{array}
\right).
\label{CSD}
\end{equation}
By explicit calculation one then finds that the neutrino mass matrix is exactly diagonalised by the TB mixing matrix in Eq.\ref{TB},
\begin{equation}
U_{\mathrm{TB}}^T m^{\nu} U_{\mathrm{TB}}=
\left( \begin{array}{ccc}
0 & 0 & 0  \\
0 & \frac{3a^2}{M_{\rm sol}} & 0 \\
0 & 0 & \frac{2e^2}{M_{\rm atm}}  
\end{array}
\right).
\end{equation}
If the charged lepton mass matrix is diagonal, the interpretation 
is that these constrained couplings $d=0$, $e=f$ with $b=a$ and $c=-a$
lead to TB mixing, with the lightest neutrino mass $m_1=0$, the second lightest neutrino identified
as the solar neutrino with mass $m_2=\frac{3a^2}{M_{\rm sol}}$ and the heaviest neutrino identified
as the atmospheric neutrino with mass $m_3=\frac{2a^2}{M_{\rm atm}}$.
While TB mixing accurately gives the good relations $t_{23}=1$ and $t_{12}=1/\sqrt{2}$,
unfortunately it also gives $\theta_{13}=0$.
This is known as constrained sequential dominance (CSD) \cite{King:2005bj}.

We can generalise the original idea of CSD to other examples of 
Dirac mass matrix with (in the notation of Eq.\ref{mD})
$d=0$ and $e=f$ as before, but now with 
$b=na$ and  $c=(n-2)a$, for any postive integer $n$, which we refer to as CSD($n$).
The original CSD in Eq.\ref{CSD} with $b=a$ and $c=-a$ is identified as the 
special case CSD($n=1$).
The motivation for CSD($n$) is that for any $n$ Eq.\ref{t12} implies $t_{23}\sim 1$ and $t_{12}\sim 1/\sqrt{2}$, 
although these results are strongly dependent on
the relative phase between the first and second column
of the Dirac mass matrix. 
CSD($n$) then corresponds to the following pattern of couplings in the Dirac mass matrix
in Eq.\ref{mD}:
\begin{itemize}
\item CSD($1$): $(m^D_{\rm atm})^T=(0,e,e)$, \ $(m^D_{\rm sol})^T=(a,a,-a)$ \cite{King:2005bj}.
\item CSD($2$): $(m^D_{\rm atm})^T=(0,e,e)$, \ $(m^D_{\rm sol})^T=(a,2a,0)$  
\cite{Antusch:2011ic}.
\item CSD($3$): $(m^D_{\rm atm})^T=(0,e,e)$, \ $(m^D_{\rm sol})^T=(a,3a,a)$
\cite{King:2013iva}.
\item CSD($4$): $(m^D_{\rm atm})^T=(0,e,e)$, \ $(m^D_{\rm sol})^T=(a,4a,2a)$ \cite{King:2013iva,King:2013xba,King:2013hoa}.
\item CSD($n$): $(m^D_{\rm atm})^T=(0,e,e)$, \ $(m^D_{\rm sol})^T=(a,na,(n-2)a)$ \cite{Bjorkeroth:2014vha}.
\end{itemize}
For the general case of CSD($n$) the Dirac mass matrix is then,
\begin{equation}
	m^D = 
	Y^{\nu}v_u=\pmatr{0 & a \\ e & na \\ e & (n-2)a }.	\label{mDn}
\end{equation}
The constrained couplings will be justified later with the help of discrete family symmetry.
For now we simply assume these couplings motivated by the desire to obtain an approximately 
maximal atmospheric angle $\tan \theta_{23} \sim e/f \sim 1$ and 
trimaximal solar angle 
$\tan \theta_{12} \sim \sqrt{2}a/(b-c) \sim 1/\sqrt{2}$.
Since experiment indicates that the bound $\theta_{13} \lesssim m_2/m_3$ is almost saturated, these schemes require certain phase choices 
$\arg (a/e)$ in order to achieve the desired reactor angle, leading to predictions for the \CP-violating phase $\delta_{CP}$, discussed below.%

In a CSD($n$) framework \cite{Bjorkeroth:2014vha}, 
the low energy effective Majorana neutrino mass matrix in Eq.\ref{2rhn}
in the two right-handed neutrino case may be written as,
\begin{equation}
	m^\nu_{(n)} = m_a 
	\left(
\begin{array}{ccc}
	0&0&0\\0&1&1\\0&1&1 
	\end{array}
\right)
	+ m_b e^{i\eta} 
	\left(
\begin{array}{ccc}
	1&n&n-2\\n&n^2&n(n-2)\\n-2&n(n-2)&(n-2)^2
	\end{array}
\right),
	\label{eq:mnu2}
\end{equation}
where $\eta$ is the only physically important phase, which depends on the relative phase between the first and second column of the Dirac mass matrix, $\arg (a/e)$.
By comparing Eqs.\ref{CSD} and \ref{eq:mnu2} for $n=1$
we identify $m_a=\frac{e^2}{M_{\rm atm}}$ and $m_b=\frac{a^2}{M_{\rm sol}}$,
which hold for any value of $n$.
This can be thought of as the minimal (two right-handed neutrino) predictive seesaw model since, for a given $n$,
only three parameters $m_a, m_b, \eta$ describe the entire neutrino sector (three neutrino masses
and the PMNS matrix).
CSD($n$) with two right-handed neutrinos always 
predicts the lightest physical neutrino mass to be zero, $m_1=0$.
It also immediately predicts TM1 mixing since,
\begin{equation}
m^\nu_{(n)} 
\left(
\begin{array}{c}
2 \\
-1\\
1
\end{array}
\right)
=
\left(
\begin{array}{c}
0 \\
0\\
0
\end{array}
\right).
\label{CSD(n)a}
\end{equation}
In other words the column vector $(2,-1,1)^T$
is an eigenvector of $m^\nu_{(n)} $ with a zero eigenvalue, i.e. it is the first column of the PMNS mixing matrix,
corresponding to $m_1=0$,
which means TM1 mixing in Eq.\ref{TMM}.

\begin{figure}[ht]
	\centering
	\includegraphics[scale=0.53]{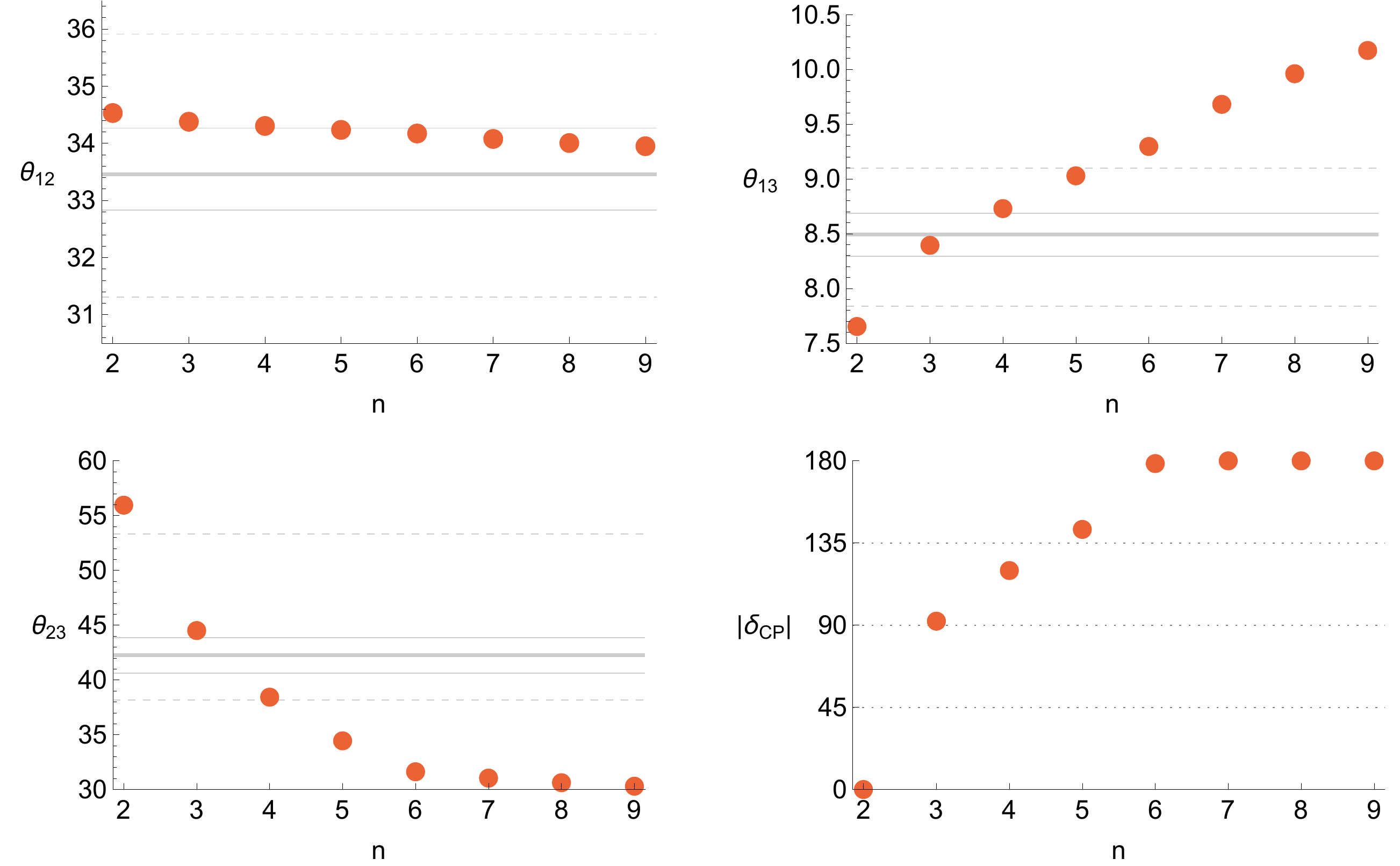}
	\caption{\footnotesize{Best-fit PMNS mixing angles and \CP-violating phase with respect to $n$, for the two right-handed neutrino CSD($n$) model. We emphasise that $|\delta_{\rm CP} |$ is a genuine prediction here since have not used the one sigma hint from experiment as an input constraint. CSD(3) and CSD(4) both yield predictions for mixing angles within the preferred range with differing predictions for the atmospheric angle $\theta_{23}\approx 45^{\circ}$ and $\theta_{23}\approx 38^{\circ}$, respectively.
Interestingly CSD(3) and CSD(4) lead to predictions for $\delta_{\rm CP} \sim -90^{\circ}$
and $\delta_{\rm CP} \sim -120^{\circ}$ with $\eta \sim 2\pi/3$ and $\eta \sim 4\pi/5$
being the best fit values of the input phase.}}
	\label{fig:mixingangles2nu}
\end{figure}

For a given choice of the positive integer \( n \), there are three real input parameters \( m_a \), \( m_b \) and \( \eta \) from which two light physical neutrino masses \( m_2 \), \( m_3 \), three lepton mixing angles, the \CP-violating phase \( \delta_{CP} \) and two Majorana phases are derived; a total of nine physical parameters 
(including the prediction $m_1=0$) from
three input parameters, i.e. six predictions for each value of $n$. As the Majorana phases are not known and 
\( \delta_{CP} \) is only tentatively constrained by experiment, this leaves five 
presently measured observables, namely the two neutrino mass squared differences and the three lepton mixing angles, from only three input parameters. 
Essentially the input parameters 
$m_a$ and $m_b$ are fixed by the two physical neutrino mass squared differences, which implies that 
the entire PMNS mixing matrix is determined by only a single parameter, namely the phase $\eta$.
The resulting best fit predictions for CSD($n$) \cite{Bjorkeroth:2014vha} are shown in Fig.\ref{fig:mixingangles2nu}
as a function of $n$. As can be seen, CSD($2$) gives a reactor angle which is too small,
\footnote{This is an example of the general result that two right-handed neutrino models with two texture zeros
and a normal hierarchy are phenomenologically excluded \cite{Harigaya:2012bw}.}
while CSD($n\geq 5$) gives a reactor angle which is too large. CSD($3$) and CSD($4$) allow
a reactor angle in the desired Goldilocks (experimentally preferred) region $\theta_{13}\sim 8.5^{\circ}$.
This value occurs for special choices of phase $\eta \sim 2\pi/3$ for CSD($3$) and
$\eta \sim 4\pi/5$ for CSD($4$) where positive values of these phases yield negative
values of \( \delta_{CP} \sim -90^{\circ}\) for CSD($3$) and \( \delta_{CP} \sim -120^{\circ}\) for CSD($4$),
with the mixing angles being independent of the sign of the phase.

\begin{figure}[t]
\centering
\includegraphics[width=0.4\textwidth]{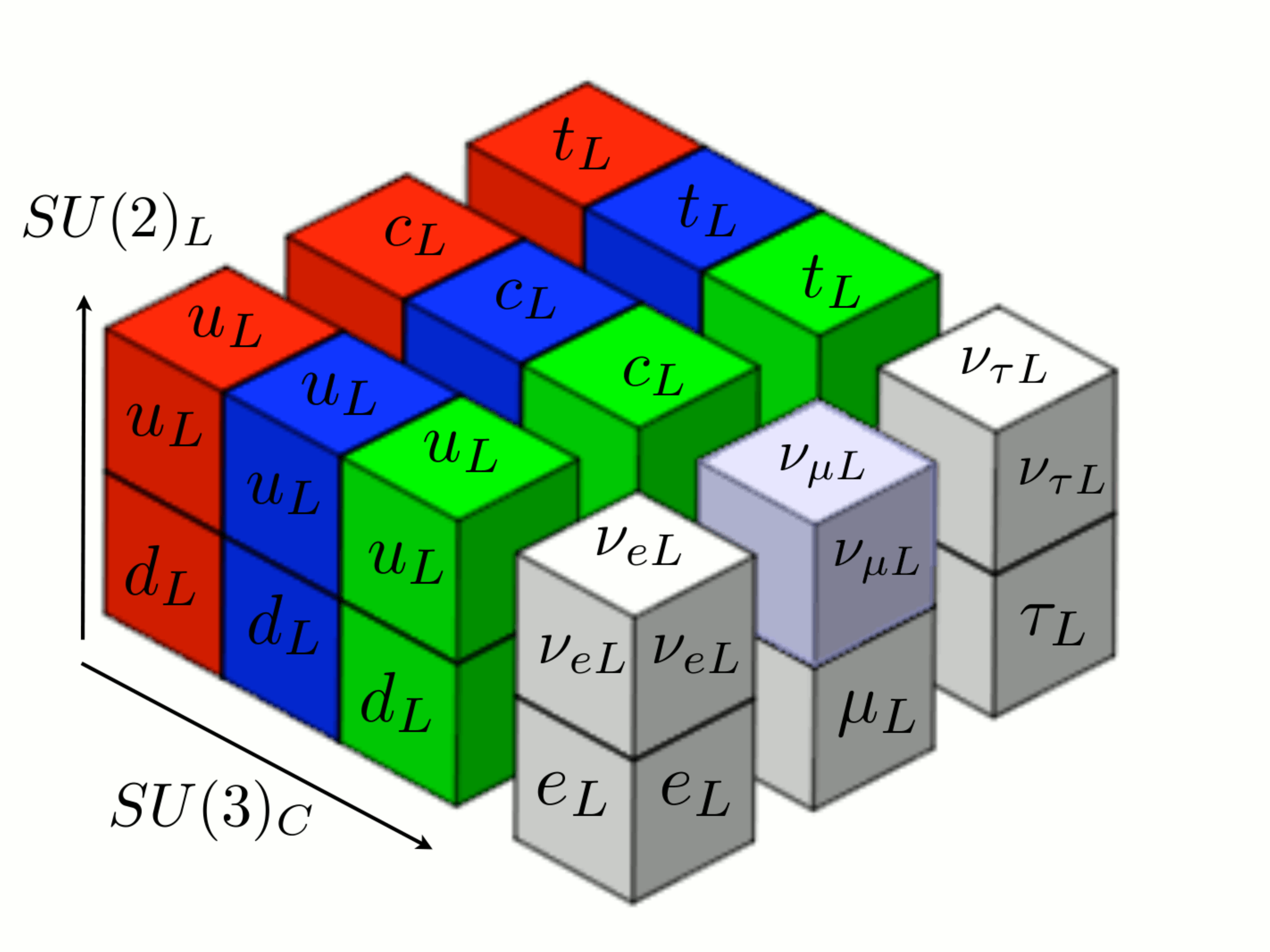}
\includegraphics[width=0.4\textwidth]{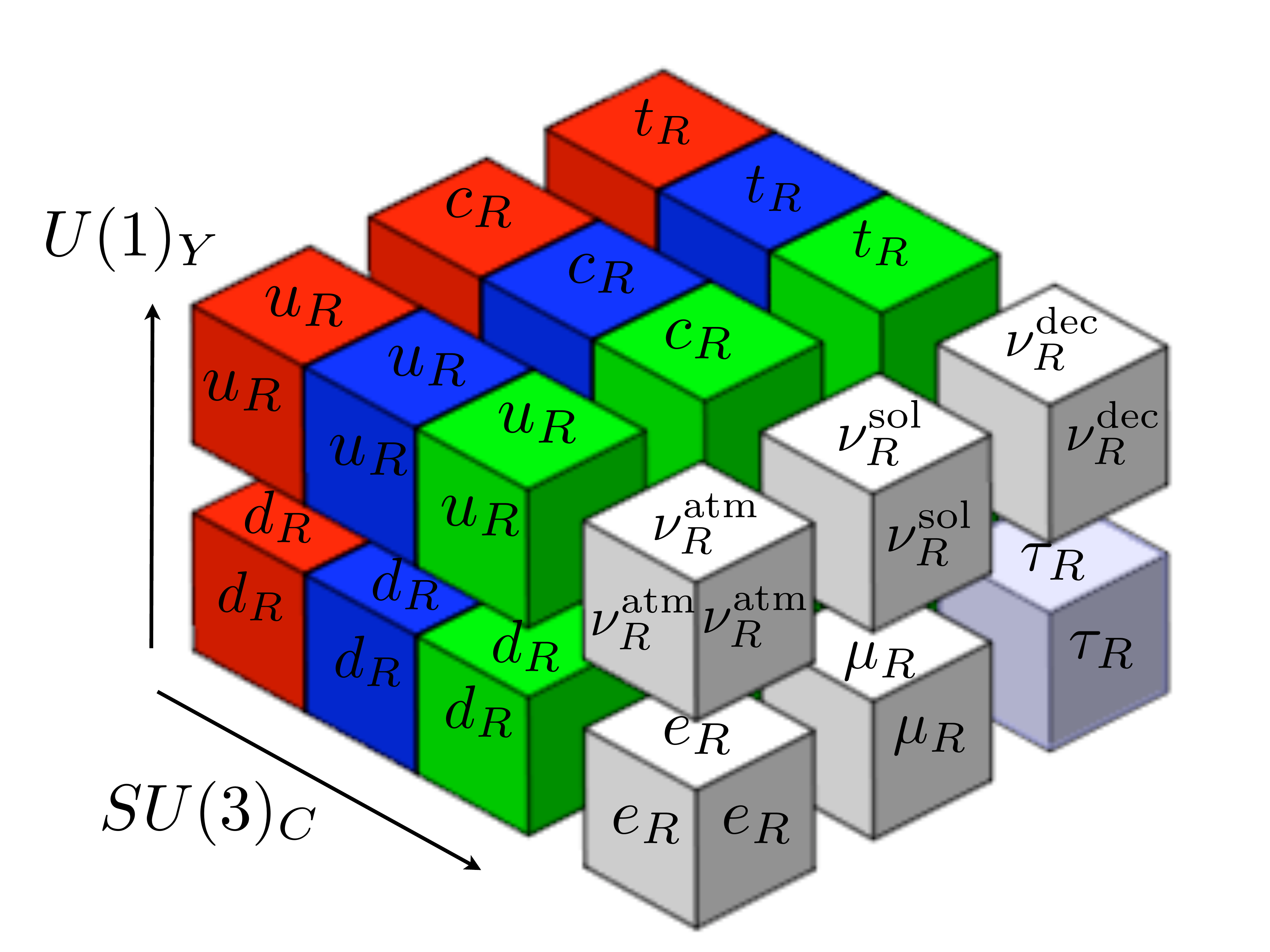}
\vspace*{-4mm}
    \caption{The Standard Model with three right-handed neutrinos defined as
    $(\nu_R^{\rm atm},\nu_R^{\rm sol},\nu_R^{\rm dec})$ which in sequential dominance are mainly responsible for the $m_3,m_2,m_1$ physical neutrino masses, respectively.} \label{SM}
\vspace*{-2mm}
\end{figure}

\subsection{See-saw mechanism with three right-handed neutrinos and sequential dominance}
\label{SD}
It is straightforward to extend the previous ideas of SD to the case of three right-handed neutrinos
\cite{King:1998jw}.
The starting assumption is that of the Standard Model supplemented by three right-handed
neutrinos with masses in the classic see-saw range TeV to M$_{\rm GUT}$
in Fig.\ref{steriles}.
Then the light neutrino mass matrix emerges from the see-saw formula
in Eq.\ref{seesaw}.
The basic idea of SD was given in the framework of 
the two-right handed neutrino model above, but can now be extended to a third,
almost decoupled, right-handed neutrino, as follows.

Extending the preceding example of two right-handed neutrinos,
it is possible to implement the see-saw mechanism with three right-handed neutrinos
using the sequential dominance (SD) mechanism \cite{King:1998jw}.
The SD assumption can be made precise as follows.
In the basis,
$M_R={\rm diag}(M_{\rm atm},M_{\rm sol},M_{\rm dec})$ 
where the Dirac mass matrix is constructed 
by extending Eq.\ref{mD} to the three columns
$m^D=(m^D_{\rm atm},m^D_{\rm sol},m^D_{\rm dec})$, 
where $m^D_{\rm atm}=(d,e,f)^T$, $m^D_{\rm sol}=(a,b,c)^T$, etc.,
the SD assumption in Eq.\ref{SD0} generalises to:
\begin{equation}
\frac{(m^D_{\rm atm})^{\dagger}m^D_{\rm atm}}{M_{\rm atm}}
\gg
\frac{(m^D_{\rm sol})^{\dagger}m^D_{\rm sol}}{M_{\rm sol}}
\gg
\frac{(m^D_{\rm dec})^{\dagger}m^D_{\rm dec}}{M_{\rm dec}}.
\label{SD}
\end{equation}
Eq.\ref{SD} 
immediately predicts a normal neutrino mass hierarchy:
\begin{equation}
 m_3\gg m_2 \gg m_1
 \end{equation}
which is the main consequence of SD.
The lightest physical neutrino mass $m_1$ is much smaller than the others since the corresponding right-handed neutrino $\nu_R^{\rm dec}$ being approximately decoupled from the see-saw mechanism.
The heaviest physical neutrino has mass $m_3$ much larger than $m_2$ since 
the atmospheric right-handed neutrino makes the dominant contribution to the see-saw mechanism.
The model approximates to the two right-handed neutrino case (as considered previously) where $m_1=0$.
The SM with three such right-handed neutrinos is depicted in Fig.\ref{SM}.

\newpage

\section{Towards a Theory of Flavour}
\label{ToF}
The flavour problem may be defined as the following collection of puzzles left unanswered by the SM:
\begin{itemize}
\item Why are there three families of quarks and leptons?
\item Why are all charged fermion masses so hierarchical with down-type quark masses being of the same order as charged lepton masses, and up-type quark masses are much more hierarchical?
\item Why are at least two neutrino masses not very hierarchical?
\item What is the origin of the neutrino mass?
\item Why are neutrino masses so tiny compared to charged fermion masses? 
\item What is the origin of fermion mixing (both CKM and PMNS matrices)?
\item Why are CKM mixing angles smaller than PMNS mixing angles apart from the Cabibbo angle
which is of the same order as the reactor angle?
\item What is the origin of \CP violation in the quark (and lepton) sectors?
\end{itemize}
These questions motivate the search for a theory of flavour beyond the SM.
In this section we explore possible directions towards a theory of flavour based on the ideas
of symmetry, in particular unification together with family symmetry.

\subsection{\label{sec:GUTs}Grand unified theories}
\begin{figure}[t]
\begin{center}
\includegraphics[width=0.93\textwidth]{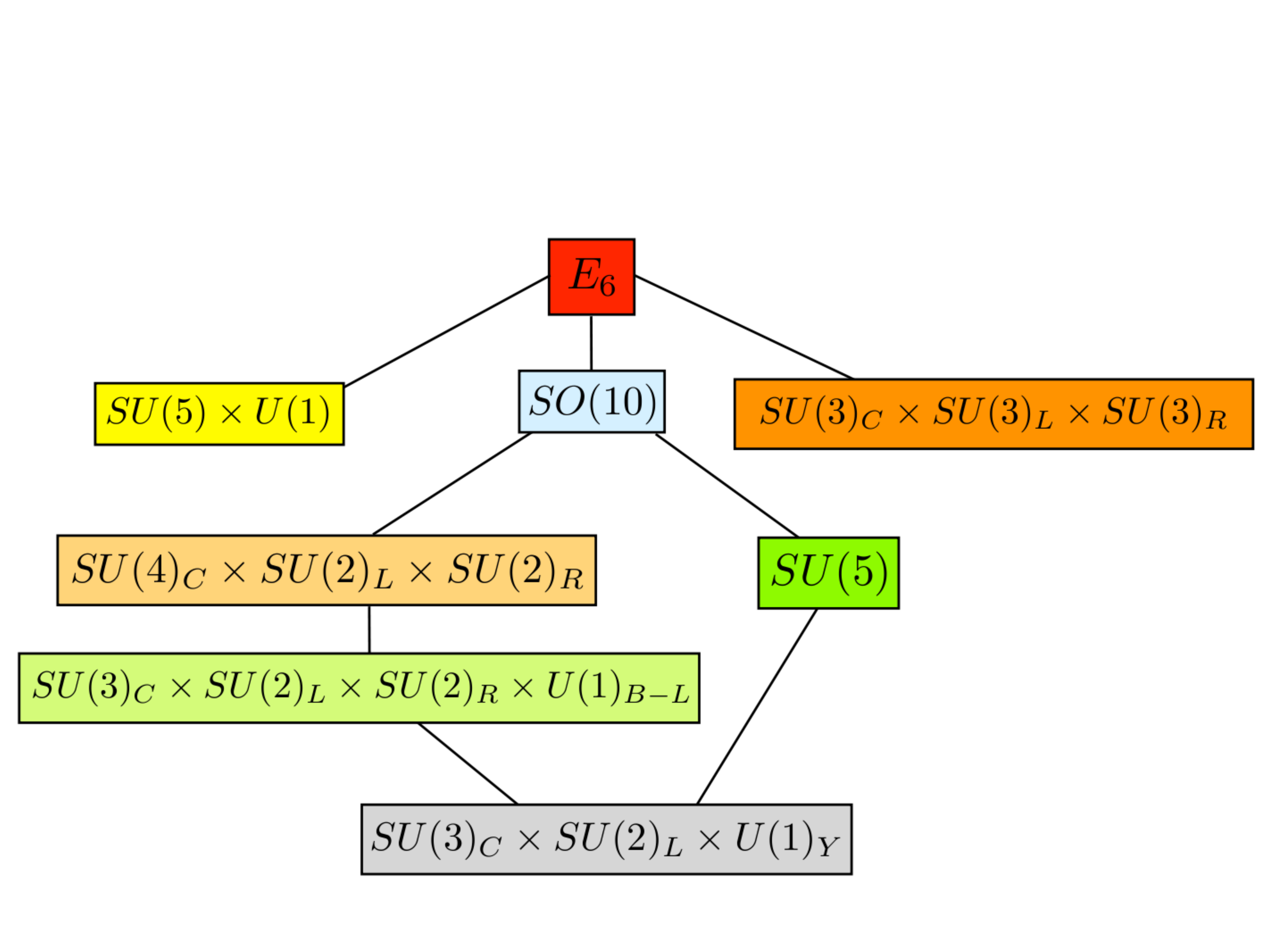}
\end{center}
\vspace*{-4mm}
    \caption{\label{GUTs}\small{Some possible candidate unified gauge groups.}}
\vspace*{-2mm}
\end{figure}

One of the exciting things about the discovery of neutrino masses
and mixing angles is that this provides additional information
about the flavour problem - the problem of understanding the origin
of three families of quarks and leptons and their masses and mixing
angles. In the framework of the seesaw mechanism, new physics beyond the
Standard Model is required to violate lepton number and generate
right-handed neutrino masses which may be as large as the
GUT scale. This is also exciting since it implies that
the origin of neutrino masses is also related to some
GUT symmetry group $G_{{\rm GUT}}$, which unifies the
fermions within each family.
Some possible candidate unified gauge groups are shown in Fig.~\ref{GUTs}.

Let us take $G_{{\rm GUT}}=SU(5)$ as an example \cite{Georgi:1974sy}. Each family of 
quarks (with colour $r,b,g$) and leptons fits nicely into $SU(5)$ representations 
of left-handed ($L$) fermions, $F=\overline{\bf 5}$ and $T={\bf 10}$
\begin{equation}
F= \left(
\begin{array}{c}d_r^c\\d_b^c\\d_g^c\\e^-\\-\nu_e \end{array}\right)_L  , \qquad
T= \left(
\begin{array}{ccccc} 0&u_g^c&-u_b^c&u_r&d_r\\
.&0&u_r^c&u_b&d_b\\ 
.&.&0&u_g&d_g\\ 
.&.&.&0&e^c\\
.&.&.&.&0
\end{array}\right)_L  \ ,
\end{equation}
where $c$ denotes \CP conjugated fermions. The $SU(5)$ representations
$F=\overline{\bf 5}$ and $T={\bf 10}$ decompose into multiplets of the SM
gauge group $SU(3)_C\times SU(2)_L\times U(1)_Y$ as $F=(d^c,L)$,
corresponding to, 
\begin{equation}
\overline{\bf 5}=(\overline{\bf 3},{\bf 1},1/3)\oplus  ({\bf 1},\overline{\bf 2},-1/2),
\end{equation}
and
$T=(u^c,Q,e^c)$, corresponding to,
\begin{equation}
{\bf 10} =(\overline{\bf 3},{\bf 1},-2/3)\oplus  
({\bf 3},{\bf 2},1/6)\oplus ({\bf 1},{\bf 1},1).
\end{equation}
Thus a complete quark and lepton SM family $(Q,u^c,d^c,L,e^c)$ is accommodated 
in the $F=\overline{\bf 5}$ and $T={\bf 10}$ representations, with
right-handed neutrinos, whose \CP conjugates are denoted as $\nu^c$, being singlets of $SU(5)$, $\nu^c={\bf 1}$. The Higgs doublets $H_u$ and $H_d$
which break electroweak symmetry in a two Higgs doublet model
are contained in the $SU(5)$ multiplets $H_{\bf 5}$ and $H_{\overline{\bf 5}}$.

The Yukawa couplings for one family of quarks and leptons are given by,
\begin{equation}
y_u H_{{\bf 5}i}T_{jk}T_{lm}\epsilon^{ijklm}+ y_{\nu}H_{{\bf 5}i}F^i\nu^c+
y_d H_{\overline{\bf 5}}^iT_{ij}F^j,
\end{equation}
where $\epsilon^{ijklm}$ is the totally antisymmetric tensor of $SU(5)$ with
$i,j,j,k,l=1,\ldots, 5$,
which decompose into the SM Yukawa couplings
\begin{equation}
y_u H_uQu^c+  y_{\nu}H_uL\nu^c+
y_d (H_dQd^c+H_de^cL).
\end{equation}
Notice that the Yukawa couplings for down quarks and charged leptons are equal
at the GUT scale. Generalising this relation to all three families we find the $SU(5)$ prediction for Yukawa matrices,
\begin{equation}
Y_d=Y_e^T,
\end{equation}
which is successful for the third family, but fails badly for the first and
second families. Georgi and Jarlskog~\cite{Georgi:1979df} suggested to
include a higher Higgs representation 
$H_{\overline{\bf 45}}$ which is responsible for the 2-2 entry of the down and charged lepton Yukawa matrices. Dropping $SU(5)$ indices for clarity,
\begin{equation}
(Y_{d})_{22} H_{\overline{\bf 45}}T_2F_2 ,
\end{equation}
decomposes into the second family SM Yukawa couplings
\begin{equation}
(Y_{d})_{22}(H_dQ_2d_2^c-3H_de_2^cL_2),
\end{equation}
where the factor of $-3$ is an $SU(5)$ Clebsch-Gordan
coefficient.\footnote{In this setup, $H_d$ is the light linear combination of the
electroweak doublets contained in $H_{\bf{\overline 5}}$~and~$H_{\bf{\overline {45}}}$.}
Assuming a hierarchical Yukawa matrix with a zero Yukawa element (texture) in
the 1-1 position, results in the GUT scale Yukawa relations, 
\begin{equation}
y_b = y_{\tau}, \quad  y_s = \frac{y_{\mu}}{3}, \quad y_d = 3y_e, 
\end{equation}
which, after renormalisation group running effects are taken into account, are consistent with the low energy masses.
The precise viability of these relations has been widely discussed in the light of recent progress in 
lattice theory which enable more precise values of quark masses to be determined, especially the strange quark mass
(see, e.g.,~\cite{Ross:2007az}). In supersymmetric (SUSY) theories with low values of the ratio of Higgs vacuum expectation values, the relation for 
the third generation $y_b = y_{\tau}$ at the GUT scale remains viable,
but a viable GUT scale ratio of $y_\mu/y_s$ is more accurately achieved within
SUSY $SU(5)$ GUTs using a Clebsch factor of $9/2$, as proposed
in~\cite{Antusch:2009gu}, which is 50\% higher than the Georgi-Jarlskog
prediction of $3$. For other Clebsch relations see~\cite{Antusch:2013rxa}.

\begin{figure}[t]
\centering
\includegraphics[width=0.60\textwidth]{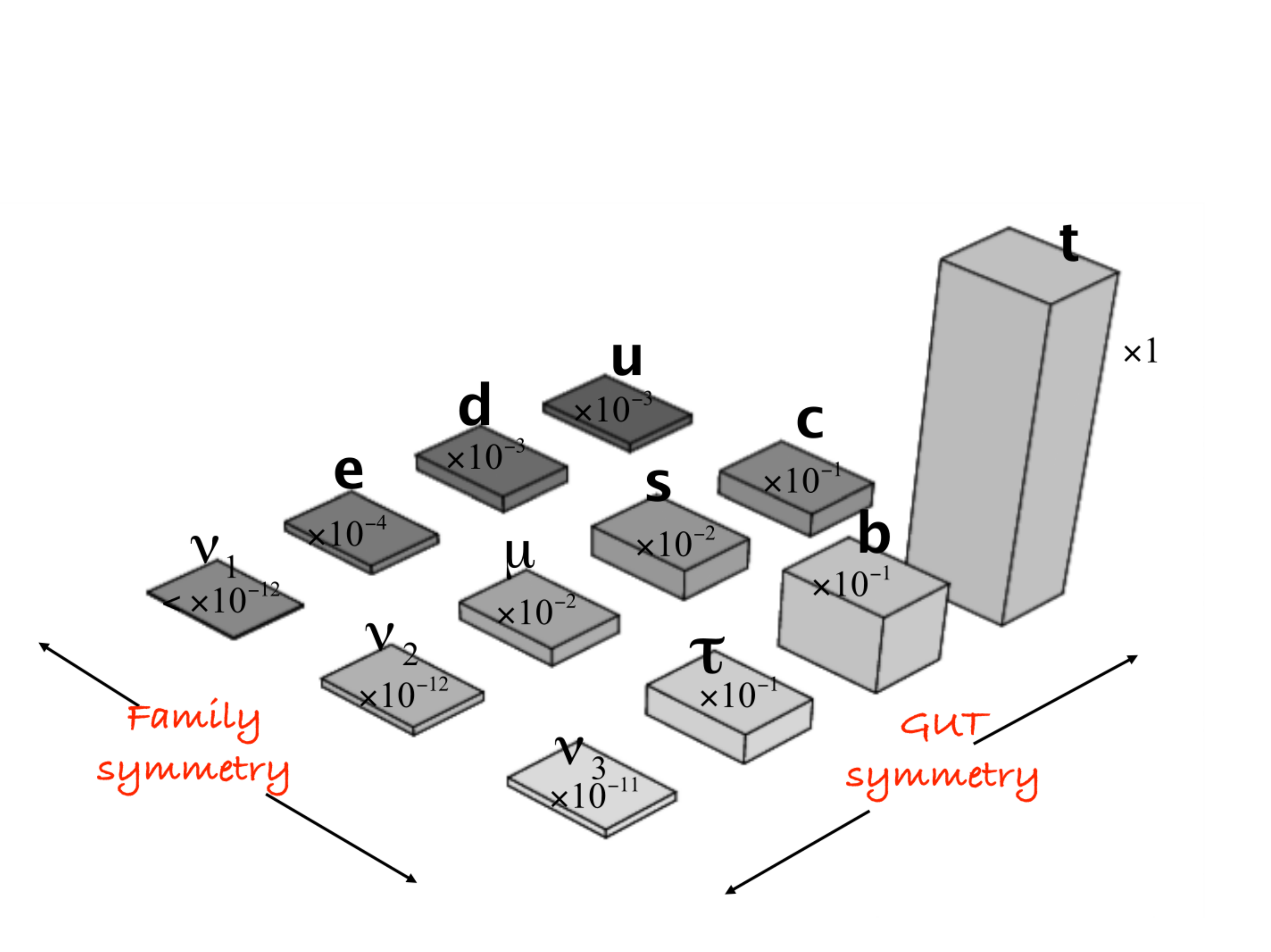}
\vspace*{-4mm}
    \caption{Quark and lepton masses lego plot (true heights need to be scaled by the factors shown)
    indicating the directions that GUT and family symmetries are acting.   } \label{masses}
\vspace*{-2mm}
\end{figure}

\subsection{Discrete family symmetry}
\label{sec:GUTxFam}
As already remarked, it is a remarkable
fact that the smallest leptonic mixing angle, 
the reactor angle, is of a similar magnitude to the largest quark mixing
angle, the Cabibbo angle, indeed they  
may even be equal to each other up to a factor of $\sqrt{2}$. Such
relationships may be a hint of a connection between leptonic mixing and quark
mixing, where such a connection might be achieved using
GUTs~\cite{Antusch:2011qg,Marzocca:2011dh}. 
For example, the Georgi-Jarlskog relations discussed above already lead to the 
left-handed charged lepton mixing angle having a simple relation with the
right-handed down-type quark mixing angle 
$\theta_{12}^{e_L} \approx \theta_{12}^{d_R}/3$ where the approximation assumes hierarchical Yukawa matrices,
with the 1-1 elements being approximately zero. If the upper $2\times 2$
Yukawa matrices are symmetric (as motivated by the successful
Gatto-Sartori-Tonin (GST) relation~\cite{Gatto:1968ss} which relates the 12
mixing $\theta_{12}^{d_{L,R}}$ to the down and strange mass by
$\theta_{12}^{d_{L,R}}  \approx\sqrt{m_d/m_s}$) then we may drop the $L,R$
subscripts and this relation simply becomes  $\theta_{12}^{e} =
\theta_{12}^{d}/3$. In large classes of models, the quark mixing originates predominantly from the down-type quark
sector, in which case this relation becomes $\theta_{12}^{e} = \theta_C/3$.
If one starts from TB mixing in the neutrino sector, resulting from some discrete family symmetry,
then, using the results in subsection~\ref{solar} such a charged lepton correction results in a reactor angle in the lepton sector of $\theta_{13}\approx \theta_C/(3\sqrt{2})$. 
This is a factor of 3 too small to account for the
observed reactor angle, but it illustrates how the reactor angle could possibly be related to the Cabibbo angle using GUTs. 
Indeed it has been suggested that perhaps the charged lepton mixing angle is exactly equal to the Cabibbo angle
in some GUT model, leading to $\theta_{13}\approx
\theta_C/\sqrt{2}$~\cite{King:2012vj,Zhang:2012mn}. However it
is non-trivial to reconcile such large charged lepton mixing with 
the successful relationships between charged lepton and down-type quark masses, and it seems more likely that
charged lepton mixing is not entirely responsible for the reactor angle.

\begin{figure}[t]
\centering
\includegraphics[width=0.60\textwidth]{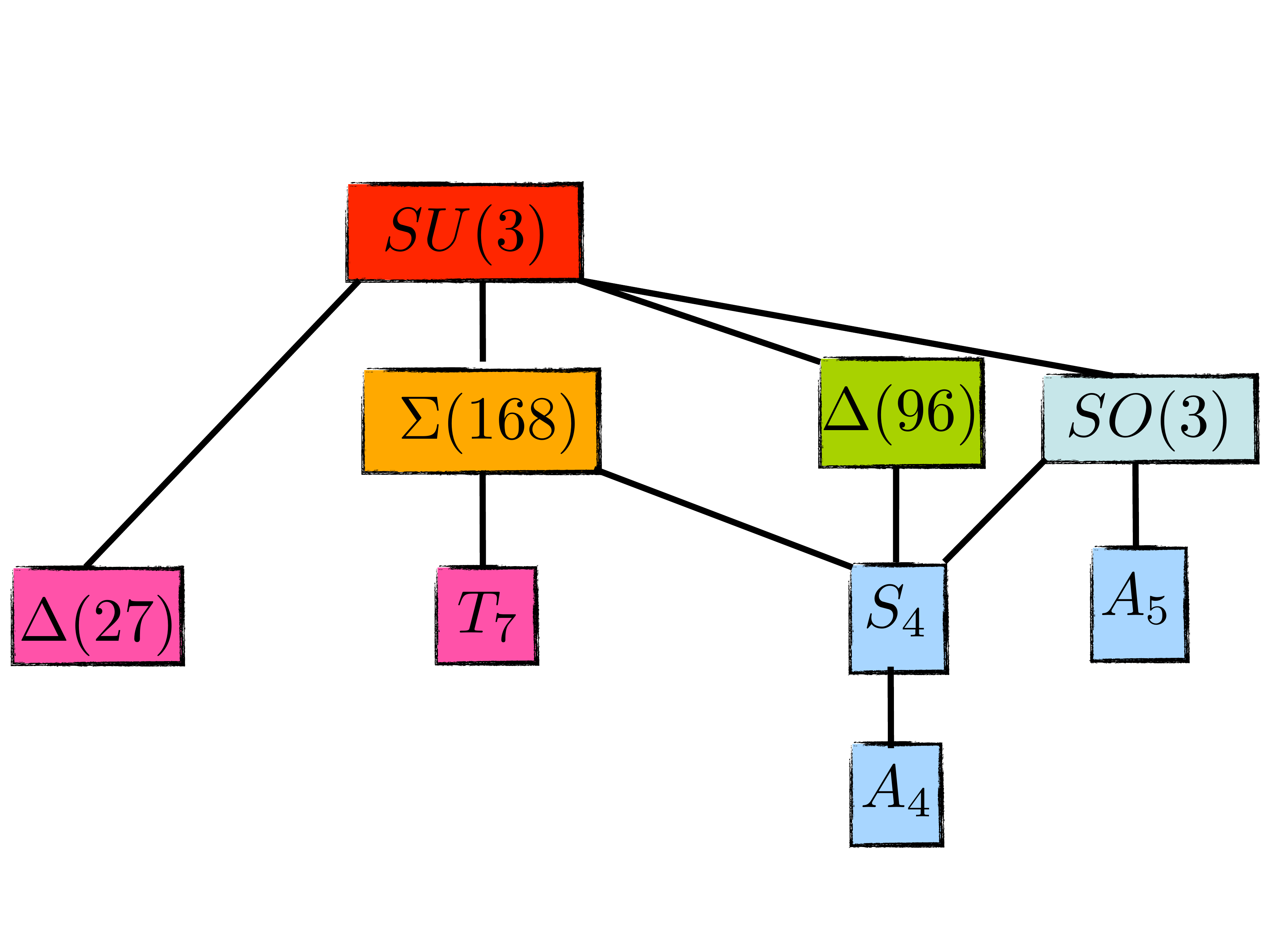}
\vspace*{-4mm}
    \caption{Some popular family symmetries and their relationships. } \label{discrete}
\vspace*{-2mm}
\end{figure}

The above discussion provides an additional motivation for
combining GUTs with discrete family symmetry in order to account for the reactor angle.
Putting these two ideas together we are suggestively led to a framework of
new physics beyond the Standard Model based on commuting
GUT and family symmetry groups,
\begin{equation}
G_{{\rm GUT}}\times G_{{\rm FAM}} .
\label{symmetry}
\end{equation}

The spectrum of quark and lepton masses may also provide some motivation for considering 
a family symmetry as well as a grand unified symmetry, acting in different directions, 
as illustrated in Fig.\ref{masses}. The (scaled) heights of the towers representing the fermion masses,
show vast hierarchies which are completely mysterious in the SM. 
Some popular family symmetries
which admit triplet representations are shown in Fig.\ref{discrete}.
The mathematics of these and other groups has recently been reviewed in \cite{Ishimori:2010au,King:2013eh,King:2014nza} to which we refer the interested reader for more details.

Here we just mention the family symmetry $A_4$ as it is the smallest non-Abelian
finite group with an irreducible triplet representation. 
$A_4$ is the symmetry group of the tetrahedron. 
There are 12 independent transformations of the tetrahedron and hence 12 group elements as follows:
\begin{itemize}
\item 4 rotations by 120 degrees clockwise (seen from a vertex) which are $T$-type 
\item 4 rotations by 120 degree anti-clockwise(seen from a vertex) which are $T$-type  
\item 3 rotations by 180 degrees which are $S$-type 
\item 1 unit operatator $\mathcal{I}$
\end{itemize}
The generators of the $A_4$ group,
can be written as $S$ and $T$ with $S^2=T^3=(ST)^3=\mathcal{I}$.
All 12 group elements can be formed by multiplying together these two generators in all possible ways.
$A_4$ has four irreducible representations, three singlets
$1,~1^\prime$ and $1^{\prime \prime}$ and one triplet. 
The products of singlets are:
\begin{equation}\begin{array}{llll}
1\otimes1=1, &1^\prime\otimes1^{\prime\prime}=1, &1^{\prime}\otimes1^{\prime}=1^{\prime\prime}
&1^{\prime\prime}\otimes1^{\prime\prime}=1^\prime.
\end{array}
\end{equation}
Later we shall sometimes work in the real basis of the triplet representation \cite{Ma:2001dn}, 
\begin{equation}\label{eq:ST}
S=\left(
\begin{array}{ccc}
1&0&0\\
0&-1&0\\
0&0&-1\\
\end{array}
\right), \ \ \ \ 
T=\left(
\begin{array}{ccc}
0&1&0\\
0&0&1\\
1&0&0\\
\end{array}
\right)\,
\end{equation}
which generate 12 real $3\times 3 $ matrix group elements after multiplying these two matrices together
in all possible ways \cite{Ma:2001dn}.
In this basis one has the following Clebsch rules for the multiplication of two triplets, 
$3\times 3 = 1+1'+1''+3_1+3_2$, with 
\begin{equation}\label{pr}
\begin{array}{lll}
(ab)_1&=&a_1b_1+a_2b_2+a_3b_3\,;\\
(ab)_{1'}&=&a_1b_1+\omega^2 a_2b_2+\omega a_3b_3\,;\\
(ab)_{1''}&=&a_1b_1+\omega a_2b_2+\omega^2 a_3b_3\,;\\
(ab)_{3_1}&=&(a_2b_3,a_3b_1,a_1b_2)\,;\\
(ab)_{3_2}&=&(a_3b_2,a_1b_3,a_2b_1)\,,
\end{array}
\end{equation}
where $a=(a_1,a_2,a_3)$ and $b=(b_1,b_2,b_3)$
are the two triplets and $\omega^3=1$.

\subsection{Klein symmetry}
The starting point for family symmetry models is to consider the Klein symmetry of the neutrino mass matrix.
First consider the phase symmetry of the diagonal charged lepton mass matrix
$M_e$, 
\begin{equation}
T^{\dagger}(M_e^{\dagger}M_e)T= M_e^{\dagger}M_e
\end{equation}
where $T={\rm diag}(1, \omega , \omega^2)$ and $\omega = e^{2\pi i/n}$.
\footnote
{Note that this is not the same basis as Eq.\ref{eq:ST} since $T$ is diagonal in this basis (but still traceless
since $1+\omega +\omega^2 = 0$).}
For example for $n=3$ clearly $T$ generates the group $Z^T_3$.
In any case, the Klein symmetry of the neutrino mass matrix, in this basis,
is given by,
\begin{equation}
m^{\nu}= S^Tm^{\nu} S, \ \ \ \ m^{\nu}= U^Tm^{\nu} U
\end{equation}
where \cite{King:2009ap}
\begin{eqnarray}
S= U_{\rm PMNS}^*\ {\rm diag}(+1,-1,-1)\ U_{\rm PMNS}^T\\
U= U_{\rm PMNS}^*\ {\rm diag}(-1,+1,-1)\ U_{\rm PMNS}^T\\
SU= U_{\rm PMNS}^*\ {\rm diag}(-1,-1,+1)\ U_{\rm PMNS}^T
\end{eqnarray}
and 
\begin{equation}
{\cal K}=\{1, S, U, SU \}
\end{equation}
is called the Klein symmetry $Z^S_2\times Z^U_2$.

\begin{figure}[t]
\centering
\includegraphics[width=0.60\textwidth]{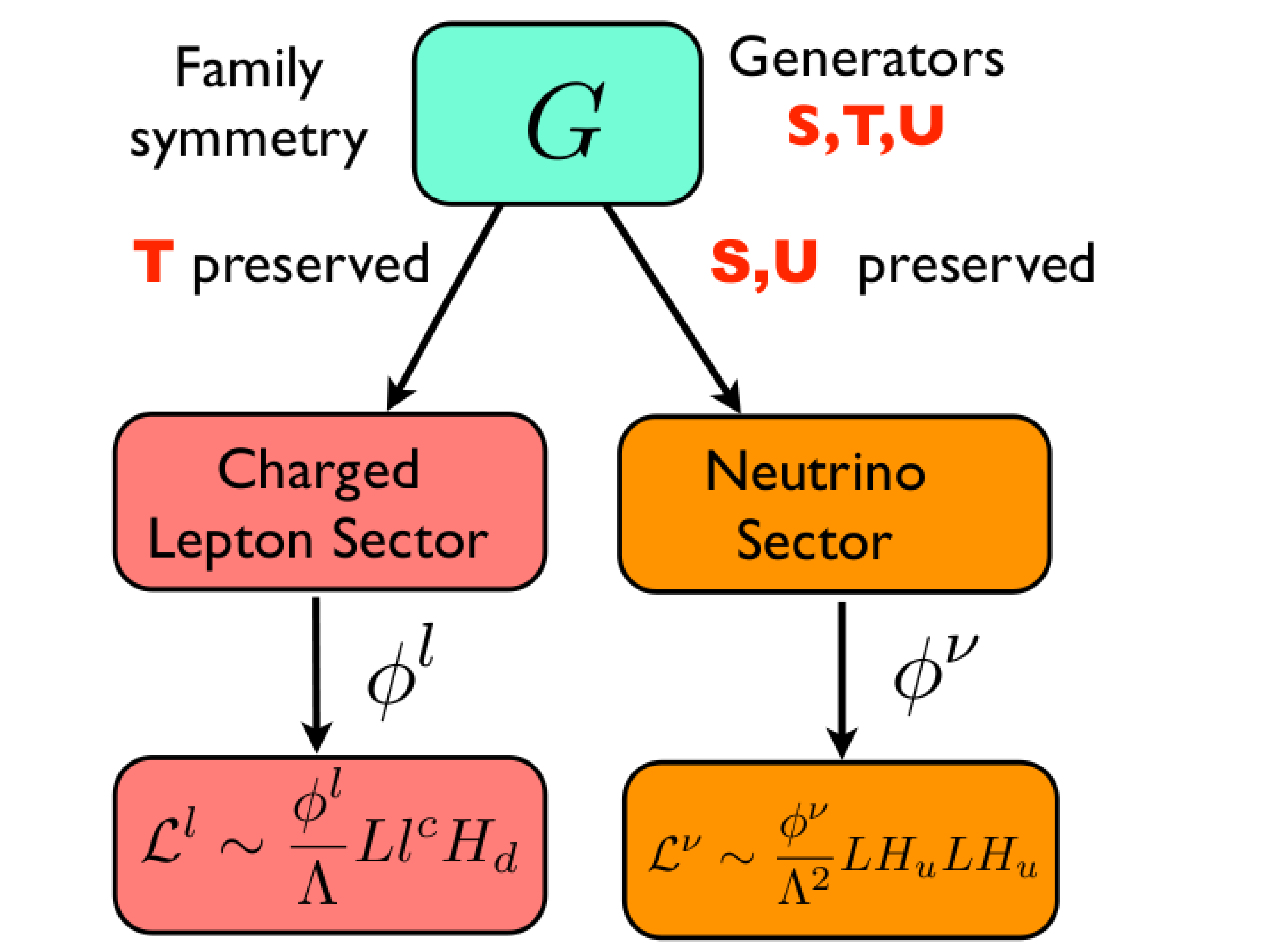}
\vspace*{4mm}
    \caption{The direct approach to models of lepton mixing.     } \label{direct}
\end{figure}

\subsection{Direct models}
The idea of direct models is that the three generators $S,T,U$ introduced above
are embedded into a discrete family symmetry $G$ which is broken by new
Higgs fields called ``flavons'' of two types: $\phi^l$ whose VEVs preserve
$T$ and $\phi^{\nu}$ whose VEVs preserve $S,U$. These flavons are segregated 
such that $\phi^l$ only appears in the charged lepton sector and $\phi^{\nu}$
only appears in the neutrino sector as depicted  in Fig.\ref{direct},
thereby enforcing the symmetries of the mass matrices.
Note that the full Klein symmetry $Z^S_2\times Z^U_2$ of the neutrino mass matrix is enforced by symmetry in the direct approach.

Following the measurement of the reactor angle, it has emerged that the 
only viable direct models are those based on $\Delta (6N^2)$
\cite{Holthausen:2012wt,King:2013vna,Fonseca:2014koa}.
Unfortunately large $N$ is required in order to achieve the desired
reactor angle. Moreover such models generally predict the \CP phase 
$\delta = 0,\pi$ resulting in the atmospheric sum rule
\cite{King:2013vna},
\begin{equation}
\theta_{23}=45^o\mp \theta_{13}/\sqrt{2}.
\end{equation}
which follows since the PMNS matrix has the TM2 form
shown in Eq.\ref{TMM}. 

\subsection{Spontaneous \CP violation}
The inclusion of discrete family symmetry and GUTs into a theory of flavour offers the possibility
of having spontaneously broken \CP symmetry.
The idea is that the high energy theory respects \CP but it becomes spontaneously broken along with 
the discrete family symmetry and GUT symmetry.
As with all types of spontaneous symmetry breaking, this offers the possibility of understanding
the origin of \CP violation, and relating \CP violation in the quark and lepton sectors.
For example, it is possible that the \CP violating phases in the quark and lepton sectors may be predicted
within this kind of approach.

\begin{figure}[t]
\centering
\includegraphics[width=0.6\textwidth]{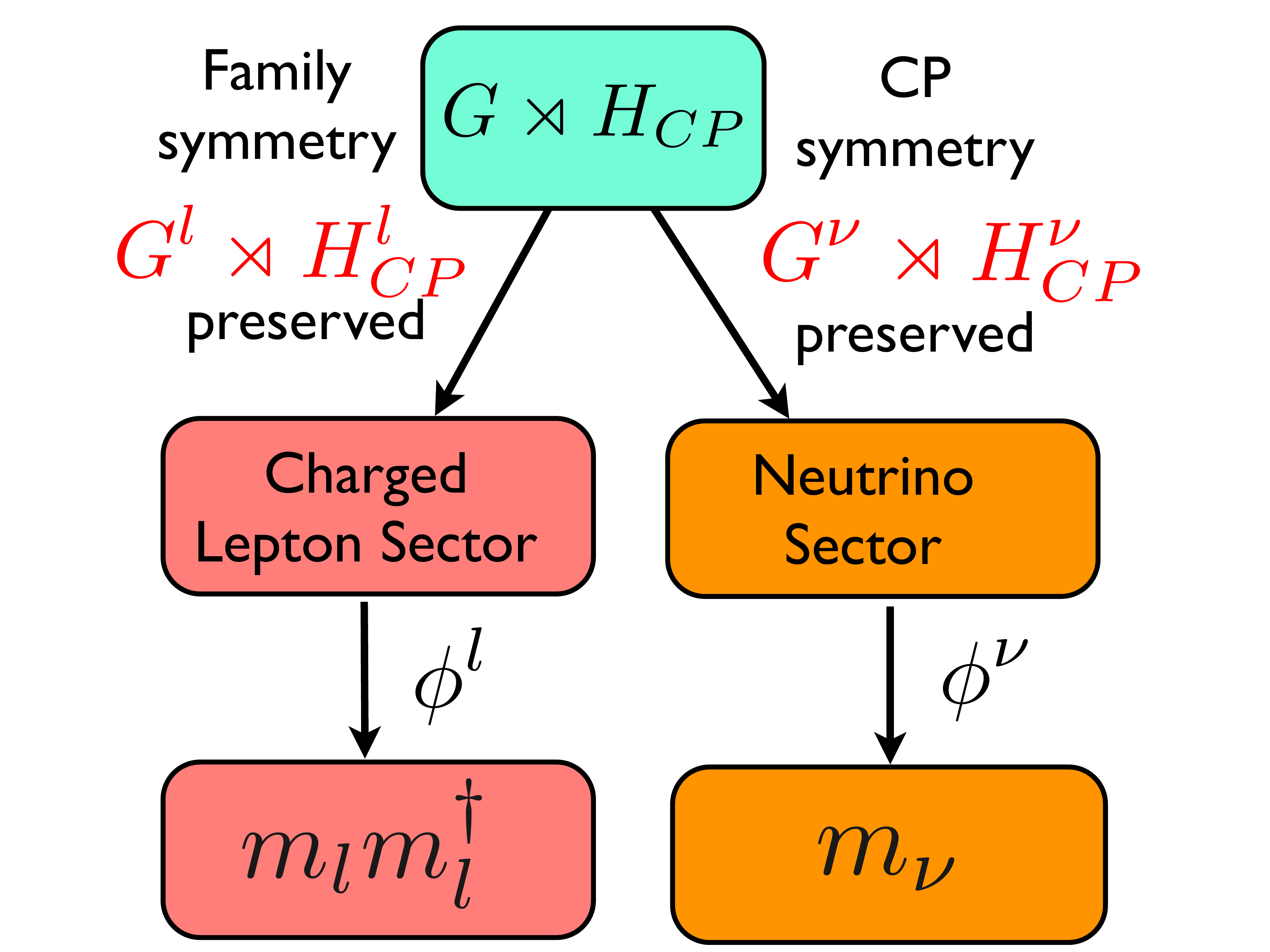}
\vspace*{4mm}
    \caption{The direct approach including \CP. The idea is that the original high energy theory conserves \CP but \CP is spontaneously broken in the low energy theory. Nevertheless one may define separate \CP 
    symmetries which are preserved in the charged lepton and neutrino sectors, which survive along with 
    preserved subgroups of the original family symmetry in each of these sectors.
    } \label{directCP}
\end{figure}

The direct approach can be generalised to the case of a conserved \CP
(see \cite{Holthausen:2012dk} and references therein)
which is spontaneously broken as shown in 
Fig.\ref{directCP}. This approach has been studied for $\Delta (6N^2)$
in \cite{King:2014rwa}. However since 
we already know that $\delta = 0,\pi$ in this case it only fixes the 
Majorana phases. 

The generalised \CP approach has also been used in
the semi-direct approach (defined below) where the phase 
$\delta$ is undetermined without \CP. Here the results are more interesting since for the smaller groups like $A_4$ and $S_4$ one generally predicts 
a discrete choice including $\delta = \pm \pi/2$ 
\cite{Feruglio:2012cw}.
However for larger groups in the series $\Delta (6N^2)$ and $\Delta (3N^2)$,
broken in a semi-direct way, the discrete predictions for $\delta$
proliferate \cite{Hagedorn:2014wha}.
Motivated by the the good experimental prospects for measuring leptonic \CP violation,
there has been considerable theoretical interest in \CP symmetry in 
different approaches to family symmetry models \cite{Hall:2013yha}.

Recently an invariant approach \CP symmetry in family symmetry models has been discussed
\cite{Branco:2015hea}.
It is worthwhile to first recap how the invariant approach works for any theory where the Lagrangian is specified.
Following \cite{Bernabeu:1986fc}, to study \CP symmetry in any model one divides
a given Lagrangian as follows $\mathcal{L}=\mathcal{L_{CP}}+\mathcal{L}_{rem}$
where $\mathcal{L_{CP}}$ is the part that automatically conserves \CP (like the kinetic terms and gauge interactions) while $\mathcal{L}_{rem}$ includes 
the \CP violating non-gauge interactions such as the Yukawa couplings.
Then one considers the most general \CP transformation that leaves $\mathcal{L_{CP}}$ invariant and check if invariance under \CP restricts  $\mathcal{L}_{rem}$ - only if this is the case can $\mathcal{L}$ violate \CP.

In the presence of a family symmetry $G$, one may check if 
a given vacuum leads to spontaneous \CP violation, as follows.
Consider a Lagrangian invariant under $G$ and \CP, containing a series of scalars which under \CP transform as 
$(\mathcal{CP}) \phi_i (\mathcal{CP})^{-1} =U_{ij}  \phi_j^*$. In order for the vacuum to be \CP invariant, the following relation has to be satisfied:
$ <0| \phi_i |0> = U_{ij} <0| \phi_j^* |0>$ \cite{Branco:1983tn}. 
The presence of $G$ usually allows for many choices for $U$. 
If (and only if) no choice of $U$ exists which satisfies the previous condition, will the vacuum violate \CP, leading to spontaneous \CP violation.
In order to prove that no choice of $U$ exists one can construct
\CP-odd invariants.

As a brief review of how to derive \CP-odd invariants, consider the Lagrangian of the leptonic part of the SM extended by Majorana neutrino masses. After electroweak breaking at low energies, the most general mass terms are as in Eq.\ref{lepton} which we rewrite in matrix form as,
\begin{equation}
\label{low}
{\cal L}^{\rm lepton}  =- \overline{e}_L m_l  e_R -  \tfrac{1}{2}  \overline{\nu}_{L} m_\nu \nu^{c}_L + H.c.\,,
\end{equation}
Due to the $SU(2)_L$ structure, the most general \CP transformation which leaves the leptonic gauge interactions invariant are (ignoring spin),
\begin{equation}
L(x)\rightarrow U L^*(x_P), \ \ e_R(x)\rightarrow V e_R^*(x_P),
\label{LCP}
\end{equation}
where $L= (\nu_L, e_L )$ are the left-handed neutrino and charged lepton fields in a weak basis,
and $x_P$ are the parity (3-space) inverted coordinates.

In order for ${\cal L}^{\rm lepton}$ to be \CP invariant under Eq.(\ref{LCP}), the terms shown in the Eq.(\ref{low}) go into their respective $H.c.$ terms and vice-versa:
\begin{equation}
U^\dagger  m_{\nu} U^* = m_{\nu}^*, \ \ \ \ 
U^\dagger m_{l} V = m_{l}^* \,.
\label{mlCP}
\end{equation}
From Eq.(\ref{mlCP}) one can infer how to build combinations of the mass matrices that will result in equations where $U$ and $V$ cancel entirely.
The condition for \CP to be conserved is \cite{Bernabeu:1986fc}:
\begin{equation}
I_1 \equiv \Tr\left( \left[H_\nu , H_l \right]^3\right) = \Tr\left( \left[H_\nu H_l -  H_l H_\nu \right]^3\right) = 0\,,
\label{hhcube}
\end{equation}
where $H_\nu \equiv m_\nu m_\nu^\dagger$ and $H_l \equiv m_l m_l^\dagger$.
This equation is a necessary and sufficient condition for Dirac \CP invariance,
since it follows from the existence of \CP transformations in Eq.\ref{mlCP}.
If the mass matrices are chosen such that $I_1=0$ then Dirac type \CP is conserved 
while if $I_1\neq 0$ then Dirac type \CP is violated.
\footnote{
This is called Dirac type \CP violation since it occurs both when neutrinos are Dirac and Majorana,
where the latter case is assumed above.
There are two further 
necessary and sufficient conditions for low energy leptonic \CP invariance 
which are peculiar to the Majorana sector \cite{Branco:1986gr}.}

As pointed out in \cite{Branco:2015hea},
once a Lagrangian is specified, which is invariant under a family symmetry $G$
and some \CP transformation, then the consistency relations \cite{Holthausen:2012dk,Feruglio:2012cw} are automatically satisfied.
In order to prove this it is sufficient to 
consider some generic 
Lagrangian invariant under a family symmetry transformation, involving some mass term $m$
(Dirac or Majorana), then define
$H=m m^{\dagger}$.
Under some $G$ transformation, 
$\rho(g)$, the mass term remains unchanged implying:
\begin{equation}
\rho(g)^{\dagger}H\rho(g)=H.
\label{cond00}
\end{equation}
Invariance of the Lagrangian under \CP transformation $U$ requires the mass term to swap with its $H.c.$, hence:
\begin{equation}
U^{\dagger}H U=H^*
\label{cond0}
\end{equation}
Taking the complex conjugate of Eq.(\ref{cond00}) we find,
\begin{equation}
(\rho(g)^{\dagger})^*H^*\rho(g)^*=H^*=U^{\dagger}HU,
\label{cond01}
\end{equation}
using Eq.(\ref{cond0}) for the last equality. 
Using Eq.(\ref{cond0}) again:
\begin{equation}
(\rho(g)^{\dagger})^*U^{\dagger}HU\rho(g)^*=U^{\dagger}HU.
\label{cond02}
\end{equation}
Hence by using once more Eq.(\ref{cond00}) for a $g'$, we find,
\begin{equation}
U(\rho(g)^{\dagger})^*U^{\dagger}HU\rho(g)^*U^{\dagger}=H=\rho(g')^{\dagger}H\rho(g').
\label{cond03}
\end{equation}
Comparing both sides of Eq.(\ref{cond03}) we see,
\begin{equation}
U\rho(g)^*U^{\dagger}=\rho(g')
\label{consistency}
\end{equation}
which is just the consistency relation \cite{Holthausen:2012dk,Feruglio:2012cw}.
In other words, if we consider Eqs.(\ref{cond00}) and (\ref{cond0}) we do not need to consider the consistency
condition separately since it always follows.

The above considerations about whether \CP is conserved or violated apply separately both to the original
theory (defined by some high energy Lagrangian, above the scale of symmetry breaking) and to the spontaneously broken
theory (defined by some low energy effective Lagrangian, below the scale of symmetry breaking).
We are mainly interested in theories which respect \CP at high energy, but where 
\CP is spontaneously broken, since these allow for the possibility of being able to predict the
amount of \CP violation (e.g. the physical \CP violating phases in some basis).

\subsection{Semi-direct models}

\begin{figure}[t]
\centering
\includegraphics[width=0.60\textwidth]{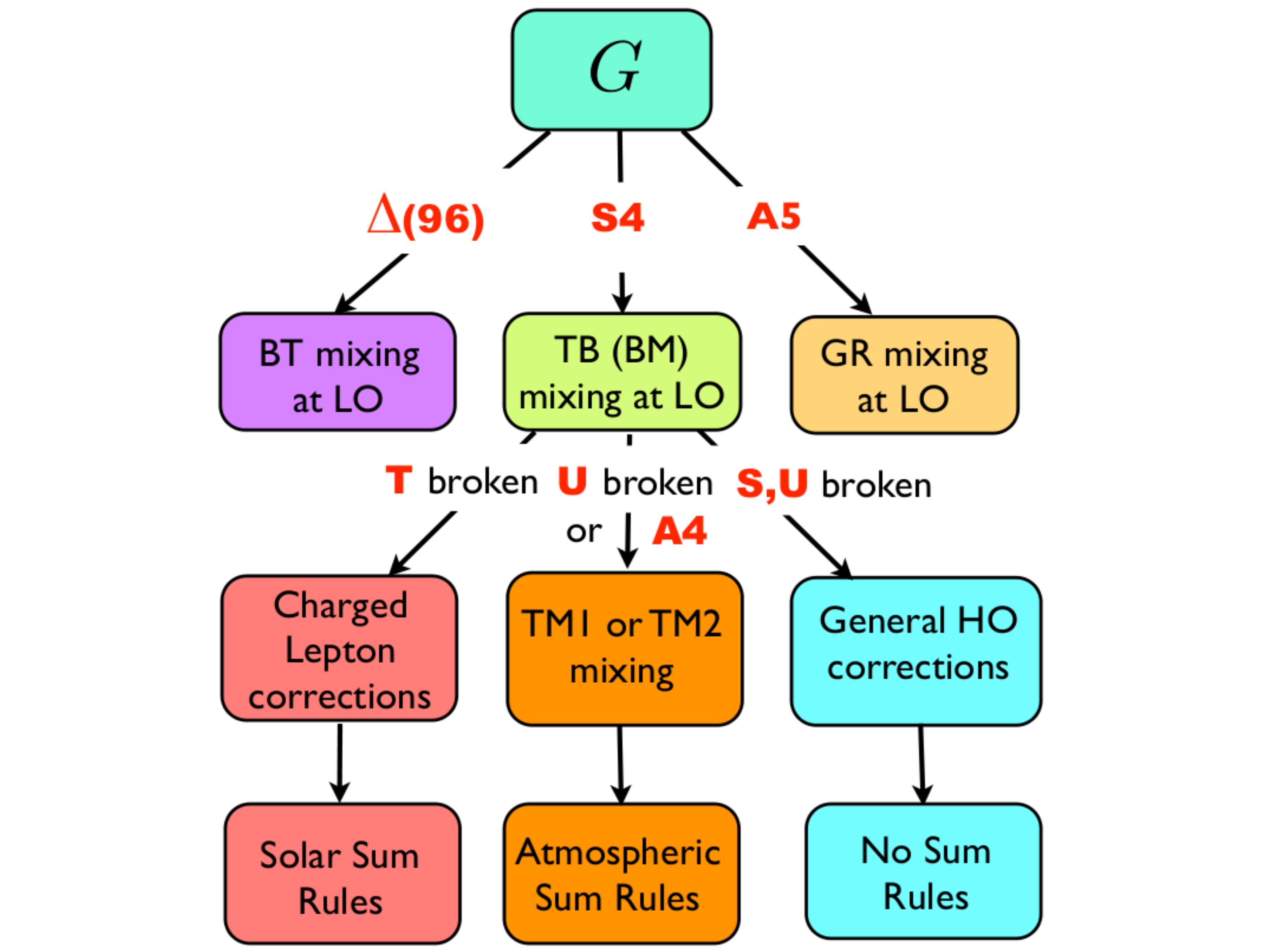}
    \caption{The semi-direct approach to models of lepton mixing.     } 
    \label{semi-direct}
\vspace*{2mm}
\end{figure}

Taking a less constrained approach to model building one may suppose that 
we start from only smaller discrete family groups such as $S_4$,
which leads to either TB or BM mixing at leading order, or $A_5$ which leads to
GR mixing at leading order as shown in Fig.\ref{semi-direct}.
Then we suppose that at higher order,
 one or more of the generators $S,T,U$ is broken, which is necessary in this
 approach since the resulting BM, TB and GR mixing patterns discussed 
 in Eqs.\ref{BM},\ref{TB},\ref{GR} are excluded. 
There are two interesting possibilities depicted in Fig.\ref{semi-direct} as follows:
\begin{enumerate} 
 \item {If the $T$ generator protecting the charged lepton
 mass matrix is broken, then we can expect charged lepton corrections
 leading to the solar sum rules discussed in section \ref{solar}.}
\item{If the $U$ generator is broken then this leads to either TM1 or TM2
mixing depending on whether $SU$ or $S$ is preserved, leading to 
atmospheric sum rules as discussed in section \ref{atmospheric}.}
\end{enumerate} 
The semi-direct approach was first used in \cite{Shimizu:2011xg,King:2011zj}
for $A_4$ where there is no $U$ generator to start with and also
$S_4$ which is broken to $A_4$ at higher order \cite{King:2011zj}.
It was subsequently generalised to von Dyck groups in
\cite{Hernandez:2012ra}.
In all cases the reactor angle is not predicted but described 
by a free parameter. This is a retreat from the original goal of predicting
lepton mixing angles using symmetry.


\subsection{Indirect models}
\label{indirectsect}

\begin{figure}[t]
\centering
\includegraphics[width=0.60\textwidth]{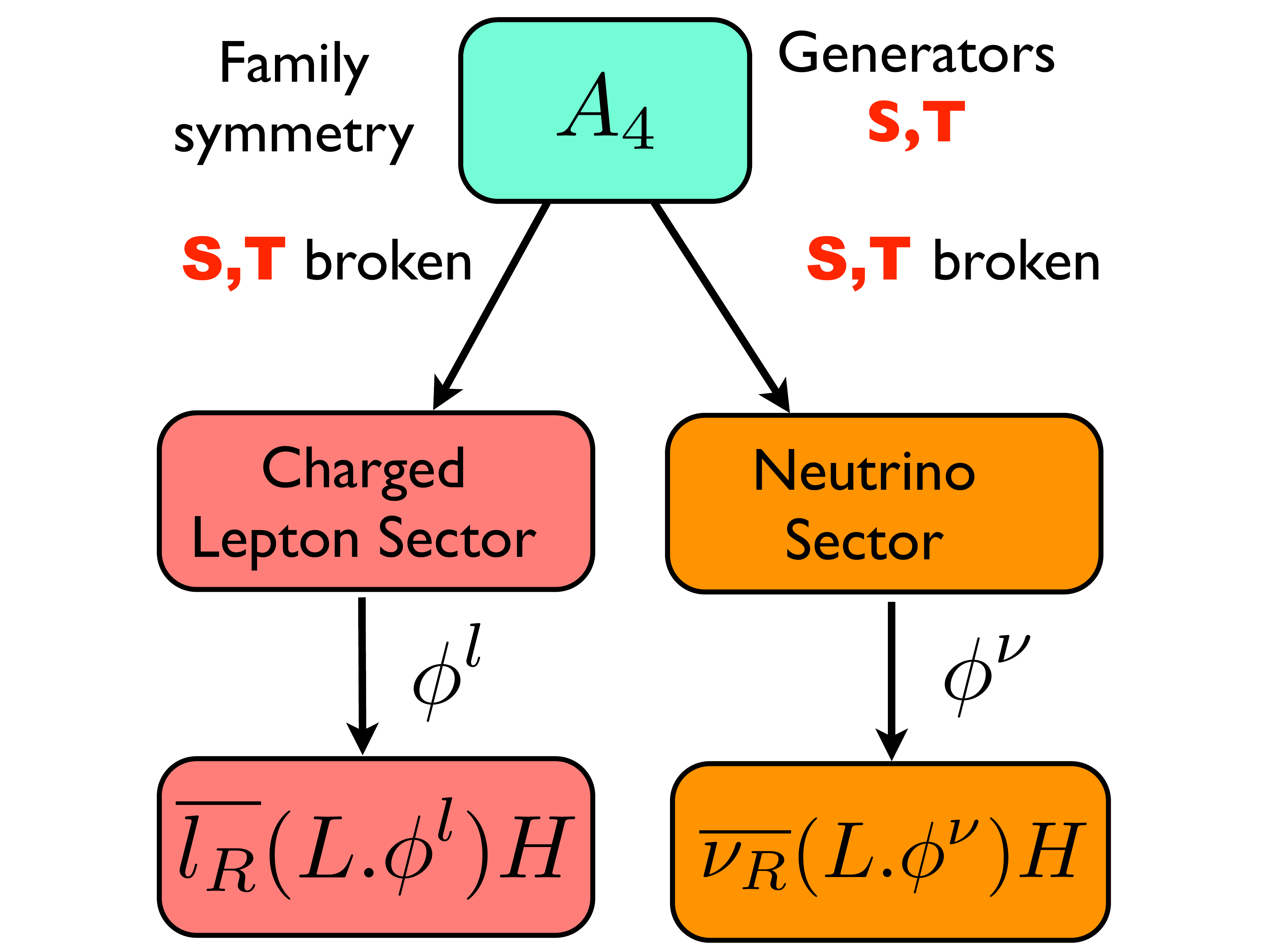}
\vspace*{4mm}
    \caption{The indirect approach to models of lepton mixing. 
    The notation is such that $\phi^l$ refers to $\phi_{e, \mu , \tau}$   
    and $\phi^{\nu}$ refers to $ \phiatm $, $ \phisol $, and 
$ \phidec $. } \label{indirect}
\end{figure}

The final logical possibility is that the family symmetry is completely broken 
in both the neutrino and charged lepton sectors as
shown in Fig.\ref{indirect} for the example of the smallest family symmetry group that admits triplet
representations,namely $A_4$.
In this approach, we allow the flavons $\phi^l$
and $\phi^{\nu}$ to have 
not only symmetry preserving vacuum alignments, but also new alignments which are orthogonal to them and break the symmetry. In the following, $\phi^l$ refers to $\phi_{e, \mu , \tau}$ which only enter the charged lepton sector and are responsible for a diagonal charged lepton mass matrix, while $\phi^{\nu}$ refers to $ \phiatm $, $ \phisol $, and  $ \phidec $ which only enter the neutrino sector and are responsible for a Dirac mass matrix of the CSD($n$) form.

The advantages of the indirect approach over the previous approaches are:
\begin{enumerate}
\item It can involve a small family symmetry group such as $A_4$ (unlike the direct approach which involves large family symmetry groups)
\item It is highly predictive since it can yield CSD($n$) where the entire PMNS matrix 
is predicted in terms of one input parameter (unlike the semi-direct approach which only predicts sum rules)
\end{enumerate}
The basic starting point is to consider some small family symmetry such as $A_4$ which admits triplet representations. The family symmetry is broken by triplet flavons $\phi_i$ whose vacuum alignment will control the structure of the Yukawa couplings. To illustrate how this works, we sketch 
a model, where the relevant operators responsible for the Yukawa structure in the neutrino sector are
\begin{equation}
	\frac{1}{\Lambda} H_u (L \cdot \phiatm) \nu_{\rm atm}^c + \frac{1}{\Lambda} H_u (L \cdot\phisol )
	\nu_{\rm sol}^c + \frac{1}{\Lambda} H_u (L \cdot\phidec)\nu_{\rm dec}^c ,
\label{Ynu_flavon}
\end{equation}
where $ L $ is the SU(2) lepton doublet, assumed to transform as a triplet under the family symmetry, while 
$ \nu_{\rm atm}^c, \nu_{\rm sol}^c, \nu_{\rm dec}^c$ are \CP conjugates of the 
right-handed neutrinos and $ H_u $ is the electroweak scale up-type Higgs field, the latter being family symmetry singlets but distinguished by some additional quantum numbers. In the charged-lepton sector,
we consider the operators,
\begin{equation}
	\frac{1}{\Lambda} H_d (L \cdot \phi_e) e^c  + \frac{1}{\Lambda} H_d (L \cdot \phi_\mu) \mu^c + \frac{1}{\Lambda} H_d (L \cdot\phi_\tau) \tau^c ,
\label{Ye_flavon}
\end{equation}
where $ e^c,\mu^c,\tau^c$ are the \CP conjugated right-handed electron, muon and tau respectively. The right-handed neutrino Majorana superpotential is typically chosen to give a diagonal mass matrix,
\begin{equation}
	M_{R} = \mathrm{diag}(M_{\rm atm},M_{\rm sol},M_{\rm dec}).
	\label{eq:mrr}
\end{equation}
Details of the origin of these operators (e.g. in terms of Majoron fields), the relative values of 
$M_{\rm atm},M_{\rm sol},M_{\rm dec}$
as well as the inclusion of any off-diagonal terms in $ M_{R} $ will all depend on the additional specifications of the model beyond our sketchy model here (we shall consider real models shortly in 
the next section~\ref{realistic}).

The idea is that CSD($n$) discussed in section~\ref{CSDn}
emerges from flavon
vacuum alignments in the effective operators involving three flavon fields $ \phiatm $, $ \phisol $, and 
$ \phidec $ which are triplets under the flavour symmetry and acquire VEVs that 
break the family symmetry completely in both the neutrino and charged lepton sectors. 
The subscripts are chosen by noting that $ \phiatm $ correlates with the atmospheric neutrino mass $ m_3 $, $ \phisol $ with the solar neutrino mass $ m_2 $, and $ \phidec $ with the lightest neutrino mass $ m_1 $, which in CSD is light enough that the associated third right-handed neutrino can, to good approximation, be thought of as decoupled from the theory \cite{King:1998jw}. 
CSD($n$) corresponds to the choice of vacuum alignments,
\begin{equation}
	\braket{\phiatm} = v_{\mathrm{atm}}\pmatr{0 \\ 1 \\ 1}, 
	\qquad \braket{\phisol} =v_{\mathrm{sol}} \pmatr{1\\n\\n-2}, 
	\qquad \braket{\phidec}= v_{\mathrm{dec}} \pmatr{0\\0\\1},
	\label{CSD(n)}
\end{equation}
where $n$ is a positive integer, and the only phases allowed are in the overall proportionality constants.
Such vacuum alignments arise from symmetry preserving alignments together with orthogonality conditions  
\cite{King:2013iva,King:2013xba}, as discussed below.
%
%

The starting point for understanding the alignments in Eq.~\ref{CSD(n)} are the symmetry preserving vacuum alignments of $A_4$,
namely:
\begin{equation}
	\pmatr{1 \\ 0 \\ 0} ,  \pmatr{0\\1\\0} , \pmatr{0\\0\\1} , \pmatr{\pm 1\\ \pm 1\\ \pm 1},
	\label{sym}
\end{equation}
which each preserve some subgroup of $A_4$ in a basis where the 12 group elements in the triplet representation are real as in Eq.\ref{eq:ST} \cite{Ma:2001dn}
(i.e. each alignment in Eq.~\ref{sym} is an eigenvector of at least one non-trivial group element with eigenvalue +1.)
The first alignment in Eq.~\ref{CSD(n)}, which completely breaks the $A_4$ symmetry,
arises from the orthogonality conditions 
\begin{equation}
\begin{pmatrix}0 \\ 1\\ 1\end{pmatrix}
\perp 
\begin{pmatrix}1 \\1\\-1\end{pmatrix},
\begin{pmatrix}1 \\0\\0\end{pmatrix}
\label{011}
\end{equation}
involving two symmetry preserving alignments selected from Eq.~\ref{sym}.
The following symmetry breaking alignment may be obtained which is orthogonal to the alignment in Eq.~\ref{011} and one of the 
symmetry preserving alignments, 
\begin{equation}
\begin{pmatrix}2 \\-1\\1\end{pmatrix}
\perp 
\begin{pmatrix}1 \\1\\-1\end{pmatrix},
\begin{pmatrix}0 \\1\\1\end{pmatrix}
\label{2,-1,1}
\end{equation}
The CSD($n$) alignment in Eq.~\ref{CSD(n)} is orthogonal to the above alignment in Eq.~\ref{2,-1,1},
\begin{equation}
\begin{pmatrix}1 \\n\\ n-2 \end{pmatrix}
\perp 
\begin{pmatrix}2 \\-1\\1\end{pmatrix}
\label{CSD(n)again}
\end{equation}
where the orthogonality in Eq.~\ref{CSD(n)again} is maintained for any value of $n$ (not necessarily integer).
To pin down the value of $n$ and show that it is a particular integer requires a further orthogonality condition.

For example, for $n=3$, the desired alignment is obtained from the two orthogonality conditions,
\begin{equation}
\begin{pmatrix}1 \\3\\1\end{pmatrix}
\perp 
\begin{pmatrix}2 \\-1\\1\end{pmatrix},
\begin{pmatrix}1 \\0\\-1\end{pmatrix}
\label{CSD(3)again}
\end{equation}
where the first condition above is a particular case of Eq.~\ref{CSD(n)again}
and the second condition involves a new alignment,
obtained from two of the symmetry preserving alignments in Eq.~\ref{sym}, 
\begin{equation}
\begin{pmatrix}1 \\0\\-1\end{pmatrix}
\perp 
\begin{pmatrix}1 \\1\\1\end{pmatrix},
\begin{pmatrix}0 \\1\\0\end{pmatrix}
\label{1,0,-1}
\end{equation}

Using Eq.~\ref{Ynu_flavon}, the vacuum alignments in Eq.~\ref{CSD(n)} make up the columns of the Dirac neutrino Yukawa matrix 
$ Y^{\nu} \propto (\braket{\phiatm}, \braket{\phisol}, \braket{\phidec}) $, giving a Dirac mass matrix
\begin{equation}
	m^D = 
	Y^{\nu}v_u=\pmatr{0 & a &0\\ e & na &0\\ e & (n-2)a &c},	\label{mDn3}
\end{equation}
which is an extension of Eq.~\ref{mDn} to include a third (decoupled) right-handed neutrino,
where $m^D=(m^D_{\rm atm},m^D_{\rm sol},m^D_{\rm dec})$
and the coefficients $ e $, $ a $, and $ c $ are generally complex. The charged-lepton Yukawa matrix is chosen to be diagonal (up to model-dependent corrections, assumed small), corresponding to the existence of three flavons $ \phi_e $, $ \phi_\mu $ and $ \phi_\tau $ in the charged-lepton sector which acquire VEVs with alignments \cite{King:2013iva,King:2013xba}
\begin{equation}
	\braket{\phi_e} = v_{e} \pmatr{1\\0\\0}, \qquad \braket{\phi_\mu} = v_{\mu} \pmatr{0\\1\\0}, 
	\qquad \braket{\phi_\tau} = v_{\tau} \pmatr{0\\0\\1}
	\label{ve}
\end{equation}
Given this choice, it is clear that $Y^e$ is diagonal, hence $ U_{e_L} $ is the identity matrix up to diagonal phase rotations, and that $ U_\mathrm{PMNS} = U^\dagger_{\nu_L} $, i.e. simply the matrix that diagonalises the neutrino mass matrix, up to charged lepton phase rotations.

\section{Realistic Theories of Flavour}
\label{realistic}
In this section we briefly review two realistic indirect models involving the family symmetry
$A_4$. The first model involves the Pati-Salam gauge group with CSD($4$),
while the second model involves $SU(5)$ GUT with CSD($3$).
We also discuss leptogenesis in these two models.

\subsection{A to Z of flavour with Pati-Salam}
\begin{figure}[t]
\centering
\includegraphics[width=0.4\textwidth]{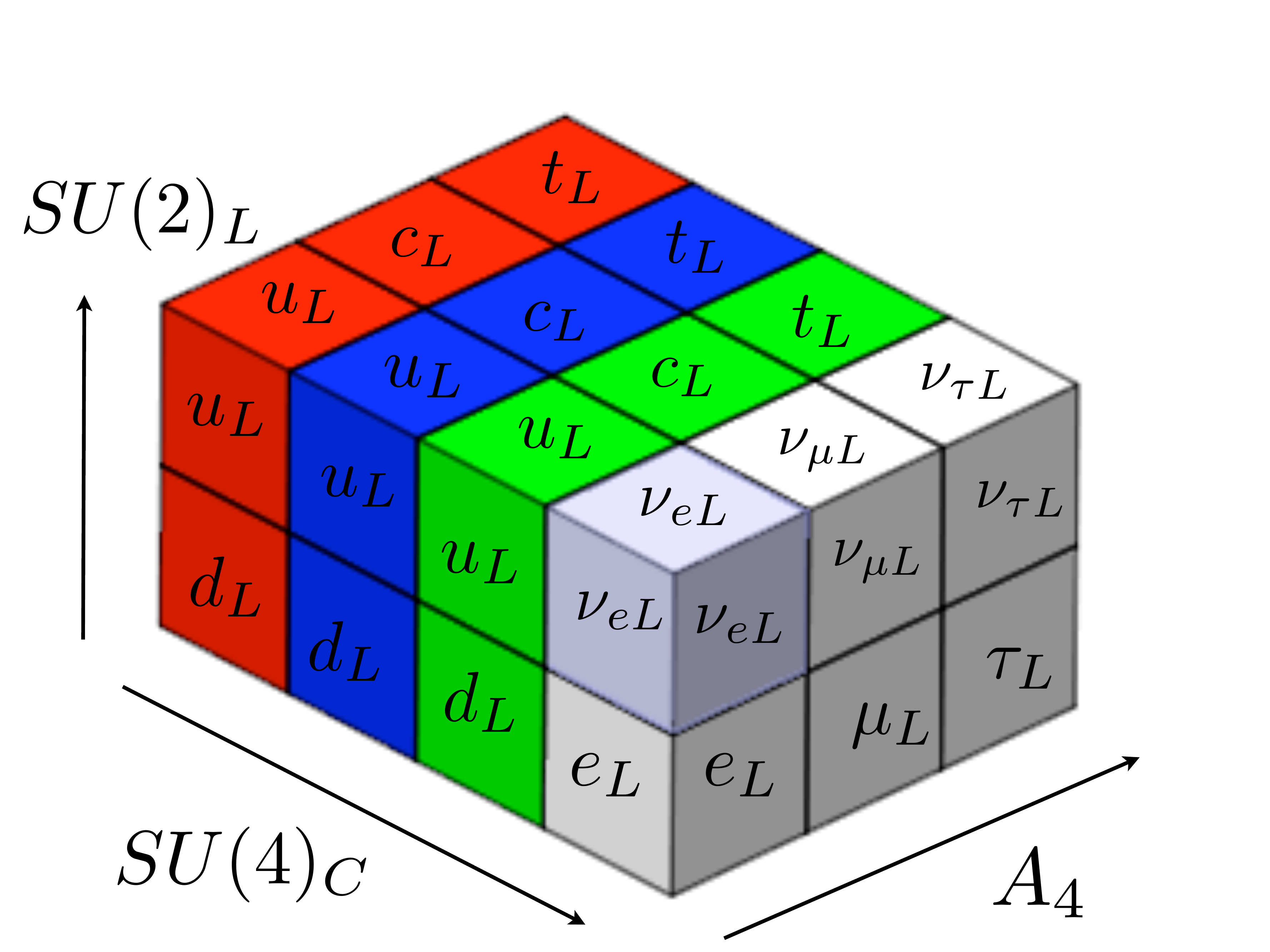}
\includegraphics[width=0.4\textwidth]{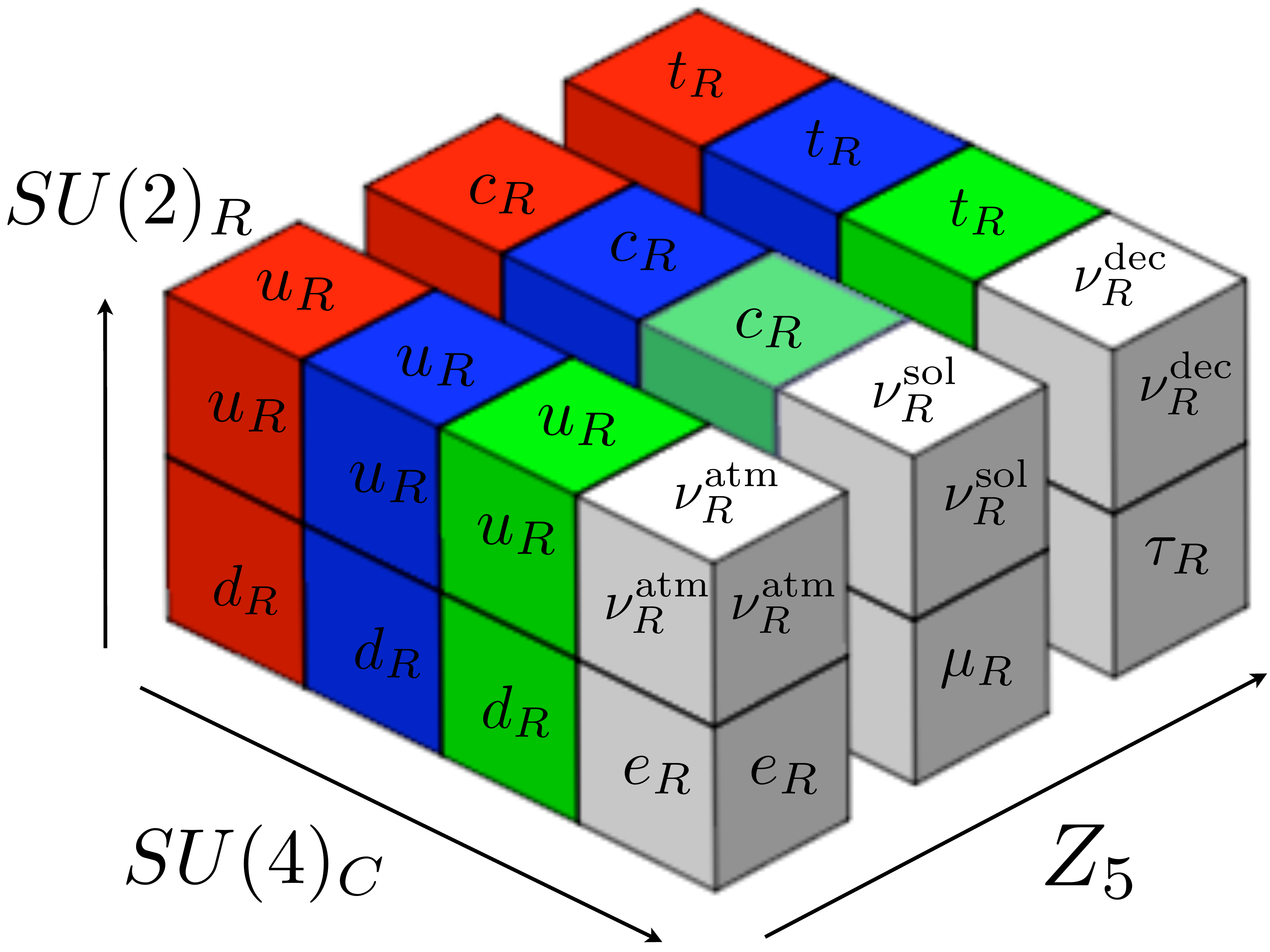}
\vspace*{-4mm}
    \caption{$A$ to $Z$ of flavour with Pati-Salam, where $A\equiv A_4$ and $Z\equiv Z_5$. 
         The left-handed families form a triplet of $A_4$ and are
         doublets of $SU(2)_L$.
        The right-handed families are distinguished by $Z_5$
and are doublets of $SU(2)_R$.
The $SU(4)_C
$ unifies the quarks and leptons with leptons as the fourth colour,
   depicted here as white.
     } \label{A2Z}
\vspace*{-2mm}
\end{figure}

As an example of an ``indirect'' model, 
an ``A to Z of flavour with Pati-Salam'' based on the Pati-Salam gauge group has been proposed 
\cite{King:2014iia} as sketched in Fig.\ref{A2Z}.
The Pati-Salam symmetry leads to $Y^u=Y^{\nu}$, where the columns of the Yukawa matrices 
are determined as in Eq.\ref{Ynu_flavon} with the flavon alignments as in 
Eq.\ref{CSD(n)} for the case $n=4$.
The first column is proportional to the alignment $(0,e,e)$
the second column proportional to the CSD(4) orthogonal alignment,
$(a,4a,2a)$ and the third column is proportional to the alignment $(0,0,c)$,
where $e\ll a\ll c$ gives the hierarchy $m_u\ll m_c\ll m_t$.
This structure predicts a Cabibbo angle 
$\theta_C\approx 1/4$
in the diagonal $Y^d\sim Y^e$ basis enforced by the 
first three alignments in Eq.\ref{sym}.
It also predicts a normal neutrino mass hierarchy with 
$\theta_{13}\approx 9^{\circ}$, $\theta_{23}\approx 45^{\circ}$
and $\delta \approx 260^{\circ}$ \cite{King:2014iia}.

The model is based on the Pati-Salam (PS) gauge group, 
with $A_4\times Z_5$ (A to Z) family symmetry,
\begin{equation}
SU(4)_{C} \times SU(2)_L \times SU(2)_R\times A_4 \times Z_5.
\label{422A4Z5}
\end{equation}
The quarks and leptons are unified in the PS representations as follows,
\begin{eqnarray}
F_i& = & (4,2,1)_i = \left(\begin{array}{cccc}u&u&u&\nu\\
d&d&d&e\end{array}\right)_i \rightarrow  (Q_i,L_i), \nonumber \\
F^c_i & = & (\bar{4},1,2)_i = 
\left(\begin{array}{cccc}u^c&u^c&u^c&\nu^c\\
d^c&d^c&d^c&e^c\end{array}\right)_i \rightarrow (u^c_i,d^c_i,\nu^c_i,e^c_i ),
\label{ql}
\end{eqnarray}
\noindent where the SM multiplets $Q_i,L_i,u^c_i,d^c_i,\nu^c_i,e^c_i$ 
resulting from PS breaking are also shown and the 
subscript $i\ (=1,2,3)$ denotes the family index.
The left-handed quarks and leptons form an $A_4$ triplet $F$, 
while the three (\CP conjugated) right-handed fields $F^c_i$ are $A_4$ singlets, distinguished by $Z_5$ charges $\alpha, \alpha^3,1$, for $i=1,2,3$, respectively.
Clearly the Pati-Salam model cannot be embedded into an $SO(10)$ Grand Unified Theory (GUT)
since different components 
of the 16-dimensional representation of $SO(10)$ would have to transform differently under $A_4\times Z_5$,
which is impossible. On the other hand, the PS gauge group and $A_4$ could emerge directly from
string theory.

The Pati-Salam gauge group is broken at the GUT scale to the SM,
\begin{equation}
SU(4)_{C}\times SU(2)_L \times SU(2)_R\rightarrow SU(3)_C\times SU(2)_L \times U(1)_Y,
\end{equation}
by PS Higgs, 
$H^c$  and $\overline{H^c}$,
\begin{eqnarray}
{H^c} & = & (\bar{4},1,2)= (u^c_H,d^c_H,\nu^c_H,e^c_H ), \nonumber \\
 {\overline{H^c}} & = & (4,1,2) = (\bar{u}^c_H,\bar{d}^c_H,\bar{\nu}^c_H,\bar{e}^c_H ).
\end{eqnarray}
These acquire VEVs in the ``right-handed neutrino'' directions, with equal VEVs
close to the GUT scale $2\times 10^{16}$ GeV,
\begin{equation}
 \langle {H^c}\rangle = \langle {\nu^c_H}\rangle
 =\langle{\overline{H^c}}\rangle=\langle{\bar{\nu}^c_H}\rangle \sim 2\times 10^{16} \ {\rm GeV},
\label{PS}
\end{equation}
so as to maintain supersymmetric gauge coupling unification.

Our starting point is to assume that the high energy theory, above the PS breaking scale,
conserves \CP symmetry.
Under a \CP transformation, the $A_4$ singlet fields $\xi, \Sigma_u, \Sigma_d$
transform into their complex conjugates,
\begin{equation}
\xi \rightarrow \xi^*, \ \ \Sigma_u \rightarrow  \Sigma_u^*, \ \ \Sigma_d \rightarrow  \Sigma_d^*,
\end{equation}
where the complex conjugate fields transform in the complex conjugate representations under
$A_4\times Z_5$. For example if $\xi \sim \alpha^4$, under $Z_5$, then $\xi^* \sim \alpha$.
Similarly if $\Sigma_u \sim 1'$, $\Sigma_d \sim 1''$, under $A_4$, then
$\Sigma_u^* \sim 1''$, $\Sigma_d^* \sim 1'$. On the other hand, 
in a particular
basis, for $A_4$ triplets $\phi \sim (\phi_1, \phi_2, \phi_3)$,
a consistent definition of \CP symmetry requires the second
and third triplet components to swap under \CP,
\begin{equation}
\phi \rightarrow (\phi_1^*, \phi_3^*, \phi_2^*).
\end{equation}
With the above definition of \CP, all coupling constants $g$ and explicit masses $m$ are real due to CP
conservation and 
the only source of phases can be the VEVs
of fields which break $A_4\times Z_5$. 
In the model of interest, all the physically interesting \CP phases will arise from
$Z_5$ breaking.

Let us now consider the
$A_4$ triplet fields $\phi$ which also carry $Z_5$ charges.
In the full model there are four such triplet fields, or ``flavons'',
denoted as $\phi^u_1$, $\phi^u_2$, $\phi^d_1$, $\phi^d_2$.
The idea is that $\phi^u_i$ are responsible for up-type quark flavour,
while $\phi^d_i$ are responsible for down-type quark flavour. 

The structure of the Yukawa matrices depends on the so-called CSD(4) vacuum alignments of these flavons, with the overall phases quantised due to $Z_5$,
\begin{equation}
\langle \phi^u_1 \rangle =
\frac{V^u_1}{\sqrt{2}}e^{im\pi/5} 
\left(
\begin{array}{c}
0 \\ 1 \\ 1
\end{array}
\right)
 \ , \qquad
\langle \phi^u_2 \rangle =
\frac{V^u_2}{\sqrt{21}} e^{im\pi/5} 
\left(
\begin{array}{c}
1 \\ 4 \\ 2
\end{array}
\right)
\ , \label{phiu}
\end{equation}
and
\begin{equation}
\langle \phi^d_1 \rangle =V^d_1 e^{in\pi/5}
\left(
\begin{array}{c}
1\\0\\0 
\end{array}
\right)
  \ , \ \qquad
\langle \phi^d_2 \rangle =V^d_2 e^{in\pi/5}
\left(
\begin{array}{c}
0\\1\\0 
\end{array}
\right)
 \ .
 \label{phid}
\end{equation}
We note here that the vacuum alignments in Eq.\ref{phid} and the first alignment in Eq.\ref{phiu} are 
fairly ``standard'' alignments that are encountered in tri-bimaximal mixing models, while the second
alignment in Eq.\ref{phiu} is obtained using orthogonality arguments, as discussed
in section~\ref{indirectsect}.
In particular we are using the vacuum alignments in Eq.\ref{CSD(n)} for the case $n=4$,
where we identify $ \phiatm $ and $ \phisol $ with $\phi^u_1$ and $\phi^u_2$.
We also use the alignments in Eq.\ref{ve} where we identify 
$\phi_e$ and $\phi_{\mu}$ with $\phi^d_1$ and $\phi^d_2$.

The model will involve Higgs bi-doublets of two kinds, $h_u$ which lead to up-type quark and neutrino Yukawa couplings and $h_d$ which lead to down-type quark and charged lepton Yukawa couplings.
In addition a Higgs bidoublet $h_3$, which is also an $A_4$ triplet, is used to give the third family
Yukawa couplings. 

After the PS and $A_4$ breaking, most of these Higgs bi-doublets 
will get high scale masses and will not appear in the low energy spectrum. In fact only two
light Higgs doublets will survive down to the TeV scale, namely $H_u$ and $H_d$.
The basic idea is that the light Higgs doublet $H_u$ 
with hypercharge $Y=+1/2$, which couples to up-type quarks and neutrinos,
is a linear combination of components of the Higgs bi-doublets of the kind $h_u$ and $h_3$,
while the light Higgs doublet $H_d$ with hypercharge $Y=-1/2$, 
which couples to down-type quarks and charged leptons,
is a linear combination of components of Higgs bi-doublets of the kind $h_d$ and $h_3$,
\begin{equation}
h_u,h_3 \rightarrow H_u, \ \ \ \ 
h_d,h_3 \rightarrow H_d.
\label{H}
\end{equation}

The renormalisable Yukawa operators, which respect PS and $A_4$ symmetries, have the following form, 
leading to the third family Yukawa couplings shown, using Eqs.\ref{ql},\ref{H},
\begin{equation}
F.h_3F_3^c  
\rightarrow 
Q_3H_uu_3^c + Q_3H_dd_3^c + 
L_3H_u\nu_3^c+L_3H_de_3^c  ,\label{3rd} 
\end{equation}
where we have used Eqs.\ref{ql},\ref{H}.
The non-renormalisable operators, which respect PS and $A_4$ symmetries, have the following form, 
\begin{eqnarray}
F.\phi^u_ih_uF_i^c  
&\rightarrow& 
Q.\langle{\phi^u_i}\rangle H_uu_i^c +
L.\langle{\phi^u_i}\rangle H_u\nu_i^c ,\label{up0} 
\label{up0} \\
F.\phi^d_ih_dF_i^c 
&\rightarrow &
Q.\langle{\phi^d_i}\rangle H_dd_i^c+
L.\langle{\phi^d_i}\rangle H_de_i^c,
\label{down0}
\end{eqnarray}
where $i=1$ gives the first column of each Yukawa matrix, while $i=2$ gives the second column
and we have used Eqs.\ref{ql},\ref{H}. Thus the third family masses are naturally larger since they correspond to renormalisable operators, while the hierarchy between first and second families arises from a hierarchy of 
flavon VEVs.
The lepton operators in Eqs.\ref{up0},\ref{down0} may be compared to the operators 
in Eqs.\ref{Ynu_flavon},\ref{Ye_flavon}.

Inserting the vacuum alignments in Eqs.\ref{phiu} and \ref{phid} into Eqs.\ref{up0} and \ref{down0},
together with the renormalisable third family couplings in Eq.\ref{3rd},
gives the Yukawa matrices of the form,
\begin{equation} \label{Y}
 Y^uv_u = Y^{\nu}v_u = 
 \left(
\begin{array}{ccc}
  0 & a   & 0  \\ 
e & 4a & 0\\  e  & 2a
 & c
\end{array}
\right)
 , \ \ \ \ 
 Y^d \sim Y^e \sim 
 \left(
\begin{array}{ccc}
   y^0_d & 0  & 0  \\ 0 & y^0_s & 0\\ 0  & 0 & y^0_b
\end{array}
\right).
\end{equation}
The PS unification predicts the equality of Yukawa matrices $Y^u = Y^{\nu}$ and $Y^d \sim Y^e$,
while the $A_4$ vacuum alignment predicts the structure of each Yukawa matrix,
essentially identifying the first two columns with the vacuum alignments in Eqs.\ref{phiu} and \ref{phid}.
With a diagonal right-handed Majorana mass matrix, $Y^{\nu}$ leads to a successful
prediction of the PMNS mixing parameters. Also
the Cabibbo angle is given by $\theta_C\approx 1/4$ \cite{King:2013hoa}.
Thus Eq.\ref{Y} is a good starting point for a theory of quark and lepton masses and
mixing, although the other 
quark mixing angles and the quark \CP phase are approximately zero.
However the above discussion ignores the effect of Clebsch factors which will alter the relationship
between elements of $Y^d$ and $Y^e$, which also include off-diagonal elements 
responsible for small quark mixing angles in the full model discussed in \cite{King:2014iia}.

In realistic unified models involving an $SO(10)$-inspired pattern of Dirac
and heavy right-handed (RH) neutrino masses, assuming the type I seesaw,
the lightest right-handed
neutrino $N_1$ is too light to yield successful
thermal leptogenesis, barring highly fine tuned solutions,
while the second heaviest right-handed neutrino $N_2$
is typically in the correct mass range.
In \cite{DiBari:2015oca} we discussed $N_2$ dominated leptogenesis in 
the A to Z model, where $N_1$ is identified with 
$ \nu_{1}^c= \nu_{\rm atm}^c$,  while $N_2$ is identified with 
$ \nu_{2}^c= \nu_{\rm sol}^c$ and $N_3$ is identified with $\nu_{3}^c= \nu_{\rm dec}^c$,
as depicted in Fig.~\ref{A2Z}.
In the A to Z model the neutrino Dirac mass matrix is equal to the up-type quark mass matrix
and has the particular constrained structure in Eq.\ref{Y}.
We showed that flavour coupling effects in the Boltzmann equations are crucial to the 
success of such $N_2$ dominated leptogenesis in this model, by helping to ensure that 
the flavour asymmetries produced at the $N_2$ scale survive
$N_1$ washout.
The numerical results, supported by analytical insight, showed that in order to achieve successful $N_2$ leptogenesis, consistent with neutrino phenomenology,
requires a ``flavour swap scenario'' whereby the asymmetry generated in the tauon flavour 
emerges as a surviving asymmetry dominantly in the muon flavour.
However successful leptogenesis requires 
a less hierarchical pattern of RH neutrino masses than naively expected, at the expense of some mild 
fine-tuning involving significant off-diagonal elements of the heavy right-handed Majorana mass matrix.
This leads to large deviations from the CSD($4$) predictions, including a NO neutrino spectrum with 
an atmospheric neutrino mixing angle 
well into the second octant and a Dirac phase $\delta_{CP}\approx 20^{\circ}$,
a set of predictions that will be tested soon in neutrino oscillation experiments,
as discussed in \cite{DiBari:2015oca}.

\subsection{Towards a complete $A_4\times SU(5)$ SUSY GUT}
\label{A4SU5}
In this section we describe a fairly complete $A_4 \times SU(5)$ SUSY GUT model which implements CSD(3) with two right-handed neutrinos~\cite{Bjorkeroth:2015ora}. This model has the following virtues:
\begin{itemize}
\item It is fully renormalisable at the GUT scale, with an explicit $SU(5)$ breaking sector and a spontaneously broken \CP symmetry.  
\item The MSSM is reproduced with R-parity emerging from a discrete $\mathbb{Z}_4^R$. 
\item Doublet-triplet splitting is achieved through the Missing Partner mechanism \cite{Masiero:1982fe}.
\item A $ \mu $ term is generated at the correct scale.
\item Proton decay is sufficiently suppressed.
\item It solves the strong \CP problem through the Nelson-Barr mechanism \cite{Nelson:1983zb, Barr:1984qx}.
\item It explains the hierarchies in the quark sector, and successfully fits all of the quark masses, mixing angles and the \CP phase, using only $\mathcal{O}(1)$ parameters.
\item It justifies the CSD(3) alignment which accurately predicts the leptonic mixing angles, as well as a normal neutrino mass hierarchy.
\item It involves two right-handed neutrinos with the lighter one dominantly responsible for the atmospheric neutrino mass.
\item There is only one physical phase in the model, called $\eta$, which is responsible for \CP violation
in both leptogenesis and neutrino oscillations.
\item A $ \mathbb{Z}_9 $ flavour symmetry fixes the phase $ \eta $ to be one of ninth roots of unity \cite{Ross:2004qn}.
\end{itemize}

Apart from $A_4\times  SU(5)$ the model also involves the discrete symmetries $\mathbb{Z}_9\times \mathbb{Z}_6\times \mathbb{Z}_4^R$. It is renormalisable at the GUT scale, but many effects, including most fermion masses, come from non-renormalisable terms that arise when heavy messenger fields are integrated out. Unwanted or potentially dangerous terms are forbidden by the symmetries and the prescribed messenger sector, including any terms that would generate proton decay or strong \CP violation. Such terms may arise from Planck scale suppressed terms, but prove to be sufficiently small. Due to the completeness of the model, the field content is too big to be listed here, but the superfields relevant for quarks, leptons and Higgs,
including flavons, are shown in Table \ref{ta:SMF}.
\begin{table}
\centering
\footnotesize
\begin{minipage}[b]{0.45\textwidth}
\centering
\begin{tabular}{| c | c c | c | c | c |}
\hline
\multirow{2}{*}{\rule{0pt}{4ex}Field}	& \multicolumn{5}{c |}{Representation} \\
\cline{2-6}
\rule{0pt}{3ex}			& $A_4$ & SU(5) & $\mathbb{Z}_9$ & $\mathbb{Z}_6$ & $\mathbb{Z}_4^R$ \\ [0.75ex]
\hline \hline
\rule{0pt}{3ex}%
$F$ 					& 3 & $\bar{5} $& 0 & 0 & 1 \\
$T_1$ 					& 1 & 10		& 5 & 0 & 1 \\
$T_2$ 					& 1 & 10		& 7 & 0 & 1 \\
$T_3$ 					& 1 & 10		& 0 & 0 & 1 \\
\rule{0pt}{3ex}%
$N_{\rm atm}^c$ 		& 1 & 1	 		& 7 & 3 & 1 \\
$N_{\rm sol}^c$ 		& 1 & 1 		& 8 & 3 & 1 \\
\rule{0pt}{3ex}%
$\Gamma$				& 1 & 1			& 0 & 3 & 1 \\[0.5ex]
\hline
\end{tabular}
\end{minipage}%
\qquad
\begin{minipage}[b]{0.45\textwidth}
\centering
\begin{tabular}{| c | c c | c | c | c |}
\hline
\multirow{2}{*}{\rule{0pt}{4ex}Field}	& \multicolumn{5}{c |}{Representation} \\
\cline{2-6}
\rule{0pt}{3ex}			& $A_4$ & SU(5) & $\mathbb{Z}_9$ & $\mathbb{Z}_6$ & $\mathbb{Z}_4^R$ \\ [0.75ex]
\hline \hline
\rule{0pt}{3ex}%
$H_5$					& 1 & 5			& 0 & 0 & 0 \\
$H_{\bar{5}}$			& 1 & $\bar{5}$	& 2 & 0 & 0 \\
$H_{45}$	 			& 1 & 45 		& 4 & 0 & 2 \\
$H_{\overline{45}}$ 	& 1 & $\overline{45}$ & 5 & 0 & 0 \\
\rule{0pt}{3ex}%
$\xi$ 					& 1 & 1			& 2 & 0 & 0 \\
$\theta_2$				& 1 & 1			& 1 & 4 & 0 \\
$\phi_{\rm atm}$		& 3 & 1 		& 3 & 1 & 0 \\
$\phi_{\rm sol}$		& 3 & 1 		& 2 & 1 & 0 \\[0.5ex]
\hline
\end{tabular}
\end{minipage}
\caption{Superfields containing SM fermions, the Higgses and relevant flavons.}
\label{ta:SMF}
\end{table}

The SM fermions are contained within superfields $F$ and $T_i$. The MSSM Higgs doublet $H_u$ originates from a combination of $H_{5}$ and $H_{45}$, and $H_d$ from a combination of $H_{\overbar{5}}$ and $H_{\overbar{45}}$. Having the Higgs doublets inside these different representations generates the correct relations between down-type quarks and charged leptons. Doublet-triplet splitting is achieved by the Missing Partner mechanism \cite{Masiero:1982fe}.

The field $ \xi $ which gains a VEV $v_\xi \sim 0.06 M_{\mathrm{GUT}} $ generates a hierarchical fermion mass structure in the up-type quark sector through terms like $ v_u T_i T_j (v_\xi/M)^{6-i-j}$, where $v_u$ is the VEV of $H_u$. It also partially contributes to the mass hierarchy for down-type quarks and charged leptons and provides the mass scales for the right-handed neutrinos as discussed later. It further produces a highly suppressed $ \mu $ term $ \sim (v_\xi/M)^8 M_{\mathrm{GUT}} $.
The resulting symmetric Yukawa matrix for up-type quarks is
\begin{equation}
Y_{ij}^u = u_{ij} \left(\frac{\braket{\xi}}{M}\right)^{n_{ij}} \sim \pmatr{\tilde{\xi}^4 & \tilde{\xi}^3 & \tilde{\xi}^2 \\ &\tilde{\xi}^2 & \tilde{\xi} \\& & 1}
\label{upYuk}
\end{equation}
where $ \tilde{\xi} = \braket{\xi}/M \sim 0.1 $. 
The up-type Yukawa matrix $ Y^u $ is highly nondiagonal while the down-type and charged lepton Yukawa matrices $ Y^d \sim Y^e $, derived from terms like $ F \phi T H $, are nearly diagonal,
\begin{equation} 
	Y^d_{LR} \sim Y^e_{RL} \sim \pmatr{ \dfrac{\braket{\xi} v_{e}}{v_{\Lambda_{24}}^2} & \dfrac{\braket{\xi} v_{\mu}}{{v_{\Lambda_{24}}} {v_{H_{24}}}} & 0 \\[2ex] 0&\dfrac{{v_{H_{24}}} v_{\mu}}{M^2} & 0 \\[2ex] 0& 0& \dfrac{v_{\tau}}{M}}
\label{downYuk}
\end{equation}
where $v_{e, \mu , \tau}$ are charged lepton flavon VEVs as in Eq.\ref{ve},
while $ v_{\Lambda_{24}} $ and $ v_{H_{24}} $ are the respective VEVs of 
heavy Higgs $ \Lambda_{24} $ and $ H_{24} $, and we include the subscripts $LR$ to emphasise the role of the off-diagonal term to left-handed mixing from $Y^d$. The off-diagonal term in $Y^e$ also provides a tiny contribution to left-handed charged lepton mixing $ \theta_{12}^e \sim m_e/m_\mu $ which may safely be neglected. It also introduces \CP violation to the CKM matrix via the phase of $\braket{\xi}$.

The relevant terms in the superpotential giving neutrino masses are,
\begin{equation}
	W_\nu = y_1 H_{5} F\frac{\phi_\mathrm{atm}}{\braket{\theta_2}} N_\mathrm{atm}^c 
	+ y_2 H_{5} F\frac{\phi_\mathrm{sol}}{\braket{\theta_2}} N_\mathrm{sol}^c 
	+ y_3 \frac{\xi^2}{M_\Gamma} N_\mathrm{atm}^c N_\mathrm{atm}^c 
	+ y_4 \xi N_\mathrm{sol}^c N_\mathrm{sol}^c,
\label{eq:neutrinomassWmodel}
\end{equation}
where the $y_i$ are dimensionless couplings, expected to be $\mathcal{O}(1)$. The alignment of the flavon vacuum is fixed by the form of the superpotential, with $\phi_{\mathrm{atm}}$ and $ \phi_{\mathrm{sol}} $ gaining VEVs according to CSD(3) in Eq.\ref{CSD(n)}:
\begin{equation}
	\braket{\phi_{\mathrm{atm}}} = v_{\mathrm{atm}} \pmatr{0\\1\\1} , \qquad\qquad \braket{\phi_{\mathrm{sol}}} = v_{\mathrm{sol}} \pmatr{1\\3\\1}.
\end{equation}

This results in a low energy effective Majorana mass matrix of the form
in Eq.\ref{eq:mnu2} for $n=3$, namely,
\begin{equation}
	m^\nu = m_a 
	\left(
\begin{array}{ccc}
	0&0&0\\0&1&1\\0&1&1 
	\end{array}
\right)
	+ m_b e^{i\eta} 
	\left(
\begin{array}{ccc}
	1&3&1\\3&9&3\\1&3&1
	\end{array}
\right),
	\label{eq:mnu3}
\end{equation}
where $\eta$ is the only physically important phase, which depends on the relative phase between the first and second column of the Dirac mass matrix in the flavour basis.
The phase $\eta$ is responsible for \CP violation in both leptogenesis and neutrino oscillations.
We identify,
\begin{align}
	m_a = \left| \dfrac{ y_1^2 v_u^2 v_{\mathrm{atm}}^2  M_\Gamma}{y_3 \braket{\theta_2}^2 v_\xi^2} \right| , \qquad
	m_b = \left| \dfrac{ y_2^2 v_u^2 v_{\mathrm{sol}}^2  }{y_4 \braket{\theta_2}^2 v_\xi} \right|.
\end{align}

The Abelian flavour symmetry $\mathbb{Z}_9$ fixes the phase $ \eta $ to be one of the ninth roots of unity, through a variant of the mechanism used in \cite{Ross:2004qn}. The particular choice $\eta = 2\pi/3$ can give the neutrino mixing angles with great accuracy. Furthermore, this phase corresponds to 
$\delta_{\mathrm{CP}} \approx -\pi/2$, consistent with hints from experimental data.

\begin{table}[ht]
\renewcommand{\arraystretch}{1.2}
\centering
\begin{tabular}{| c | c  c  c | c  c  c  c  c  c |}
\hline
\rule{0pt}{4ex}%
$n$ 	& \makecell{$m_a$ \\ {\scriptsize (meV)}} & \makecell{$m_b$ \\ {\scriptsize (meV)}} & 
\makecell{$\eta$  \\ {\scriptsize (rad)}}  	& \makecell{$\theta_{12}$ \\ {\scriptsize ($^{\circ}$)}} & \makecell{$\theta_{13}$ \\ {\scriptsize ($^{\circ}$)}}  & \makecell{$\theta_{23}$ \\ {\scriptsize ($^{\circ}$)}} & \makecell{$\delta_{\mathrm{CP}}$ \\ {\scriptsize ($^{\circ}$)}} & \makecell{$m_2$ \\ {\scriptsize (meV)}} & \makecell{$m_3$ \\ {\scriptsize (meV)}} \\ [2ex] \hline 
\rule{0pt}{4ex}%
3 	& 26.57		& 2.684		& $ \dfrac{2\pi}{3} $	& 34.3		& 8.67		& 45.8		& -86.7		& 8.59		& 49.8 \\[1.7ex]
\hline
\end{tabular}
\caption{Best fit parameters and predictions for an $ A_4 \times SU(5) $ SUSY GUT with CSD(3) and a fixed phase $ \eta = 2\pi/3 $, as described in \cite{Bjorkeroth:2015ora}.
The spectrum is NO with 
lightest neutrino mass $m_1=0$ and hence the remaining Majorana phase (predicted but not indicated)
will be practically impossible to measure.}
\label{tab:model}
\end{table}

The relevant best fit parameters from our model are given in Table \ref{tab:model}, along with the model predictions for the leptonic mixing angles and neutrino masses, for $ \tan \beta = 5 $.

Using the above estimates, in \cite{Bjorkeroth:2015tsa} we estimated the baryon asymmetry of the Universe
(BAU) for this model resulting from $N_1$ leptogenesis:
\begin{equation}
	Y_B \approx 2.5 \times 10^{-11}\sin \eta \left[\frac{M_1}{10^{10} ~\mathrm{GeV}} \right].
\label{BAU}
\end{equation}
Using $\eta = 2\pi/3$ and the observed value of $ Y_B $ fixes the lightest right-handed neutrino mass:
\begin{equation}
	M_1 \approx 3.9 \times 10^{10} ~\mathrm{GeV}.
\end{equation}
Note that the phase $\eta$ controls the BAU via leptogenesis in Eq.\ref{BAU}.
The phase $\eta$ also controls the entire PMNS matrix, including all the lepton mixing angles as well as all low energy \CP violation.
The single phase $\eta$ is the therefore the source of all \CP violation in this model,
including both \CP violation in neutrino oscillations and in leptogenesis, providing a direct link between these two phenomena in this model. We not only have a correlation between the sign of the BAU and the sign of 
low energy leptonic \CP violation, but we actually know the value of the 
leptogenesis phase: it is $\eta = 2\pi/3$ which leads to the observed excess of matter over antimatter
for $M_1 \approx 4.10^{10}$ GeV
together with an observable neutrino oscillation phase $\delta_{\mathrm{CP}} \approx -\pi/2$.

\section{F-theory origin of SUSY GUTs with discrete family symmetry}
\label{F}

F-theory models have attracted considerable interest over the recent years~\cite{Beasley:2008kw}.
For example, SUSY GUTs based on $SU(5)$
have been shown to emerge naturally from F-theory.
However, in the F-theory context, the $SU(5)$ GUT group
is only one part of a larger symmetry. The other parts
manifest themselves at low energies as Abelian and/or non-Abelian discrete symmetries,
which can be identified as family symmetries,
leading to significant constraints in the effective superpotential
(for a review see e.g.~\cite{Leontaris:2012mh}).

In \cite{Antoniadis:2013joa} non-Abelian fluxes were conjectured to give rise to 
 non-Abelian discrete family symmetries in the low energy effective theory. 
 The origin of such a symmetry is the non-Abelian $SU(5)_{\perp}$
which accompanies $SU(5)_{GUT}$ at the $E_8$ point of enhancement. Whether a non-Abelian
symmetry survives in the low energy theory will depend on the geometry of the compactified space and the
fluxes present. The usual assumption is that the $SU(5)_{\perp}$ is first broken to 
a product of $U(1)_{\perp}$ groups which are then further broken by the action of discrete symmetries associated with the monodromy group. Instead it was conjectured in  
\cite{Antoniadis:2013joa} that  {\em non-Abelian fluxes} can break $SU(5)_{\perp}$ first to a non-Abelian discrete group $S_4$ then to a smaller group such as $A_4$, $D_4$ and so on
which act as a family symmetry group in the low energy effective theory~\cite{Karozas:2014aha}.
This could provide the origin of the $A_4\times  SU(5)$ SUSY GUT model discussed in
section~\ref{A4SU5}.

\section{Conclusion}
\label{conclusion}
In conclusion, although the reactor angle has been accurately measured, which rules out simple patterns of lepton mixing such as BM, TB and GR, it is still possible to have simple patterns of lepton mixing with the first or second column of the TB matrix preserved, namely TM1 or TM2, with atmospheric sum rules.
It is also possible to maintain BM, TB and GR mixing for {\it neutrinos} with
the reactor angle is due to charged lepton corrections, leading to solar sum rules.

Although adding right-handed neutrinos is a very simple and minimal thing to do, the number of right-handed (sterile) neutrinos is undetermined by anomaly cancellation, and their mass spectrum is completely unknown. The classic see-saw mechanism would correspond to having three right-handed neutrinos with masses in the range TeV-M$_{\rm GUT}$. 

Sequential dominance (SD) continues to provide an elegant and natural way to understand neutrino mixing angles, with the dominant right-handed neutrino couplings providing the atmospheric mixing angle,
the sub-dominant solar right-handed neutrino couplings providing the solar mixing angle and the decoupled right-handed neutrino couplings being irrelevant.
The main predictions of SD are a normal neutrino mass hiearchy and the bound on the 
reactor angle $\theta_{13} \lesssim m_2/m_3$, which indicated the potential largeness
of the reactor angle a decade before it was measured. Spurred on by 
the success of SD, recent
versions of constrained sequential dominance (CSD) 
have been proposed which explain 
why the reactor angle bound is saturated. A particular class of such models called CSD($n$) give a successful
desciption of the PMNS matrix in terms of a small number of input parameters
with good fits of lepton mixing angles for $n=3,4$.  

Turning to theories of flavour, a very promising approach is the combination of GUT and family symmetry. The large lepton mixing angles suggest some sort of discrete family symmetry at work, although not in the most simple direct way imagined before the reactor angle was measured. 
In particular the direct symmetry approach in which the symmetries of the mass matrices are directly embedded into the family symmetry, drives us to family symmetry groups in the $\Delta (6N^2)$ series with large $N$ values necessary in order to explain the reactor angle. 

One possibility is that only part of the symmetries of the mass matrices can be found in the family symmetry group, which is called the semi-direct approach. This allows
smaller family groups such as $S_4,A_4,A_5$ whose generators $S,T,U$
may only partly survive. For example if $T$ is broken but the Klein symmetry 
$S,U$ survives in the neutrino sector this would correspond to BM, TB or GR neutrino mixing but with charged lepton corrections, leading to solar sum rules.
If $T$ is preserved but $U$ is broken then this corresponds to TM1 or TM2 mixing
with atmospheric sum rules. 

An attractive alternative is the indirect approach where a small family symmetry such as $A_4$
is completely broken.
In this case new vacuum alignments are possible which can be used to give
interesting Yukawa couplings corresponding to different types of CSD($n$),
leading to highly predictive models. 
We have given two examples of such models,
namely an A to Z of flavour with Pati-Salam gauge group and a rather complete $A_4\times SU(5)$ SUSY GUT of flavour, which could originate from F-theory.
The A to Z model allows $N_2$ leptogenesis at the expense of large deviations from the 
CSD($4$) predictions, but still remains very predictive, with NO neutrino masses,
an atmospheric angle in the second
octant and a Dirac phase $\delta_{CP} \sim 20^{\circ}$, predictions that will be tested soon
by neutrino oscillation experiments.

The $A_4\times SU(5)$ SUSY GUT is consistent with $N_1$ leptogenesis
arising from the minimal (two-right handed neutrino) predictive seesaw model, 
and accurately reproduces the CSD($3$) predictions with $\eta = 2\pi /3$
being the 
only source of \CP violation for low energy neutrino physics as well as 
in the early Universe. This model therefore provides a direct link between 
matter-antimatter asymmetry and  \CP violation in neutrino oscillation experiments.
The neutrino masses and PMNS matrix (nine observables) are fixed in this model 
by three input parameters leading to the predictions in Table~\ref{tab:model}.
For example
the model predicts a NO spectrum with maximal atmospheric angle $\theta_{23}\sim \pi/4$ and leptonic 
\CP violation $\delta_{CP} \sim -\pi/2$ in agreement with current experimental hints.
Experiment will soon decide if this model is on the right track.

In conclusion, the discovery of neutrino mass and mixing continues to offer tantalising clues
that may help to unravel the mystery of fermion flavour, mass, mixing and \CP violation.
The SM is clearly unable to answer such questions, or provide an explanation of the origin of neutrino mass,
dark matter or matter-antimatter asymmetry. 
The history of physics suggests that the answer to these questions 
will involve symmetry. The largeness of atmospheric and solar mixing,
which resemble tri-bimaximal mixing, motivates the use of 
non-Abelian discrete family symmetries, where such approaches must and can 
allow Cabibbo sized reactor mixing as we have discussed.
The combination of GUTs and discrete family symmetry, together with spontaneous \CP violation,
continues to provide promising and testable candidate theories of flavour capable of answering the intruiging puzzles left in the wake of the SM. We have seen that such theories may provide a link between
matter-antimatter asymmetry and \CP violation in neutrino oscillation experiments,
accompanied by a set of precise predictions which are readily testable by neutrino oscillation experiments.

\vspace{0.1in}
SFK acknowledges partial support from the STFC Consolidated ST/J000396/1 grant and 
the European Union FP7 ITN-INVISIBLES (Marie Curie Actions, PITN-
GA-2011-289442). I would also like to congratulate my colleagues Takaaki Kajita who is the scientist in charge of  the University of Tokyo node of ITN-INVISIBLES and Arthur B. McDonald (Sudbury Neutrino Observatory)
on the joint award of the 2015 Nobel Prize for Physics 
``for the discovery of neutrino oscillations, which shows that neutrinos have mass''.

\end{document}